\documentclass[12pt,preprint]{aastex}

\usepackage{epsf}

\newcommand{\tx}[1]{\textrm{#1}}

\newcommand{\kms}{km~$\tx{s}^{-1}$}
\newcommand{\dv}{$r^{1/4}\,$}

\newenvironment{inlinefigure}{
\def\@captype{figure}
\noindent\begin{minipage}{0.999\linewidth}\begin{center}}
{\end{center}\end{minipage}\smallskip}

\newenvironment{inlinetable}{
\def\@captype{table}
\noindent\begin{minipage}{0.999\linewidth}\begin{center}}
{\end{center}\end{minipage}\smallskip}

\shorttitle{Dark Matter in Galaxy Clusters}
\shortauthors{Sand, Treu, Smith \& Ellis}

\begin{document}
\title{The Dark Matter Distribution in the Central Regions of Galaxy
Clusters: Implications for CDM}

\author{David J. Sand, Tommaso Treu\altaffilmark{1,}\altaffilmark{2}, Graham P. Smith \& Richard S. Ellis}
\affil{California Institute of Technology,
Astronomy, mailcode 105--24, Pasadena, CA 91125}
\email{djs@astro.caltech.edu}
\altaffiltext{1}{Hubble Fellow}
\altaffiltext{2}{Current Address: Department of Physics \& Astronomy, UCLA, Box 951547, Los Angeles, CA 90095-1547}

\begin{abstract}

We have undertaken a spectroscopic survey of gravitational arcs in a
carefully chosen sample of six clusters each containing a dominant
brightest cluster galaxy. We use these systems to study the relative
distributions of dark and baryonic material in the central
regions. Three clusters present both radial and tangential arcs and
provide particularly strong constraints on the mass profiles, whereas
the other three display only tangential arcs and act as a control
set. Following Sand et al.\ (2002), we analyze stellar velocity
dispersion data for the brightest cluster galaxies in conjunction with
the arc redshifts and lens models to constrain the dark and baryonic
mass profiles jointly. For the systems containing radial arcs we find
that the inner dark matter density profile is consistent with a 3-D
distribution, $\rho_{DM}\propto r^{-\beta}$, with logarithmic slope
$\langle\beta\rangle=0.52^{+0.05}_{-0.05}$ (68\% CL).  Similarly, we
find that the tangential arc sample gives an upper limit, $\beta<$0.57
(99\% CL).  Taking the 6 clusters together, the mean dark matter
distribution is inconsistent with the standard Navarro, Frenk \& White
(1997) value, $\beta$=1.0, at $>$99~\% confidence.  In addition, we
find considerable cosmic scatter in the $\beta$ ($\Delta\beta\sim$0.3)
values of the radial arc sample.  We find no evidence that systems
with radial arcs preferentially yield flatter dark matter profiles as
might be expected if they were a biased subset.  We discuss the
validity of our 1-D mass reconstruction method and verify its
conclusions by comparing with results of a more rigorous ray-tracing
code that does not assume axial symmetry. Our results extend and
considerably strengthen the earlier conclusions presented by Sand et
al. (2002) and suggest the relationship between dark and visible
matter in the cores of clusters is much more complex than anticipated
from recent simulations.

\end{abstract}

\section{Introduction}

The cold dark matter (CDM) paradigm has been extremely successful in
explaining observations of the universe on large scales at various
epochs, from that of the cosmic microwave background, through high
redshift studies of the $Ly\alpha$ forest to the distribution of
galaxies and clusters in local surveys (e.g.\  Percival et al.\ 2001,
Spergel et al.\ 2003; Croft et al.\ 2002; Bahcall et al.\ 2003). A
primary tool for making the necessary predictions is that of N-body
simulations which are now able to resolve structures on highly
non-linear scales so that the properties of dark matter (DM) halos can
be predicted on $\sim$kpc scales.

A central prediction arising from CDM simulations is that the density
profile of DM halos is universal in form across a wide range of mass
scales from dwarf galaxies to clusters of galaxies (e.g.\ Navarro,
Frenk \& White 1997; hereafter NFW97).  Within a scale radius,
$r_{sc}$, the DM density aymptotes to $\rho\propto r^{-\beta}$ while
external to $r_{sc}$, $\rho\propto r^{-3}$. The value of the
logarithmic inner slope, $\beta$, is still a matter of
debate. However, in nearly all studies, $\beta$ ranges between 1
(NFW97) and 1.5 (Moore et al.\ 1998; Ghigna et al.\ 2000; hereafter
referred to as the ``Moore'' slope for convenience). Recent work by
Power et al.\ (2003) and Fukushige et al.\ (2003) has suggested that
with proper account of the timestep, force accuracy and particle
number, the inner slope does not converge to a power law, as predicted
from lower resolution simulations, but instead becomes progressively
shallow at smaller radii.  Power et al.\ found $\beta$=1.2 at their
innermost reliable location.  Further work in this area will allow for
even more precise predictions of the form of DM halos.

An observational verification of the NFW97 (or Moore) form, via a
convincing measurement of $\beta$ and its scatter over various mass
scales, has proved controversial despite the motivation that it offers
a powerful test of the CDM paradigm. A major observational hurdle is
the importance of convincingly separating the baryonic and
non-baryonic components. Indeed, observations may guide the
interpretation of the numerical simulations, both because the
inclusion of baryons into simulations is difficult (e.g.\ Frenk 2002)
and because it is expensive computationally to simulate a sufficient
number of halos (with proper convergence) to characterize the expected
scatter in halo shapes.

Most of the observational effort has been directed via dynamical
studies of low surface brightness and dwarf galaxies as these are
thought to be DM dominated at all radii. However, analyses of the
various datasets have given conflicting values of $\beta$ and many of
the assumptions used have been questioned (see discussion by Simon et
al. 2003 and Swaters et al.\ 2003). Some studies have provided
evidence for cores of roughly constant density (e.g.\ de Blok et al.\
2001; de Blok \& Bosma 2002; Simon et al.\ 2003) whereas others find
their data are consistent with $\beta$=1 (e.g.\ van den Bosch \&
Swaters 2001; Swaters et al.\ 2003). Steep inner profiles with
$\beta\approx$1.5 seem to be ruled out.

In order to test the simulations convincingly, observations should not
be confined to mass scales probed by dwarf galaxies. Accordingly,
several attempts have been made to constrain the DM profiles of more
massive systems. Observations of spiral and early-type galaxies tend
to favor inner slopes that are shallower than predicted by CDM
simulations (Treu \& Koopmans 2002; Koopmans \& Treu 2003; Borriello
\& Salucci 2002; Borriello, Salucci, \& Danese 2003; Jimenez, Verde \&
Oh 2003), although the dominance of stellar mass at small scales makes
it difficult to achieve an accurate measurement of the dark halo
component.

More effort has been devoted to galaxy cluster mass scales. Most
common has been the use of X-ray observations of the hot intracluster
medium under the assumption of hydrostatic equilibrium. Whether
hydrostatic equilibrium is maintained in the inner regions, where
there are often irregularities and ``cooling flows'', remains an
important question (see Arabadjis, Bautz \& Arabadjis 2003).  Within
the context of the hydrostatic equilibrium assumption, many studies
have considered only a limited range of DM profiles, comparing, for
example, NFW (or Moore) fits with those of a non-singular isothermal
sphere (e.g.\ Schmidt, Allen, \& Fabian 2001; Allen, Schmidt, \&
Fabian 2002; Pratt \& Arnaud 2002). In general, X-ray analyses have
led to wide ranging results, with $\beta$ ranging from $\simeq$0.6
(Ettori et al.\ 2002) through $\simeq$1.2 (Lewis, Buote \& Stocke
2003) to $\simeq$1.9 (Arabadjis, Bautz \& Garmire 2002).  However,
when using just X-ray data alone, it is difficult to account for the
stellar mass of a central brightest cluster galaxy (BCG), which leads
to complications in interpreting the shape of the DM density profile
at small radii (Lewis et al.\ 2003).  In fact, although the stellar
component is small in terms of the total mass of the system, it can
dominate the mass density at small radii, and can mimic a cuspy DM
halo if it is not taken into proper account.

Gravitational lensing offers a particularly promising probe of the
total mass profile. Projected mass maps of the inner regions of
clusters constrained by strongly lensed features of known redshift
have been compared with CDM predictions (Tyson, Kochanski, \&
Dell'Antonio 1998, Smith et al.\ 2001; hereafter S01).  By using weak
lensing, and stacking a sample of clusters, Dahle, Hannested, \&
Sommer-Larsen (2003) found an inner DM slope roughly in agreement with
CDM predictions (albeit with large uncertainties).  Recently a
combined strong and weak lensing analysis of Cl0024 has confirmed the
prediction of CDM numerical simulations that the DM density profile
falls off like $\rho\propto r^{-3}$ at large radii, strongly ruling
out a density profile that falls off like an isothermal mass
distribution (Kneib et al.\ 2003).  A combined strong and weak lensing
analysis has also been used to constrain the inner DM slope in the
cluster MS2137-23 (Gavazzi et al.\ 2003; hereafter G03), although the
precise value of the slope depends on the assumed stellar
mass-to-light ratio of the BCG.

As with the X-ray studies, gravitational lensing alone is unable to
separate the baryonic (luminous) and non-baryonic (dark)
components. Given the observational evidence that BCGs often lie at
the bottom of the cluster potential in regular, non-interacting
systems (e.g. S01, G03, Jones et al.\ 1979), the dynamics of the
stellar component offers a valuable route to resolving this
problem. In practice, the stellar kinematics of the BCG provides an
additional measure of the total mass at small radii. In work by
Dressler (1979), the velocity dispersion profile of the BCG in Abell
2029 was found to rise significantly at large radii and this was taken
as evidence that the cluster DM halo was being probed.  More recently,
Kelson et al. (2002) have measured an extended velocity dispersion
profile in the BCG in Abell 2199, for which they concluded that the
best-fitting DM density profile for the cluster was shallower than
NFW.  Miralda-Escud\'e (1995) first suggested that a combination of
lensing and stellar velocity dispersion measurements could separate
the luminous and dark components in the inner regions of clusters (see
also Natarajan \& Kneib 1996 for a lensing + dynamics analysis of
Abell 2218). This article highlighted the system MS2137-23 which, at
the time, was unique in containing both radial and tangential
gravitational arcs.

In an earlier paper (Sand, Treu \& Ellis 2002; hereafter STE02) we
combined a simple axisymmetric lensing model of MS2137-23 with stellar
velocity dispersion measurements of the BCG to place strong
constraints on the inner slope of the DM density profile.  The
resulting $\beta$ value was markedly inconsistent with
$\beta\geqslant$ 1 and we demonstrated carefully how the combination
of lensing and dynamics offers superior constraints to those provided
by either method alone.

The goal of this paper is to extend the results of STE02 to a larger
sample of six galaxy clusters.  In addition to MS2137-23 we consider
two additional systems containing both radial and tangential
gravitational arcs.  The three other clusters contain only a
tangential arc and analysis of this subsample offers a valuable
control from which we expect to deduce whether selecting the rarer
systems with radial arcs might bias our conclusions towards flatter
inner slopes. As a further test on the robustness of our results, we
check our lensing model by dropping the assumption of radial symmetry.

A plan of the paper follows.  In \S 2 we discuss how the sample of
clusters was chosen.  In \S 3 we discuss the archival Hubble Space
Telescope and further infrared imaging observations and how we derived
the location of the critical lines and the surface photometry of the
BCGs. In \S 4 we present spectroscopic measurements made with the Keck
telescope which delivered the redshifts of the gravitational arcs and
the stellar velocity dispersion profile of the BCGs.  We discuss our
analysis of the DM density profiles in the context of the assumed mass
model for both the radial+tangential and tangential-only arc
subsamples in \S 5. In \S 6 we present a thorough discussion of
possible systematic uncertainties associated with our method. In \S7
and \S8 we discuss and summarize our results, respectively.

Throughout this paper, we adopt $r$ as the radial coordinate in
3-D space and $R$ as the radial coordinate in 2-D projected space.
We assume H$_0$=65~km s$^{-1}$Mpc$^{-1}$, $\Omega_m=0.3$ and
$\Omega_{\Lambda}$=0.7.

\section{Sample Selection}

The aim of this project is to combine constraints from the velocity
dispersion profile of a BCG with those from gravitational lensing to
measure the slope of the inner DM density profile in galaxy clusters,
as described in STE02.  Two important simplifying assumptions inherent
to our method are that the BCG lies at the bottom of the cluster
potential, and that the BCG is a purely pressure supported system,
whose dynamics can be described by the Jeans' equation. For this
reason, a suitable galaxy cluster for this project must have a
dominant, relatively isolated central galaxy (coincident with the
cluster's center of mass) with nearby strong lensing features and no
indications of significant substructure or a significantly elongated
potential. Radial gravitational arcs -- albeit uncommon -- are
particularly valuable since they constrain directly the derivative of
the total enclosed mass (e.g. STE02).

In order to find a sample of suitable targets, we undertook an
exhaustive search of the Hubble Space Telescope (HST) Wide Field and
Planetary Camera 2 (WFPC2) archive for radial gravitational arcs in
galaxy clusters.  In summary, all galaxy cluster pointings in the
redshift range $0.1<z<1.0$ were retrieved from the HST archive,
$\sim$150 different clusters in all.  This is the first search of its
kind and has yielded $\sim$15 candidate radial arc systems.  Dozens of
smaller lensed features have been uncovered as well, due to the high
angular resolution of HST.

We performed spectroscopic follow-up at the Keck Telescope of many
candidate lensing systems, with emphasis on systems with both radial
and tangential arcs.  Spectroscopic confirmation is particularly
important for radial arc candidates. In fact, since they normally
occur in the very inner regions of galaxy clusters, they can easily be
confused with optical filaments associated with cluster cooling
flows. Indeed, several radial features proved to be contaminant
optical filaments at the cluster redshift.  A description of the
search, the complete catalog and spectroscopic identifications will be
described in a follow-up paper (Sand et al.\ 2004, in preparation).

In this paper we focus on a sample of six spectroscopically confirmed
lensing clusters, for which we have also obtained a stellar velocity
dispersion profile of the BCG (Table~1). The sample includes three
galaxy clusters with radial and tangential arcs and three clusters
with just tangential arcs, one of which does not have HST imaging
(MACS 1206; Ebeling et al.\ 2004, in preparation).

\section{Imaging Data and Analysis}

This section describes the two measurements that are to be made from
the imaging data to determine the cluster mass distribution: the
surface brightness profile of the BCG and the positions of the lensing
critical lines as inferred from the location of symmetry breaks in the
giant arcs.  In addition, two of the six BCGs in our sample (RXJ 1133
and Abell 383) have obvious dust lanes.  K-band images of the cluster
centers were used to correct for the dust lanes and obtain the
unreddened surface brightness profile of the BCG as described in \S
3.2.  Table~1 is an observation log of all the optical/NIR
observations.

\subsection{Optical Data}

Archival HST imaging is available for five of the six clusters.  A
gunn I-band image in good seeing conditions ($0\farcs7$ FWHM) was
obtained for the final cluster, MACS1206, using the Echelle
Spectrograph and Imager (ESI; Sheinis et al.\ 2002) at the Keck-II
Telescope.  Figure~1 shows the inner regions of the six clusters and
their accompanying gravitational arcs.  Superimposed on the images are
the spectroscopic slit positions that will be described in \S 4.

Since our sample results from an extensive HST archive search, no
specific observing strategy is common to all clusters.  The HST
observing strategies fall into two categories: 1) multiple-orbit
observations separated by integer pixel dithers (MS 2137-23, Abell
383, and Abell 963) and 2) single orbit SNAP observations comprising
two CR-SPLIT, undithered exposures (Abell 1201, RXJ 1133).
Accordingly, two different data reduction procedures were employed.

The clusters with dithered exposures were pipeline processed with the
DITHER package (Fruchter \& Hook 2002) in {\sc iraf} to remove cosmic
rays, correct for the undersampling of the point spread function, and
to shift and combine the frames.  The effective resolution for the
images (F606W and F702W) was $\sim0\farcs15$.  The undithered targets
were processed by first applying the {\sc iraf} task {\sc warmpix}
using the table data supplied by the WFPC2 website for the dates of
the observations.  The images were then combined with the task {\sc
crrej} to remove cosmic ray hits.  A small number of residual cosmic
rays were removed with the {\sc iraf} task {\sc lacosmic} (van Dokkum
2001).

The single, gunn I band exposure of MACS 1206 was reduced in a
standard way with cosmic ray removal being performed by the {\sc iraf}
task {\sc lacosmic} (van Dokkum 2001).  Photometric calibration, good
to 0.03 mag, was obtained from two photometric standard star fields
(Landolt 1992).

\subsection{Near--infrared Data}

Dust features in BCGs are common (see e.g.\ Laine et al.\ 2002), but
hinder attempts at measuring structural parameters.  To correct for
internal dust extinction in Abell 383 and RXJ 1133, we observed the
two BCGs with NIRC on the Keck I Telescope in the $K_{s}$ band (see
Table~1 for the observing log).  The data were reduced in a standard
manner using {\sc iraf} tasks to dark subtract, linearize,
flat--field, align and combine the individual frames.  The
flat--fields were created from a rolling median of the adjacent
science frames.  The point--spread--function of both final reduced
frames has ${\rm FWHM}\simeq0.6''$.

A dust correction is obtained as described in Treu et al.\ (2001) and
Koopmans et al.\ (2003).  Briefly, we first assume that dust has a
neglible effect at large radii and that any intrinsic BCG color
gradient is small.  Then we smooth the HST image to the resolution of
the K-band image, and compute an extinction map in the observer frame:
\medskip
\begin{equation}
\label{eq:extinct}
E_{HST,K}(x,y) = \mu_{HST}(x,y) - \mu_{K}(x,y)  - \mu_{HST,K}(\infty),
\end{equation}
where $\mu_{HST}(x,y)$ and $\mu_{K}(x,y)$ are the surface brightness
in a given pixel and $\mu_{HST,K}(\infty)$ is the color at large BCG
radii.  Adopting the Galactic extinction law of Cardelli, Clayton, \&
Mathis (1989; $R_{V}=3.1$) we find the following relations between the
absorption coefficients in the individual bands and the color excess
(Eqn 1.). For Abell 383 $A_{F702W}=1.199E_{F702W,K}$ and
$A_{K}=0.199E_{F702W,K}$. For RXJ1133 $A_{F606W}=1.164E_{F606W,K}$ and
$A_{K}=0.164E_{F606W,K}$.

The correction removes any visible trace of the dust lane.  However,
it reveals that the BCG in Abell 383 has a very close, compact
companion.  For the purpose of surface photometry analysis, the close
companion is easily dealt with by fitting it simultaneously to the
BCG. The companion is $\sim0\farcs7$ from the center of the BCG and is
$\sim$5.5 mag fainter with $R_{e}\approxeq0\farcs5$.  The spectrum of
the companion and the BCG cannot be distinguished in the ESI spectrum
so no relative velocity can be measured.  Assuming that the relative
masses of the two galaxies is proportional to their flux ratio, the
companion galaxy should not effect our later dynamical analysis.

\subsection{Surface brightness fitting}

Total magnitudes and effective radii ($R_{\rm e}$) were measured from
our (dust corrected) optical images by fitting two dimensional \dv
surface brightness profiles as described in STE02 using the software
developed by Treu et al.\ (1999, 2001). For the purpose of the
fitting, \dv models were convolved with artificial Point Spread
Functions (PSFs). Tiny Tim (Krist 1993) PSFs were used for the HST
images, while gaussian PSFs were adopted for the ground based
images. Note that uncertainties in the artificial PSF have negligible
impact on the determination of the effective radii which are always
much larger than the PSF HWHM (c.f. Treu et al. 2001).  Figure~2 shows
the measured surface brightness profile along with the best fitting
\dv fit (PSF convolved).

Observed magnitudes were corrected for galactic extinction using the
$E(B-V)$ values and extinction coefficients calculated by Schlegel,
Finkbeiner \& Davis (1998). Finally, observed magnitudes were
transformed to rest frame absolute magnitudes through the standard
filter that best matches the observed bandpass through a K-color
correction as in STE02 and Treu et al.\ (1999,2001). Typical error
estimates on the transformation are of order 0.05 mag.  All BCG
photometric results (both rest and observed frame) are listed in
Table~2.

\subsection{Critical line determination}

Crucial to the simple lensing method that we describe in
\S~\ref{sec:method} is the location of the lensing critical line
(either radial or tangential).  Formally, the critical lines of a lens
model are those regions where the magnification of the images diverge,
although this does not occur in practice due to the extended size of
the source. In addition to being strongly magnified, objects near the
radial critical line will be distorted strongly in the direction
radial to contours of constant density, while objects near the
tangential critical line will be distorted tangentially.  Typically,
giant bright arcs are the result of multiple highly magnified merging
images. Two merging images with opposite parity often bracket a
critical line.

In this work, the critical line position was chosen by visual
inspection, either in between two merging images or near strongly
distorted arcs.  No prior lensing analysis was done, although for
those systems that have published lens models this extra information
was taken into account.  The critical line positions including
conservative estimates of the uncertainties are listed in
Table~\ref{tab:geo}.  Note that the radial critical line uncertainties
are larger due to contamination by the bright BCGs and the radial
nature of these arcs.  In contrast, the tangential critical line
uncertainties are within a factor of $\sim$2-3 of the seeing disk.

\section{Spectroscopic Data and Analysis}

All spectroscopic measurements -- yielding arc redshifts and/or a BCG
velocity dispersion-- were made with either the Low Resolution Imager
and Spectrograph (LRIS) on Keck I (Oke et al.\ 1995) or ESI on Keck II
(Sheinis et al.\ 2002).  Table~\ref{tab:specsum} summarizes the
spectroscopic observations for each cluster.

\subsection{Data Reduction}

The LRIS data were reduced in a standard way with bias-subtraction,
flat-fielding and cosmic ray rejection.  Wavelength calibration was
performed using calibration arc lamps and unblended sky lines.  The
instrumental resolution for the 600/5000 grism (blue arm, 560
dichroic) used for the velocity dispersion measurement of Abell 963
was measured to be 175 km s$^{-1}$ from unblended night sky lines. For
the ESI observations, a set of IRAF tasks (EASI2D) were developed for
the specific goal of removing echelle distortions while preserving the
two-dimensional shape of the spectrum. The instrumental resolution of
the reduced 2D spectra for the $1\farcs25\times20''$ slit was measured
to be 32 km s~$^{-1}$ from unblended night sky lines and the spatial
scale ranges from 0.12 to 0.17 arcsec/pixel from the bluest to the
reddest order. 

EASI2D consists of the following steps: 1) Bad column interpolation,
debiasing, and an initial flat-fielding are performed on the entire
two dimensional spectrum. 2) The curved echelle orders are mapped
using multi-hole exposures (spaced a constant 2.68 arcseconds apart
(Goodrich, R \& Radovan, M, private communication)) and multiple
stellar exposures at prescribed positions along the slit.  3) Each
order is rectified using the {\sc iraf} task {\sc transform}
(conserving counts).  Arc lamps and twilight sky flats are rectified
along with science frames for further calibrations.  4) Sky flats are
used to correct for the non-uniform sensitivity (slit function) in the
spatial direction of the individual echelle orders.  5) After
rectification, each exposure of each order is separated and cosmic ray
cleaned using {\sc LACOSMIC}.  6) After rectification, each order is
separated and wavelength calibrated individually. 7) Sky subtraction
is performed on each order interactively with a low order polynomial
fit along the spatial direction to blank regions of the
slit. Alternatively, sky subtraction can be achieved by subtracting
appropriately scaled dithered science exposures from each other. This
second method is preferable beyond $\sim$ 7000\AA\, where sky emission
lines are strongest. 8) If needed, one dimensional spectra can be
extracted from the two-dimensional spectra of each order and combined
on a single spectrum. This step was typically undertaken only for the
kinematic template star spectra (see \S~\ref{sec:kine}).

\subsection{Redshift measurements and stellar kinematics}\label{sec:kine}

Table~1 (BCGs), Table~\ref{tab:geo} (gravitational arcs) and
Figure~\ref{fig:spectra} details all of the redshift measurements
made.  Many of these measurements were dependent on the high spectral
resolution of ESI, since many emission lines were buried in the OH sky
background.  All arc redshift identifications are based on the
detection of the [OII]3726,3729 doublet in emission.  Note that, for
the purpose of this work, we are only using the ``northern'' arc of
Abell 963 (Ellis, Allington-Smith \& Smail 1991).  Abell~383 has also
been studied extensively by S01 (see also Smith 2002 and Smith et al.\
2004, in prep.), who obtained a spectroscopic redshift for the
tangential arc in this cluster.  We add to S01's study by measuring
spectroscopic redshifts for both the radial arc and a different
portion of the tangential arc.  These new data provide more stringent
constraints on the gravitational potential of this cluster (\S6.1).

We now describe the measurement of the line of sight velocity
dispersion profile of the BCGs. The 2D spectra were summed into
spatial bins corresponding approximately to the seeing during the
observation, thus ensuring that each velocity dispersion measurement
is approximately independent, and increasing the signal-to-noise ratio
per spatial bin.  In Abell 383, an entire side of the spectrum was
avoided due to the interloping galaxy (\S 3.2) and the presence of the
dust lane.  Also, we avoided the side of the BCG in RXJ 1133 effected
by the dust lane.  For Abell 1201 two slightly different position
angles were used.

Following well established procedures (Franx 1993; van Dokkum \& Franx
1996; Treu et al.\ 1999,2001; Kelson et al.\ 2000; van Dokkum \& Ellis
2003; Gebhardt et al.\ 2003), the velocity dispersion for each spatial
bin is measured by comparing stellar templates (appropriately
redshifted and smoothed to the instrumental resolution of the galactic
spectra) broadened by Gaussian line profiles with the galactic
spectrum.  The fit is performed using the Gauss-Hermite Pixel Fitting
Software (van der Marel 1994) in pixel space to allow for easy masking
of emission lines and regions of high night sky residuals.  All
velocity dispersion measurements were taken from spectral regions
around the G band absorption feature.  Also measured with the
Gauss-Hermite Pixel Fitting Software was the relative velocity profile
of each BCG.  There is no evidence of rotation (within the
uncertainties) in any of the BCGs and so we will assume in our
analysis that the systems are completely pressure supported.

For each BCG, the velocity dispersion profile was measured using all
of the available stellar templates with a variety of continuum fits.
The stellar template that yielded the lowest $\chi^{2}$ was adopted as
the best fit.  Table 5 contains the tabulated velocity dispersion
measurements obtained.  Listed uncertainties are the sum in quadrature
of a random component (taken from the output of the Gauss-Hermite
Pixel Fitting Software) and a systematic component due to template
mismatch (the rms of the velocity dispersion obtained from all
templates).

In order to measure accurate velocity dispersions, spectra with
sufficient S/N are required.  In general, the minimum S/N needed
depends both on the instrumental resolution and the velocity
dispersion to be measured.  A higher S/N is needed as the velocity
dispersion becomes comparable to and less than that of the
instrumental resolution (e.g.\ Treu et al.\ 2001; Jorgensen, Franx \&
Kj{\ae}rgaard 1995).  Since the typical central velocity dispersion of
a BCG is $\sigma\sim$ 300-400 \kms (e.g.\ Fisher, Illingworth, \&
Franx 1995) compared to the 32 (175) \kms\ resolution of the
$1\farcs25$ ($1\farcs50$) ESI (LRIS) slit the velocity dispersion
measurement should be reliable down to low S/N.  In order to verify
this numerically for the case of the ESI configuration, high S/N
template spectra were broadened by Gaussian line profiles to known
velocity dispersions ($\sigma= $50, 100, 150, 200, 250, 300, 350 \kms)
and Poisson noise was added (S/N= 5, 7, 10, 12, 15, 20) for a hundred
different realizations.  The velocity dispersion of these broadened,
noisy spectra was then recovered using an area around the G band
absorption feature with the Gauss Hermite Pixel Fitting software with
both the same template (K0III) and a template with a different
spectral type (G9III template used on a broadened K0III spectrum).
Using the K0III template on the broadened K0III spectrum recovers the
expected velocity dispersion with formal uncertainties less then
10$\%$ down to S/N of 5 (the exception being the 50 \kms measurement
with S/N=5, which had an average formal uncertainty of 8 \kms).  Using
the G9III template on the K0III broadened spectrum did lead to
systematic offsets from the input velocity dispersion of up to
6.5$\%$, comparable to the formal uncertainty in the measurement.
Thus, we will take into account possible uncertainties with regard to
template mismatch in the data both in our final uncertainties and when
deriving mass profiles.

\section{Analysis and Results}

We now combine the observed photometric and spectroscopic properties
of the BCG and giant arcs to constrain the luminous and DM
distribution in the central region of the clusters.  In particular,
the goal of this analysis is to determine the range of inner DM
density slopes permissible and to compare these to predictions from
numerical CDM simulations.  The key to this method is to combine the
constraints on the mass from the velocity dispersion profile with that
from lensing to get a better overall measurement than can be made with
each individual technique.  The method used is identical to that
employed by STE02, with improved numerical accuracy.  There is no
change in our basic conclusions on MS2137-23 from that work.

\subsection {Mass Model and overview of the fitting procedure}\label{sec:method}

We adopt a spherically symmetric two component mass model comprising
the BCG and cluster DM halo.  We assume that the BCG is coincident
with the bottom of the cluster potential.  To describe the luminous
BCG component we used a Jaffe (1983)
\begin{equation}
\label{eq:jaffe}
\rho_L(r)=\frac{M_{L} r_{J}}{4 \pi r^{2} (r_{J} + r)^{2}},
\end{equation}
mass density profile with total mass $M_{L}$ and $R_{e}$ =
0.76$r_{J}$.  A Jaffe profile reproduces well the actual surface
brightness profiles of the BCGs in our sample and has the extra
advantage of giving analytic solutions to the surface brightness
profile and line of sight velocity dispersion.  We assume that a
single mass-to-light ratio describes well the stellar component of the
mass.  For a given stellar $M_{*}/L$ both $M_{L}$ and $r_{j}$ can be
deduced from the observed surface photometry.  We also investigated
the effects on the inner DM slope by changing the luminous mass
density profile to a Hernquist profile (Hernquist 1990) in
\S\ref{sec:test} as a robustness check of our results.  Note that the
PSF convolved Hernquist and Jaffe luminous matter distributions
bracket the observed data in all of our clusters (see \S~6.3 for a
discussion).

The cluster DM halo is modeled as
\begin{equation}
\label{eq:gnfw}
\rho_d(r)=\frac{\rho_{c} \delta_{c}}{(r/r_{sc})^{\beta}\left[1+(r/r_{sc})\right]^{3-\beta}},
\end{equation}
which is a generalization of the numerically simulated CDM halos, with
$\rho_{c}$ being the critical density and $\delta_{c}$ a scaling
factor.  This density profile asymptotes to $r^{-\beta}$ at $r \ll
r_{sc}$ and $r^{-3}$ at $r \gg r_{sc}$.  For values of $\beta = 1,
1.5$, the DM density profile is identical to that found by NFW and
nearly identical to that of M98, respectively; and thus using this
general form for the DM halo allows for direct comparison to numerical
results.

Considering the observations made, there are four free parameters in
our mass model: (1) the stellar mass-to-light ratio, $M_{*}/L$, (2)
the inner slope of the DM profile $\beta$, (3) the DM density scale
$\delta_{c}$, and (4) the DM scale radius $r_{sc}$.  In general,
$r_{sc}$ is much larger (greater than 100 kpc) than our most distant
mass probe, tangential gravitational arcs (Bullock et al.\ 2001;
Tasitsiomi et al.\ 2003; see also Wu \& Xue 2000).  Given this, the
location of the critical lines then only depend slightly on $r_{sc}$,
since it is the projected mass that is important.  In our modeling we
set $r_{sc}$=400 kpc leaving only three free parameters.  The value of
$r_{sc}$ was chosen as a typical value seen in galaxy cluster
numerical simulations for a typical cluster with virial radius of
$\sim$2 Mpc (see discussion in \S~6.3).  Much larger values of
$r_{sc}$ seem to be ruled out on observational grounds as well
(Gavazzi et al. 2003; Kneib et al. 2003; Wu \& Xue 2000).  In \S 6.3
we show that allowing $r_{sc}$ to vary within reasonable bounds as
proscribed by CDM simulations has a small effect on our $\beta$
measurement, but that the effect is comparable to our other
systematics.

We discuss in detail the analysis of both the lensing and velocity
dispersion data in the Appendix.  We briefly describe the method here.
By comparing the observed position (and its uncertainty) of
gravitational arcs with the predicted position of the arcs (given a
set of free parameters, \{$M_{*}$/L,$\beta$, $\delta_{c}$\}) a
likelihood function can be calculated over the appropriate parameter
space.  Similarly, the observed velocity dispersion profile (which
depends on the mass enclosed at a given radius and the relative
contribution from luminous and dark matter) for a BCG in a cluster can
be compared with that expected for a given set of free parameters
(taking into account the seeing and spatial binning of the
observations) and another likelihood function can be calculated.  The
total likelihood for a given set of free parameters is simply the
product of the lensing likelihood and the stellar kinematics
likelihood.  In the next two subsections we discuss how we use these
likelihood functions to place confidence limits on the inner DM
density slope, $\beta$.

\subsection{Radial Arc Results}

The three clusters in our sample with radial and tangential arcs
(hereafter known as the radial arc sample) allow for strong
constraints to be placed on the DM density profile.  The tangential
arc provides a measurement of the projected mass enclosed at a given
radius and the radial arc gives a measurement of the derivative of the
projected mass enclosed at its radius.

Assuming that all underlying distributions are normal, we can use a
$\chi^{2}$ minimization technique to obtain confidence contours on our
parameter estimates.  This involves simply taking the $\chi^{2}$
difference at any point in parameter space with respect to the minimum
$\chi^{2}$.  Confidence contours in the $M_{*}$/L-$\beta$ plane allow
one to visualize acceptable $\beta$ values as a function of the mass
of the BCG.  After marginalization with respect to $\delta_{c}$,
the 68, 95, and 99$\%$ confidence contours in the $M_{*}$/L-$\beta$
plane were placed at a $\Delta \chi^2$=2.30, 6.17, and 9.21
respectively.  Figure 3 (top row) shows the contours obtained for
our radial arc sample.

To constrain the inner slope $\beta$ alone, we further marginalized
with respect to $M_{*}$/L. The resulting probability distribution
function (PDF) for $\beta$ for all three clusters is shown in
Figure~\ref{fig:radpdf}. We adopt the peak of the distribution as best
estimate of the parameters. Confidence intervals are obtained by
integrating the PDF above a threshold such that the total area under
the curve is 68\% (95\%) of the total. Doing this, we found
$\beta$=$0.57^{+0.11}_{-0.08}$ ($^{+0.25}_{-0.17}$) for MS2137-23,
$0.38^{+0.06}_{-0.05}$ ($^{+0.12}_{-0.11}$) for Abell 383, and
$0.99^{+0.18}_{-0.14}$ ($^{+0.28}_{-0.28}$) for RXJ 1133.

Note immediately that the intervals for the individual clusters do not
overlap at the 68\% level. Therefore we conclude that there is
significant intrinsic scatter in the inner slopes of the DM halos.  To
assess the scatter in $\beta$ values that we find in the radial arc
sample, we calculate the standard deviation without account of the
corresponding PDF and find $\Delta\beta\sim$0.3.  The scatter and its
possible consequences will be discussed in \S~\ref{sec:scatdisc}.

Having noted the existence of significant intrinsic scatter, we can
determine the average inner slope of DM by looking at the joint radial
PDF, obtained as the product of the three individual distributions
(shown in Figure~\ref{fig:radpdf} as a solid line; note that this
measure is analogous to the weighted average).  We find that the
average inner slope and related uncertainty are
$\beta=0.52^{+0.05}_{-0.05}$ ($^{+0.11}_{-0.10}$). Assuming that our
sample of clusters is representative of the entire cluster population
this means the average slope is inconsistent at $>99\%$ CL with both
the NFW and Moore profile.

\subsection{Tangential Arc Results}

Before discussing the radial arc results any further we consider the
issue of sample selection bias in more detail. Are radial arc clusters
a representative subsample of relaxed clusters as far as DM inner
slopes are concerned? It is well known that total density
distributions that are steeper than $\rho \propto r^{-2}$ do not
produce radial arcs (e.g.\ Hattori, Kneib \& Makino 1999).  Thus, if
there is a wide range in the distribution of inner slopes, by
selecting radial arc systems we might be rejecting the more cuspy
systems.  This bias (hereafter the radial arc selection bias) might be
exacerbated by the fact that the radial arcs in our sample are buried
in the BCG, a steep density profile in its own right.  We investigate
how robust our results on $\beta$ are with respect to our choice of
luminous density profile in \S~\ref{sec:test}.

A clean and powerful way to address this issue is to obtain a control
sample of tangential arc-only systems (hereafter the tangential arc
sample). This will enable us to determine if the radial arc systems
appear to be outliers in the general cluster population. At the same
time this tangential arc sample will provide an additional -- albeit
less precise -- measurement of the DM inner slope.

The bottom row in Figure~3 and \ref{fig:tanpdf} display the results
for the tangential arc sample.  This was subject to the same analysis
as for the radial arc sample with the exception that we adopted a
prior to ensure that the DM profile is monotonically declining with
radius ($\beta \geqslant 0$). Note in fact that the results always go
toward $\beta$=0 for the tangential arc sample, at variance with the
results for the radial arc sample.  In fact, the shapes of the
confidence contours in the $M_{*}$/L-$\beta$ plane are markedly
different from the analogous contours for the radial arc sample (see
\S~\ref{sec:firstsum}).  We calculated upper limit confidence levels
on $\beta$, since the shape of the probability distribution function
lends itself to this type of interpretation.  The 68\% (95 and 99\%)
upper limits are $\beta = $0.29 (0.62,0.82), 0.40 (0.67, 0.77), and
0.43 (0.80, 0.97) for Abell 1201, MACS 1206, and Abell 963
respectively.  The joint tangential arc distribution has 68, 95, and
99\% confidence upper limits of $\beta$=0.20, 0.43, and 0.57
respectively.

Is the radial arc sample probing an outlier population of galaxy
clusters due to the fact that radial arcs cannot form in systems with
density distributions steeper than $\rho \propto r^{-2}$?  In the
following we assume that the joint distribution for each sample is a
fair representation of the underlying distribution, despite the sample
size.  As can be seen from Figure \ref{fig:tanpdf}, the radial arc
sample does not have a shallower DM density profile than the
tangential arc sample, as would be expected if there was a radial arc
bias.  To compare the two samples we convolved the radial and
tangential arc sample probability distribution functions in order to
compute the probability distribution function for the variable
$\beta_{r} - \beta_{t}$, where the subscripts represent the radial and
tangential arc sample values of $\beta$.  Due to the one-sided nature
of the tangential arc probability distribution function, it is
appropriate to use upper limits to quantify the confidence region of
the variable $\beta_{r} - \beta_{t}$.  The value of $\beta_{r} -
\beta_{t}$ is less than 0.45 and 0.57 with 68 and 95\% confidence,
respectively.  The probability that $\beta_{r} - \beta_{t}$ is less
than 0 (as would be expected if there was a radial arc bias) is
$\sim2\%$.  There is no indication of radial arc bias, and the radial
and tangential arc samples are reasonably consistent given the small
number of systems.

\subsection{Summary of results}\label{sec:firstsum}

We have presented new measurements of the inner slope ($\beta$) of DM
halos in clusters of galaxies, considering a sample of three radial
arc systems and a sample of three tangential arc systems in carefully
chosen relatively relaxed clusters.

The main results from the radial arc systems are: (i) the average
$\langle\beta\rangle=0.52\pm0.05$ is much smaller than that suggested
by numerical DM only simulations (either NFW or Moore); (ii) our
precision allows us to determine a first measurement of the intrinsic
scatter in $\beta$, which we estimate to be $\Delta\beta\sim$0.3; and
(iii) individual clusters can be as cuspy as NFW (RXJ1133). The
results from the tangential arc sample confirm and reinforce our
findings: (i) the upper limit to the average slope is $\beta$=0.57
(99\% CL), again much smaller than numerical simulations (NFW or
Moore); (ii) although with larger uncertainties, the results from the
tangential arc sample are statistically consistent with those from the
radial arc sample, confirming that our results are not affected by a
radial arc selection bias.

Before moving on to discuss in detail the comparison with numerical
simulations and consider the broader implications of these results (\S
\ref{sec:disc}), we need to address two further issues. First, we
would like to discuss in greater detail our method, understanding at
least qualitatively some of its features. This will hopefully provide
an element of physical intuition in addition to the statistical
anlysis. Secondly, we need to make sure that systematic uncertainties
are not dominating our error budget, which so far includes only random
uncertainties. The first point is the subject of the remainder of this
section. Section~\ref{sec:sys} is devoted to a careful analysis of all
known systematics and related uncertainties on $\beta$.

The joint fitting of the lensing and velocity dispersion data greatly
enhances our ability to distinguish between DM profiles (see STE02).
The top left panel of Figure~7 illustrates why that is the case.  The
hatched boxes represent the velocity dispersion measurement for
MS2137-23 and their 1-$\sigma$ uncertainties.  The solid black curve
shows the best fitting velocity dispersion profile model obtained with
our combined lensing and velocity dispersion analysis.  The dashed
curve shows a velocity dispersion profile for a set of free parameters
that agrees extremely well ($\Delta\chi^{2}<1$; $\beta$=1.30) with the
gravitational lensing measurements alone, but does not match the
measured velocity dispersion profile of the BCG.  This special case
(where the $M_{*}$/L=0 indicates that the luminous component is a
massless tracer of the potential) clearly shows how mass models with
too steep an inner profile cannot both match the velocity dispersion
profile measurement and reproduce the positions of the gravitational
arcs.  The remaining panels in Figure~6 plot both the observed and
best-fitting velocity dispersion profile for each of the six clusters.

Our best-fitting mass models produce density profiles that are
remarkably similar in their makeup (see Figure~7).  On $\lesssim$10
kpc scales, the matter distribution is dominated by the luminous, BCG
component, with the DM component dominating at larger radii.  Dubinski
(1998) has found a similar result by numerically simulating the
formation of a BCG in the presence of a cuspy DM halo.  As can be seen
from Figure~7, the velocity dispersion measurement of the BCG allows
us to probe the matter distribution where luminous matter is
important, while the gravitational arcs probe regions where DM
dominates.  The measurement techniques complement each other.

It is appropriate to assess the goodness-of-fit of our best-fitting
models.  While $\Delta\chi^{2}$ is distributed as a $\chi^{2}$
distribution with three degrees of freedom (representing the three
free parameters in our model), the best-fitting model (with
$\chi^{2}_{min}$) is distributed as a $\chi^{2}$ distribution with
$N-3$ degrees of freedom, where N is the number of data points and
three again represents the number of free parameters (see e.g. Press
et al. 1997). The total best-fitting $\chi^{2}$ for each cluster is:
Abell 383, $\chi^{2}/dof$=8.3/2; MS2137-23, $\chi^{2}/dof$=8.9/7; RXJ
1133, $\chi^{2}/dof$=1.0/2; Abell 1201, $\chi^{2}/dof$=6.6/6; Abell
963, $\chi^{2}/dof$=2.9/3; MACS 1206, $\chi^{2}/dof$=1.4/1.  Of the
total, the contribution from the gravitational lensing portion of the
$\chi^{2}$ is never more than 0.2, meaning that the bulk is due to the
velocity dispersion profile (see Figure 6 for the best-fitting
velocity dispersion profiles).  The one cluster with a relatively high
$\chi^{2}$ is Abell 383.  However, given the simplicity of our mass
model, this relatively high $\chi^{2}$ should not be alarming.  In
section 6 we explore in detail possible systematic effects in our
current analysis, any of which could be responsible for a less than
perfect fit to the data.  Since all of these systematic checks
indicate that $\Delta\beta\lesssim$0.2, we are confident in the
robustness of our results.

What causes the difference in the confidence contour shapes in the
$M_{*}$/L-$\beta$ plane between the radial arc sample and the
tangential arc sample?  Both samples do not allow a steep DM inner
density profiles because of their inability to match the observed BCG
velocity dispersion profiles described in the previous paragraph.
However, it seems as if the radial arc sample is capable of
pinpointing the DM inner density slope, while the tangential arc
sample can give just an upper limit.  Due to the functional form of
the radial eigenvalue (see Eqn. A5), radial arcs cannot form in total
density profiles steeper than $\rho \propto r^{-2}$.  For our mass
model, as $M_{*}$/L increases and $\beta$ becomes small, the above
criteria for radial arc formation is not met unless the radial arc
position is pushed out radially (where DM will have a larger
contribution and thus soften out the effects of the cuspy luminous
distribution) to a point where it is incompatible with its observed
position.  It is for this reason that low values of $\beta$ are not
allowed in the radial arc analysis and the DM inner density slope can
be pinpointed.

The summary of the results presented in this subsection are at odds
with predictions of CDM simulations and have claimed to measure the
intrinsic scatter in the inner DM slope, $\beta$.  In order for such
results to be taken seriously, it is imperative that our method is
tested thoroughly with respect to our simplifying assumptions.  It is
our goal in the next section to test systematically all of our
assumptions before we discuss the implications of our results for CDM.

\section{Systematics}\label{sec:sys}

During the course of our analysis, many simplifying assumptions were
made.  As an exploratory study aiming is to obtain tight constraints
on the DM density profile, we strived to simplify our model and its
inputs while still extracting the correct inner density profile.
However, it is possible that these simplifications are giving
systematically different values of the DM inner density slope than a
more complex modeling.  To judge the robustness of our method we have
performed a battery of tests.  First, in \S~6.1, we explore the fact
that we neglected both ellipticity and substructure in our lensing
treatment.  Second, in \S~6.2, we look at possible complications in
our analysis of the BCG dynamics (e.g.\ orbital anisotropy and
template mismatch).  Finally, in \S~6.3 we report on tests run to
check our results depending on changes in our luminous mass model and
due to possible uncertainties in our measurements.  Abell 383 has the
tightest constraints in our sample (see \S~5.2) and by using it to
illustrate our test results (in \S~6.2 and \S~6.3) we demonstrate the
impact our assumptions have on our determination of $\beta$.

\subsection{Impact of Cluster Substructure and Ellipticity}\label{sec:lenstool}

Our cluster sample has been selected to comprise relaxed systems with
no obvious signatures of strong ellipticity and/or bi--modality in
their underlying mass distributions.  Nevertheless, previous analyses
of two of the clusters (Abell 383 and MS 2137-23) does reveal that
they are not perfectly circular in projection (e.g.\ S01;
Miralda-Escud\'e 2002).  Our simple lensing method deliberately 
does not attempt to fit the detailed positions of all the
multiply--imaged features of the clusters, concentrating instead on
the positions of relevant critical lines, estimated from visual
inspection of symmetry breaks in the observed multiple--images.  In
this section we exploit sophisticated two--dimensional lens models to
investigate whether the simplifying assumptions in our
one--dimensional models introduce any systematic bias into our
results.  Qualitatively, the key differences between the models
discussed in this section and those upon which our main analysis is
based is that in this section we include the ellipticity and
substructure (arising from bright cluster ellipticals) of the clusters
in the models, and also fit the models to all of the observed
multiple--image systems.

We use the {\sc lenstool} ray--tracing software to construct a
detailed model of each of the clusters in the radial arc sample.  The
details of this method are explained elsewhere, and we refer the
interested reader to the relevant articles (Kneib 1993; Kneib et al.\
1996; Smith 2002; Smith et al.\, 2004, in prep.).  Briefly, we use the
observed positions, redshifts, shapes and orientations of the observed
multiple--image systems to constrain a model of the \emph{total}
surface mass density in each cluster core.  We stress that we do not
attempt to decompose the best--fit total matter distributions into
their respective dark and luminous components.  Each model therefore
consists of the minimum number of analytic matter components (each one
parametrized as a truncated pseudo--isothermal elliptical
mass-distribution -- Kassiola \& Kovner 1993; Kneib et al.\ 1996)
required to fit the observables.  In practice, each model contains a
central dominant cluster--scale mass component that is centered on
each BCG, plus an additional central mass component for the BCG, and
a small number ($\le4$) of smaller mass components to account for
contributions from likely cluster members that lie adjacent to the
observed multiple--image systems.  We briefly describe each model:

{\bf MS 2137-23} -- This cluster has been extensively modeled by
several authors (Mellier et al.\ 1993; Hammer et al.\ 1997; G03).  We
adopt STE02's spectroscopic redshifts for the dominant tangential and
radial arcs as constraints on our {\sc lenstool} model.  A four
component model is able to produce an acceptable fit to these
constraints ($\chi^2/{\rm dof}\simeq1$).  These components comprise
the cluster--scale potential, the BCG and two galaxies lying $3''$
North-West of the BCG, adjacent to the radial arc (Fig.~1).  This
model predicts a central fifth image of the galaxy that appears as the
giant tangential arc that is in broad agreement with that predicted by
G03's model.  However when subtracting a model of the BCG from the
\emph{HST} frame, we are unable to confirm G03's claimed detection of
the fifth image.  We therefore do not include this image as a
constraint on the model.  The ellipticity of the best--fit fiducial
model is $\epsilon=(a^2-b^2)/(a^2+b^2)=0.18$.

{\bf Abell\,383} -- S01 constructed a detailed lens model of this
cluster, based on their spectroscopic redshift of the brightest
component (B0a) of the tangential arc (see also Smith 2002).  We build
on this analysis to add the spectroscopic redshifts of B1a/b and the
radial arc into S01's model.  These new spectroscopic redshifts
constrain the volume of parameter space occupied by the family of
acceptable models.  The best--fit model lies within the family of
models identified by S01 as providing an acceptable fit to the data.
Figure 8 illustrates the resulting lensing model for this cluster,
showing the derived tangential critical curves for $z=1,3$.  The
detailed multiple-image interpretation of this cluster is described by
Smith et al. (2004, in prep).

{\bf RXJ\,1133} -- We identify the tangential and radial arcs as
comprising two images each of the background source.  The compact high
surface--brightness feature in the radial arc has a profile shape and
FWHM similar to a point source.  The origin of this point source is
unclear since it does not appear in the other lensed images.  It is
likely a foreground star, or possibly a transient event in the lensed
galaxy that only manifests itself in the radial arc due to time
delays.  To obtain an acceptable fit to the lensing constraints, we
use a five component model: cluster--scale potential, BCG, two dwarf
galaxies adjacent to the fifth image and the moderately bright likely
cluster member $7''$ away from the BCG directly opposite these two
dwarfs.  The ellipticity of the best--fit fiducial model is
$\epsilon=0.19$.

After obtaining the best--fit fiducial lens model for each cluster, we
systematically explored the parameter space of each model to determine
the family of acceptable models ($\Delta\chi^2\le1.0$).  We then
compare this family of models with the family of acceptable models
with $\Delta\chi^2\le1.0$ (projected from 3D to 2D) from the analysis
presented in \S 5.2 (this is equivalent to the 68\% confidence
interval for one parameter).  Fundamentally, the lensing constraints
contain information about the enclosed mass at the position of the
tangential arcs and the derivative of the projected mass enclosed at
the position (i.e.\ symmetry break between the two components) of the
radial arc.  We extract azimuthally averaged projected density
profiles from these two sets of models for each cluster and compare
$M(\leqslant R_{\rm tangential})$ and $d(M/R)/dR(R_{\rm radial})$,
taking due account of the uncertainties in the determinations of
$R_{\rm tangential}$ and $R_{\rm radial}$.

We plot the results of this comparison in Fig.~9.  The difference in
$M(\leqslant R_{\rm tangential})$ between the two methods is never
more that 8\% of that of the {\sc lenstool} result.  Note that the 1D
method presented in this work is robust in its measurement of $\beta$
when the position of the tangential critical line is shifted
$\pm0\farcs5$ (see \S 6.3).  A shift in the tangential critical line
position is equivalent to changing $M(\leqslant R_{\rm tangential})$,
and so we defer to that subsection for a discussion of how a
mismeasurement of $M(\leqslant R_{\rm tangential})$ effects our
conclusions on the DM density profile.

We concentrate here on the radial arc comparison, focusing on the
discrepancy identified in MS\,2137-23 (the most discrepant in the
sample).  In the following we will assume that any correction
necessary in the method is solely a correction that must be made to
the DM component.  This is certainly a very conservative estimate
since the luminous BCG mass component contributes significantly on the
scales of the radial gravitational arcs.  For simplicity, we assume
that we are dealing with power-law surface density profiles,
$\Sigma(R)\propto R^{\gamma}$.  This implies that $M(<R)\propto
R^{\gamma+2}$ and $\rho(r)\propto r^{\gamma-1}$.  In MS2137-23
$\gamma=$-0.29 and -0.50 for the method in this work and the {\sc
lenstool} results respectively, at the position of the radial critical
line.  A systematic mismeasurement of $\Delta\gamma=$0.2 will cause a
similar sized mismeasurement in the value of $\beta$,
$\Delta\beta=$0.2. This is roughly twice the size of the random error
component ($\beta=0.57^{+0.11}_{-0.08}$) quoted in \S 5.2 for
MS2137-23.  Note that the correction implied from the {\sc lenstool}
analysis is in the direction of lower $\beta$ values, even further
away from predictions made by CDM numerical simulations.  We conclude
that any systematic effect due to the axisymmetric lens modeling in
this work can affect our $\beta$ measurement by of order the random
error components we have calculated.

To aid comparison of our empirical measurements with future
observational and theoretical studies, we list here the values of
$\gamma$, assuming that $\Sigma(R)\propto R^{\gamma}$.  Using the 1-D
approach presented in this work, $\gamma(R_{radial})=-0.29\pm0.03$ for
MS2137-23, $-0.43\pm0.05$ for RXJ1133 and $-0.36\pm0.03$ for Abell
383.  With the 2D {\sc lenstool} analysis
$\gamma(R_{radial})=-0.50\pm0.01$ for MS2137-23, $-0.33\pm0.10$ for
RXJ1133, and $-0.43\pm0.02$ for Abell 383.

\subsection{Velocity Dispersion Measurements \& Modeling}\label{sec:vdtest}

In \S 4.3 we presented the velocity dispersion measurements such that
the final uncertainty tabulated in Table 5 is the addition in
quadrature of a random component (taken from the output of the
Gauss-Hermite Pixel Fitting Software) and a systematic component
associated with template mismatch.  Template mismatch is due to the
fact that we used a single stellar spectral type to determine the
kinematics of the BCGs.  We quantified the effect of template mismatch
in \S 4.3 by taking the rms deviation among the different stellar
templates used to represent a possible systematic offset.  While we
incorporated this uncertainty into our mass modeling analysis, it is
nonetheless necessary to understand how robust our results are to
systematic offsets of the velocity dispersion profile.  For each BCG
we shifted the measured velocity dispersion profile up and down by the
systematic uncertainty (typically $\sim$15-20 \kms) and reran our
analysis to determine the impact on our results.  The typical shift in
the $M_{*}$/L-$\beta$ plane is about $\Delta\beta\sim\pm$0.15, and we
conclude that template mismatch can not greatly alter our final
results.

In our dynamical modeling of the BCGs we assumed isotropic orbits (see
Appendix~A.2) for the constituent stellar tracers.  This assumption is
justified on several grounds.  From an observational point of view,
Kronawitter et al.~(2000) found that in their sample of galaxies the
best-fitting models were nearly isotropic with typical $\alpha\simeq
0.3$ ($\alpha$ is the anisotropy parameter, see Appendix Eqn. A8) at
$R_{e}/2$, fallin to $\alpha=0$ at larger radii.  There was little
indication of any tangential ($\alpha<0$) anisotropy in that study.
Gerhard et al.\ (2001) obtain complementary results from an extended
sample.  Similar conclusions have been obtained theoretically (e.g.\
van Albada 1982), with strong radial anisotropy leading to instability
(Merritt \& Aguilar 1985; Stiavelli \& Sparke 1991).

Nonetheless, it is still instructive to rerun the analysis with
anisotropy, especially radial anisotropy.  Using an anisotropy radius,
$r_{a}$ (implementing the Osipkov-Merritt parametrization, Eqn. A8),
equal to $0.5r_{j}$, approximately at the point where observations
have indicated that orbits are somewhat radially anisotropic, we have
investigated the effects on our confidence contours.  Note that with
this parametrization, stellar orbits become more and more radially
anisotropic with increasing radius.  The confidence contours in the
$M_{*}$/L-$\beta$ plane move towards lower acceptable values of
$\beta$ ($\Delta\beta\sim-$0.20 for $r_{a}=0.5r_{j}$; see Fig. 11),
increasing the disagreement with predictions of N-body simulations.
We conclude that modest radial anisotropy will only strengthen our
claim that the observed DM profiles are shallower than predicted
theoretically.

Likewise, we have introduced a constant tangential anisotropy of
$\alpha=-0.5$.  Observationally, tangential anisotropy in the inner
regions of giant ellipticals is very rare (e.g.~ Kronawitter et al.\
2000; Gerhard et al. 2001).  As expected, the results indicate a
slight steepening of the DM halos ($\Delta\beta\sim$+0.20; see
Fig. 11).  Given the extreme case presented here, we conclude that our
results are robust to slight tangential anisotropy in the BCG.

In this work we have used Gaussian line profiles to represent the
line-of-sight velocity dispersions of the BCGs.  This approach
provides a good low-order fit to galactic spectra, but in the outer
parts of galaxies deviations from Gaussian can be of order $\sim$10\%
leading to systematic mismeasurements of rotation velocities and
velocity dispersions of the same order (van der Marel \& Franx 1993).
Higher order velocity moments are routinely measured for nearby
galaxies giving information on their orbital structure
(e.g. Kronawitter et al. 2000; Gerhard et al. 2001; Carter, Bridges,
\& Hau 1999).  In the inner regions of galaxies, these studies have
indicated that deviations from a Gaussian line profile are small,
especially on the scales probed in this work ($\lesssim$0.5$R_{e}$).
However, to make these measurements, high signal-to-noise is needed,
and this makes it hard to measure these parameters at even modest
redshift due to cosmological surface brightness dimming.  For this
reason, we were unable to measure deviations from Gaussian line
profiles in even the central regions of the BCGs.  Any systematic
introduced due to the Gaussian line profiles used must be small on the
scales we are probing.  Earlier in this section we have shown that our
results are robust to orbital anisotropies (which would lead to
deviations from Gaussian line profiles).  Miralda-Escud\'e (1995)
suggested that in the rapidly rising portion of the velocity
dispersion profile expected from BCGs at large radii (see e.g. Fig. 6)
that deviations from Gaussian line profiles should be expected, even
in systems with isotropic orbits.  At the moment, this can only be
verified in nearby BCGs where higher-order moments could be measured
to high radii, beyond the scales probed in this work.

\subsection{Other Assumptions and Measurement Uncertainties}\label{sec:test}

We have subjected our data to several additional tests.  We changed
the luminous mass model to a Hernquist profile, systematically altered
the seeing by 30\%, adjusted $R_{e}$ of the BCG by 10\%, modified the
scale radius of the DM halo, $r_{sc}$, and shifted the radial and
tangential critical lines.  All tests were performed by changing one
parameter at a time.  We report the results of these tests below.

(1) We replaced the Jaffe luminous density profile with a Hernquist
profile to see how robust our constraints on $\beta$ are with respect
to our choice of the Jaffe density profile for the BCG.  We have
chosen the Jaffe and Hernquist profiles because they give analytic
solutions to the surface brightness profile and line of sight velocity
dispersion.  As can be seen from Figure~2, the Jaffe and Hernquist
profile bracket the data at low radii ($<$ 1'').  The Hernquist
luminous mass density profile goes like $\rho \propto r^{-1}$ at small
radii.  A Jaffe density profile is slightly more cuspy than an \dv
profile, while a Hernquist profile is slightly less cuspy.  We
obtained confidence contours in rough agreement with the original mass
model.  However, a Hernquist profile give a much larger best-fitting
$\chi^{2}$ ($\Delta\chi^{2}\sim10$ with respect to the best-fitting
Jaffe luminous density profile results).  Since we are bracketing the
true surface brightness profile with our Jaffe and Hernquist
parameterizations and we get roughly equivalent results on $\beta$, we
are confident that are choice of the Jaffe luminous distribution is
not biasing our results towards shallow DM halo profiles.

(2) We argued in \S~5.1 that our final results are not very sensitive
to the scale radius, $r_{sc}$, that we assume for the DM density
profile, although there should be some dependence due to projection
effects.  Since it is the goal of this work to test the predictions of
CDM, it is important that a range of $r_{sc}$ compatible with
numerical simulations are checked to make sure that this possible
systematic is not large.  Using the parameterization of the
concentration parameter, $c_{vir}$, adopted by Bullock et al. (2001)
for a $\sim10^{15}M_{\sun}$ halo with $R_{vir}\sim$ 2 Mpc, we expect
$r_{sc}$ to lie between 240 and 550 kpc (68\% CL).  Recently,
Tasitsiomi et al. (2003) have simulated fourteen cluster-sized DM
halos having cuspy profiles with mean $r_{sc}$ of $450\pm300$ kpc.
These values seem reasonable observationally, as well.  For example,
G03's best fitting NFW profile for MS2137-23 using weak lensing had
$r_{sc}=67^{+300}_{-24}$ kpc.  As a test, we briefly considered
$r_{sc}$ as a fourth free parameter, taking a flat prior on $r_{sc}$
between 100-800 kpc, in accordance with the range of $r_{sc}$ found by
Tasitsiomi et al. (2003).  After marginalization with respect to the
other free parameters, this analysis caused a shift of
$\Delta\beta\sim0.15$ towards steeper values of $\beta$, giving us
confidence that for reasonable values of $r_{sc}$ our constraints on
$\beta$ are robust.

(3) Since the seeing is one of the measured inputs in the velocity
dispersion portion of the analysis (see Eqn. A11 in the Appendix), we
changed the seeing value by $\pm30\%$ to determine how robust our
conclusions are to mismeasurements in the seeing.  We found shifts of
$\Delta\beta\sim$0.05 and thus concluded that even significant
mismeasurements in the seeing do not effect our final results.

(4) Additionally, we perturbed the positions of the effective radius
of the BCG surface photometry fits by $\pm10\%$. As mentioned earlier,
mismeasurements in the luminous mass distribution could possibly alter
the shape of the inferred DM halo.  However, changing $R_e$ had a
negligible effect on the measured confidence contours.

(5) Finally, although the visual measurement of the critical lines
agrees to within 1-$\sigma$ with those obtained from the 2D averaged
results of the LENSTOOL analysis, we still tested to see how sound our
results are to perturbations in the critical line positions.  In our
tests, both radial and tangential critical lines were perturbed by
$\pm0\farcs5$ from their reported positions.  We found that changing
the tangential critical lines by this amount had a negligible effect,
while adjusting the position of the radial critical lines produced
shifts on order $\beta\la$0.1.  We conclude that our results are not
extremely dependent on the exact location of the critical lines.

\subsection{Summary of Systematics}

In this section, we have performed a wide variety of tests to our
model and mass measurement technique, with no test indicating that our
method is incomplete or lacking.  The main conclusions of these tests
of the systematics can be summarized as follows:

(1) In \S~\ref{sec:lenstool}, we explored the consequences of our
    axisymmetric gravitational lens models by comparing our results
    with the sophisticated 2D ray-tracing software, {\sc lenstool}.  A
    comparison suggests that at most our constraints on the inner DM
    density logarithmic slope are shifted by $\Delta\beta\sim$0.2.

(2) In \S~\ref{sec:vdtest} we checked the robustness of our method in
    the face of possible systematics associated with the dynamics of
    the BCGs.  Stellar template mismatch and orbital anisotropies can
    at most shift $\beta$ by $\sim$0.2.

(3) In \S~\ref{sec:test} we performed a battery of tests to check our
    luminous mass model and our sensitivity to the observations.  The
    most serious effect is associated with our assumed value of
    $r_{sc}$.  For reasonable values of this parameter, the inner DM
    density logarithmic slope is shifted by $\Delta\beta\sim$0.15.

Figure~\ref{fig:systests} plots the results for those tests performed
in \S~\ref{sec:vdtest} and \S~\ref{sec:test} that produce the largest
changes in our estimation of $\beta$.

\section{Discussion}\label{sec:disc}

\subsection{Comparison with Simulations}

The results of N-body simulations indicate that we should expect DM
inner density profiles of between $\beta=1$ and 1.5 even with the
current refinements in modern N-body work that pay special attention
to issues of convergence (e.g.\ Power et al.\ 2003; Fukushige et al.\
2003).  We have found a range of acceptable values of the inner
logarithmic slope, with $\langle\beta\rangle=0.52^{+0.05}_{-0.05}$,
for our radial arc sample and $\beta<$0.57 (at $>$99\% confidence) for
all three clusters in our tangential arc sample.  We detect scatter in
our radial arc sample, which we will discuss in \S~7.2.

So what can account for the apparent discrepancy between observations
and simulations?  There are two questions that must be addressed.  Do
the scales probed in the observations correspond to those resolved in
the simulations?  Second, what effect do baryons have in our
comparison?

In this work, we are only able to probe the mass distribution out to
the distance of the tangential gravitational arc, which for our sample
is $<$100 kpc.  The original work by NFW97 had a gravitational
softening radius of $\sim$20 kpc (for their largest mass, galaxy
cluster sized halos), although it is not clear that they achieved
proper convergence down to this radius.  Subsequent higher resolution
work (e.g.\ Ghigna et al.\ 2000; Klypin et al.\ 2001) focused on the
issue of convergence and reported that their results for the DM
density profile were reliable down to scales of $\sim$50 kpc at the
cluster scale, both groups found $\beta\sim$1.5.  Even more recently,
Power et al. (2003) and Fukushige et al.\ (2003) have performed
extremely high resolution simulations, with density profile results
reliable down to $\sim$5 kpc ($\sim$0.002 $R_{vir}$).  Note that all
of these works used different criteria for convergence.  It seems safe
to say that modern N-body simulations are becoming reliable down to
the $\sim$10 kpc scale for galaxy clusters, which is comparable to the
scales being probed in this study.  It is also comparable to $R_{e}$
in a typical giant BCG, and so it is clear that baryons should play a
more central role in further investigations.

It must be emphasized that these simulations include only
collisionless CDM particles.  It is unclear how baryonic matter,
especially in regions where it may dominate the total matter density
may affect the DM distribution.  Several possible scenarios have been
presented in the literature and the following discussion is not
exhaustive.  One possible situation, known as adiabatic contraction,
is that as baryons sink dissipatively into the bottom of the total
matter potential they are likely to steepen the underlying DM
distribution simply through gravitational processes (Blumenthal et
al. 1986).  This situation would only exacerbate the difference
between our observed shallow slopes and those expected from N-body
simulations.  It has also been suggested by Loeb \& Peebles (2003)
that if stars form at high redshift ($z>$6) before large structures
form that they can be treated in a similar manner as the underlying DM
particles.  This scenario would suggest that instead of separating
dark and luminous components of the matter distribution for comparison
with simulations that we should compare the total mass distribution
observed with the DM distribution found in N-body simulations.
However, Loeb \& Peebles (2003) admit that this scenario is not
strongly motivated and that some dissipative process must still take
place within the baryonic material.  Recently, Dekel et al. (2003)
have suggested that DM halos must be as cuspy as NFW due to the merger
process, unless satellite halos are disrupted at large radii, possibly
due to baryonic feedback.  One final scenario describes a situation in
which baryonic material is initially concentrated in small clumps of
mass ($\geq$0.01\% of the total mass) with dynamical friction causing
the final DM halo shape to flatten due to these clumps (El-Zant et
al. 2003; El-Zant, Shlosman \& Hoffman 2001).  The issue of baryons
must be looked into further and it is possible that the DM ``core''
problem cannot be resolved until baryonic material can be properly
incorporated into the numerical experiments.

We have gone to great lengths in this paper to disentangle the
luminous BCG component from the overall cluster DM in our mass model
so that we could compare directly our results with those of N-body
simulations.  Albeit with considerable scatter, the average DM density
profile is too shallow, especially if adiabatic contraction describes
well the interaction between dark and baryonic matter.  Does this
indicate that something may be wrong with the $\Lambda$CDM paradigm of
structure formation?  Dark matter only simulations do not appear to be
sufficient, especially as they begin to probe down to scales where
complicated gaseous physics play a significant role.  While work has
been done to model the formation of cDs and BCGs (see e.g.\ Dubinski
1998; Nipoti et al.\ 2003) these have mainly focused on the accretion
of galaxies to form cD-like objects, and have not been concerned with
the resulting effect on the DM halo.

\subsection{Is the DM slope universal?}\label{sec:scatdisc}

This is a question that begs to be asked after looking at Figure 4.
We detect an intrinsic scatter in $\beta$ values of
$\Delta\beta\sim$0.3 in the radial arc sample.  Unfortunately, the
tangential arc sample can give only an upper limit on $\beta$, and
thus does not provide any further measure of the scatter.

The most recent and highest resolution N-body simulations of eight
galaxy clusters performed by Fukushige et al.\ (2003) did show some
signs of run-to-run inner slope variations, and although they claimed
that this argued against a ``universal'' inner DM profile, they did
not quantify the scatter.  Several claims have been made that the DM
density profile is dependent on the DM halo mass (e.g.\ Ricotti 2002;
Jing \& Suto 2000), however, these studies have focused on the
difference in slopes between different mass scales (e.g.\ individual
galaxies versus clusters of galaxies) while the clusters in this study
are all roughly the same mass.  It is plausible that the formation
history of any given cluster sized halo can cause a natural cosmic
scatter in $\beta$ (e.g.\ Nusser \& Sheth 1999).  Ultimately,
numerical simulations should be able to reproduce not only the
observed mean slope of the DM density profile in galaxy clusters, but
also its measured scatter.

\section{Summary}

We have performed a joint gravitational lensing and dynamical analysis
in the inner regions of six galaxy clusters in order to constrain the
inner DM density slope $\beta$.  By studying the velocity dispersion
profile of the BCG, we were able to disentangle luminous and DM
components of the total matter distribution in these clusters on
scales $<100$ kpc.  The main results of the paper can be summarized as
follows:

1) The average inner slope of the 3 systems with both radial and
   tangential arcs is $\langle\beta \rangle=0.52^{+0.05}_{-0.05}$. The
   3 clusters with only tangential arcs provide an upper limit of
   $\beta<0.57$ (99\%CL). The measured slopes are thus inconsistent at
   high confidence level ($>99$\%CL) with the cusps ($\beta=1-1.5$)
   predicted by dark matter only cosmological numerical simulations.

2) The agreement of the results from the radial arc sample and the
   tangential sample shows that the shallow slopes found for MS2137-23
   (Sand et al.\ 2002) and the other radial arc systems are not the
   result of a selection effect.

3) Our precise measurements allow us to give a first estimate of the
   intrinsic scatter of the inner density DM slope
   ($\Delta\beta$$\sim$0.3) of clusters of galaxies. The analysis of a
   larger sample of systems to better characterize the intrinsic
   scatter of the inner slope would provide a further observational
   test for future numerical simulations.

4) Our method is robust with respect to known systematic effects,
   including those related to the choice of the mass model, the
   description of orbital anisotropy in the dynamical models, and the
   simplifying assumptions inherent to our axisymmetric lensing
   analysis. A comprehensive and detailed analyis of these effects
   shows that the related systematic uncertainties on $\beta$ are
   smaller than 0.2. Therefore, even for the system with the smallest
   random uncertainties (Abell 383) systematic errors do not dominate
   the error budget.

In conclusion, our results are in marked disagreement with the
predictions of dark matter only cosmological simulations. The
inclusion of baryons in the models via a simple adiabatic contraction
mechanism would further steepen the theoretical dark matter halo
making the disagreement even more pronounced. Therefore, a more
sophisticated treatment of baryons in the simulations appears
necessary if one wants to reconcile the CDM paradigm with the present
observations.

\acknowledgements

We would like to thank the referee for a thorough and constructive
report.  Also, we would like to acknowledge useful discussions with
Neal Dalal, Chuck Keeton, Matthias Bartelmann and Massimo Meneghetti.
We acknowledge financial support for proposal number HST-AR-09527
provided by NASA through a grant from STScI, which is operated by
AURA, under NASA contract NAS5-26555.  DJS would like to acknowledge
financial support from NASA's Graduate Student Research Program; NASA
grant No. NAGT-50449.  DJS is grateful for statistical advice given by
Benjamin D. Wandelt.  We are grateful for the use of the Gauss-Hermite
Pixel Fitting Software developed by R.P. van der Marel.  We thank Edo
Berger, Alicia Soderberg and Shri Kulkarni for the NIRC observations.
We would also like to thank Jean-Paul Kneib and Pieter van Dokkum for
personally recommending clusters MACS 1206 and Abell 1201,
respectively.  We are grateful to the MACS collaboration, in
particular Harald Ebeling, Alastair Edge and Jean-Paul Kneib, for
access to MACS 1206.  GPS thanks Jean-Paul Kneib for sharing his {\sc
lenstool} ray tracing code.  DJS would like to thank Melissa Enoch for
providing CPU time.  We would like to thank all those who have worked
to make both LRIS and ESI such fantastic instruments.  Finally, the
authors wish to recognize and acknowledge the cultural role and
reverence that the summit of Mauna Kea has always had within the
indigenous Hawaiian community.  We are most fortunate to have the
opportunity to conduct observations from this mountain.

\appendix

\section{Analysis Technique}

\subsection{Lensing}

\label{app:lens}

Given our simple, spherically symmetric two-component mass model, we
adopted a simple lensing analysis using only the positions and
redshifts of the gravitational arcs in our sample.  Our method is a
generalization of that used by Bartelmann (1996).

Since the extent of the galaxy cluster (lens) is much less than the
distance from the observer to the lens and the lens to the source, we
make the thin-screen approximation in our gravitational lensing
calculations.  The total surface mass density is the sum of the
luminous and DM components: $\Sigma_{tot}=\Sigma_{DM}+\Sigma_{L}$.
The surface mass density of our chosen DM halo profile, $\Sigma_{DM}
(R)$, is

\begin{equation}
\label{eq:sddm}
\Sigma_{DM} (R)=2 \rho_{c} r_{sc} \delta_{c} x^{1-\beta} \int_0^{\pi/2}d\theta \sin\theta (\sin\theta + x)^{\beta-3},
\end{equation}
where x = R/$r_{sc}$ (Wyithe, Turner \& Spergel 2001).  The surface
mass density of the luminous component is $\Sigma_{L}=I(R) M_{*}/L$,
where I(R) is given in Jaffe's (1983) original paper.

Using standard gravitational lensing nomenclature (see, e.g.,
Schneider, Ehlers, \& Falco 1992) we describe $\Sigma_{tot}$ in terms
of the critical surface mass density, $\Sigma_{cr}$,

\begin{equation}
\label{eq:kappa}
\kappa (R) = \frac{\Sigma_{tot} (R)}{\Sigma_{cr}},
\end{equation}
where
\begin{equation}
\label{eq:sdcrit}
\Sigma_{cr} = \frac{c^{2}}{4 \pi G}\frac{D_{s}}{D_{l} D_{ls}},
\end{equation}
and $D_{l}$, $D_{ls}$, and $D_{s}$ are the angular diameter distance
to the lens, between the lens and source and to the source,
respectively.  Another convenient quantity when describing the mapping
from the source plane to the lens plane is dimensionless and
proportional to the mass inside projected radius x,

\begin{equation}
\label{eq:promenc}
m (x)=2  \int_0^{x}dy \kappa (y) y. 
\end{equation}

With these definitions, the two eigenvalues of the Jacobian mapping
between the source and image plane read:

\begin{equation}
\label{eq:dlessmass}
\lambda_{r}= 1-\frac{d}{dx}\frac{m(x)}{x}, \lambda_{t}=1-\frac{m(x)}{x^{2}}.
\label{eq:eigen}
\end{equation}

The root of these two equations describes the radial and tangential
critical curves of the lens.  Since the magnification of the source is
equal to the inverse of the determinant of the Jacobian, the radial
and tangential critical curves are where the magnification of the
source formally diverges.  While this does not happen in practice (due
to the spatial extent of the source), it guarantees that when an image
of a source lies near a critical line it is strongly distorted (in the
radial direction in the case of radial arcs and tangentially for the
case of tangential arcs).  Merging lensed images merge across critical
lines, which provides a simple way of approximating their position by
visual inspection.  Thus highly distorted image pairs are an excellent
way of approximating the position of the critical line for a lens.
For the simple, axisymmetric lens model explored in this work, both
the radial and tangential critical lines are circular with the radial
critical curve always lying inside that of the tangential critical
curve. Our sample of clusters (approximately round clusters with
little visible substructure) was chosen specifically with this concern
in mind (tests of our lens model are described in
\S~\ref{sec:lenstool}).  However, this is not always the
case. Therefore, extreme caution must be exercised when applying our
simple lens model to other samples of clusters.

Given our mass model (\S 5.1), the measured redshifts of the arcs and
clusters, and a set of free parameters \{$M_{*}$/L,$\beta$,
$\delta_{c}$\}, we can compute the predicted position of the arcs
(assuming they lie very close to their associated critical line), by
finding the root of the appropriate eigenvalue from
Eq.~\ref{eq:eigen}.  By comparing the predicted position of the arcs
with the actual position taken from the images, we can calculate the
likelihood function,
\begin{equation}
\label{eq:like_lens}
P(M_{*}/L,\delta_{c},\beta) \propto exp \{-\frac{1}{2} \Sigma_{i} \left[ \frac{y_{i} - \tilde{y}_{i}(M_{*}/L,\delta_{c},\beta)}{\Delta_{i}} \right]^{2} \},
\end{equation}
assuming that our underlying distributions are normal.  Here, y is the
distance of the arc from the center of the cluster potential (as
measured from the center of the BCG), $\Delta_{i}$ is our assigned
uncertainty to the position of the critical line and the sum in the
exponential is over all the critical line arcs with known redshift.

\subsection{Dynamics}

\label{app:dyn}

In addition to the gravitational arc redshifts, we have also measured
extended velocity dispersion profiles for all of the BCGs in our
sample. This is used as an additional constraint on our mass model,
using a joint likelihood analysis.

We compute the model velocity dispersion starting from the spherical
Jeans Equation (Binney \& Tremaine 1987):
\begin{equation}
\label{eq:sJeqn}
\frac{d\rho_{*} (r) \sigma_{r}^{2} (r)}{dr}+\frac{2 \alpha (r) \rho_{*} (r) \sigma_{r}^{2} (r)}{r} = -\frac{G M_{enc} (r) \rho_{*} (r)}{r^{2}},
\end{equation}
where G is Newton's gravitational constant, M(r) is the
three-dimensional mass enclosed, $\sigma_{r}$ is the radial velocity
dispersion. The anisotropy parameter $\alpha(r)$ is defined as,
\begin{equation}
\label{eq:aniseqn}
\alpha (r) \equiv 1 - \frac{\sigma_{\theta}^{2} (r)}{\sigma_{r}^{2} (r)} \equiv \frac{r^{2}}{r^{2} + r_{a}^{2}},
\end{equation}
where $\sigma_{\theta}$ is the tangential component of the velocity
dispersion. The final definition introduces the Opsikov-Merritt
(Osipkov 1979; Merritt 1985a,b) parameterization of anistropy that we
mainly use in our dynamical models. By default, we use isotropic
orbits (i.e $r_a=\infty$) which appears to be a realistic description
of the inner regions of early-type galaxies. However, in \S 6.2 we
explore the consequences of anisotropic velocity dispersion tensors on
DM density profiles we measure, by considering $r_a \geqslant 0$ models and
models with constant tangential anisotropy ($\alpha<0$).

Using our parameterization of the anisotropy, we can readily derive
the radial velocity dispersion (Binney 1980)
\begin{equation}
\label{eq:sigma_3d}
\sigma_{r}^{2}(r) = \frac{G \int_r^{\infty} dr' \rho_{*}(r') M_{enc}(r') \frac{r_{a}^{2}+r'^{2}}{r'^{2}}}{(r_{a}^{2}+r^{2}) \rho_{*}(r)}
\end{equation}
and the projected velocity dispersion 
\begin{equation}
\label{eq:sigma_2d}
\sigma_{p}^{2}(R) = \frac{2}{(M_{*}/L)I(R)} \int_R^{\infty} dr' \left[1-\frac{R^{2}}{r_{a}^{2}+r'^{2}}\right]\frac{\rho_{*}(r') \sigma_{r}^{2}(r') r'}{(r'^{2}-R^{2})^{1/2}},
\end{equation}
with I(R) being the surface brightness profile (modeled as a Jaffe
profile with parameters derived from surface photometry).

Before comparing the model with the observations it is necessary to
take two furher steps.  First, we must account for the atmospheric
seeing, which blurs spectroscopic measurements in the spatial
direction.  This can be modeled as:
\begin{equation}
\label{eq:sigma_see}
\sigma_{s}^{2}(R) = \frac{\int d^{2}R' P(R-R') I(R') \sigma_{p}^{2}(R') }{\int d^{2}R' P(R-R') I(R') }
\end{equation}
where we assume a Gaussian point-spread function, P(R-R') (see
discussion in Binney \& Merrifield 1998; Eqn. 4.6-4.8).  

Second, we must account for the non negligible slit width and spatial
binning used. This was calculated numerically such that,

\begin{equation}
\label{eq:sigma_bin}
\sigma_{b}^{2}(R) = \frac{\int_{A} dA' I_{s}(R') \sigma_{s}^{2}(R') }{\int_{A} dA' I_{s}(R) },  
\end{equation}

where A is the area of the slit used for a given measurement and
$I_{s}(R)$ is the seeing corrected intensity at a given projected
radius.

With an understanding of the observational setup and seeing conditions
one can calculate the expected velocity dispersion for a given set of
free parameters, \{$M_{*}$/L,$\beta$, $\delta_{c}$\}.  Analogous to
the likelihood technique employed in the lensing appendix subsection,
one can construct a likelihood for the velocity dispersion profile of
the BCGs by comparing the expected velocity dispersion for a given set
of free parameters with the measured velocity dispersion,

\begin{equation}
\label{eq:like_vd}
P(M_{*}/L,\delta_{c},\beta) \propto exp \{-\frac{1}{2} \Sigma_{i} \left[ \frac{\sigma_{i} - \tilde{\sigma}_{i}(M_{*}/L,\delta_{c},\beta)}{\Delta_{i}} \right]^{2} \}.
\end{equation}
Here, $\sigma$ is the velocity dispersion in a given bin and
$\Delta_{i}$ is the uncertainty in the measurement. 

With both the lensing and velocity dispersion likelihoods calculated,
it is now possible to find constraints on the inner dark matter
density slope, $\beta$.  Since the two techniques are independent, the
total likelihood for a given set of free parameters is just the
product of the lensing and velocity dispersion likelihoods.

\clearpage

\begin{table*}[t]
\begin{center}
\caption{Optical/NIR Imaging Log\label{imglog}}
\begin{tabular}{lccccc}
\tableline\tableline
Cluster & $z_{clus}$&Date & Telescope/ & Filter&Exposure\\
&&&Instrument&&time (ks)\\
\tableline
MS2137-23&0.313&May 28-31,1995&HST/WFPC2&F702W&22.2\\
Abell 383&0.189&Jan 25, 2000&HST/WFPC2&F702W&7.5\\
&&Dec 17, 2002&Keck/NIRC&$K_{s}$&1.0\\
Abell 963&0.206&May 7, 2000&HST/WFPC2&F702W&7.8\\
RXJ 1133&0.394\tablenotemark{a}&Feb 20, 2001&HST/WFPC2&F606W&1.0\\
&&Dec 17, 2002&Keck/NIRC&$K_{s}$&0.8\\
MACS 1206&0.440\tablenotemark{a}&April 13, 2002&Keck/ESI&I&0.3\\
Abell 1201&0.169&April 7, 2001&HST/WFPC2&F606W&0.8\\
\tableline
\end{tabular}
\tablecomments{Imaging Observation Log of the clusters in our
sample.}
\tablenotetext{a}{New spectroscopic measurement}
\end{center}
\end{table*}

\clearpage

\begin{table*}
\scriptsize
\begin{center}
\caption{BCG Photometric Properties\label{tab:phot}}
\begin{tabular}{lcccccc}
\tableline\tableline
Cluster&Filter&$R_{e}$&M &$SB_{e}$&K-color&$(1-b/a)_{e}$\\
&&(arcsec/kpc)&(mag)&(mag arcsec$^{-2}$)&Correction&\\
\hline
MACS 1206&gunn I&$2.08\pm0.17$&$17.48\pm0.07$&$22.46\pm0.23$&$0.81\pm0.03$&$0.35\pm0.05$\\
&V&$12.75\pm1.04$&$-23.93\pm0.08$&$21.57\pm0.23$\\
MS 2137-23&F702W&$5.02\pm0.50$&$16.48\pm0.07$&$23.58\pm0.34$&$0.49\pm0.03$&$0.17\pm0.01$\\
&V&$24.80\pm1.68$&$-24.38\pm0.09$&$22.76\pm0.34$\\
RXJ 1133&F606W&$5.18\pm0.12$&$18.00\pm0.06$&$24.96\pm0.33$&$0.41\pm0.03$&$0.18\pm0.05$\\
&B&$29.73\pm0.69$&$-23.44\pm0.07$&$23.89\pm0.33$\\
Abell 383&F702W&$13.75\pm0.60$&$14.67\pm0.06$&$22.95\pm0.25$&$0.60\pm0.04$&$0.19\pm0.03$\\
&V&$46.75\pm2.04$&$-24.78\pm0.07$&$22.72\pm0.25$\\
Abell 1201&F606W&$15.01\pm0.10$&$15.44\pm0.08$&$24.81\pm0.21$&$0.10\pm0.05$&$0.32\pm0.02$\\
&V&$46.68\pm0.31$&$-24.23\pm0.09$&$24.18\pm0.21$\\
Abell 963&F702W&$11.04\pm0.14$&$15.08\pm0.05$&$23.67\pm0.27$&$0.59\pm0.03$&$0.36\pm0.02$\\
&V&$40.19\pm0.51$&$-24.38\pm0.06$&$23.39\pm0.27$\\

\tableline
\end{tabular}
\tablecomments{Photometric properties derived from our 2D
surface brightness profile fitting.  The first line for each cluster
is in the observed filter while the second is in a rest filter.  The
K-color correction and ellipticity at the effective radius are listed
as well.}
\end{center}
\end{table*}

\clearpage

\begin{table*}
\begin{center}
\caption{Gravitational Arc Properties\label{tab:geo}}
\begin{tabular}{lcccc}
\tableline\tableline
Cluster&$R_{rad}$&$R_{tan}$&$z_{radial}$&$z_{tan}$\\
&(arcsec)&(arcsec)\\
\hline
MACS 1206&-&$21.3\pm0.4$&-&1.035\tablenotemark{a}\\
MS 2137-23&$4.5\pm0.3$&$15.35\pm0.20$&1.502\tablenotemark{b}&1.501\tablenotemark{b}\\
RXJ 1133&$3.2\pm0.5$&$10.9\pm0.3$&1.544\tablenotemark{a}&1.544\tablenotemark{a}\\
Abell 383&$1.90\pm0.6$&$15.7\pm0.4$&1.010\tablenotemark{a}&1.009\tablenotemark{a}\\
Abell 1201&-&$2.2\pm0.3$&-&0.451\tablenotemark{c}\\
Abell 963&-&$11.9\pm0.2$&-&0.77\tablenotemark{d}\\

\tableline
\end{tabular}
\tablenotetext{a}{New spectroscopic measurement}\tablenotetext{b}{Sand et al.\ 2002}
\tablenotetext{c}{Edge et al.\ 2003}
\tablenotetext{d}{Ellis, Allington-Smith \& Smail 1991}
\tablecomments{Geometric properties of the gravitational arcs and
BCGs, along with the distance scale for each cluster based on the
adopted cosmology.  The BCG ellipticity at approximately the effective
radius, $R_{e}$, and the positions of the gravitational arcs with
respect to the BCG center are listed.}
\end{center}
\end{table*}

\clearpage

\begin{table*}
\scriptsize
\begin{center}
\caption{Spectroscopic Observation Log\label{tab:specsum}}
\begin{tabular}{lccccccc}
\tableline\tableline
Cluster &Date &Target &Instrument &Exposure &Seeing&Pos. Angle& Slit\\ 
&  & &  &time (ks) & (``)&(degrees)&size (``)\\ 
\tableline 
MS2137-23&July 21,2001&Arcs/BCG&ESI&5.9&0.6&0&1.25 \\ 
Abell 383&Dec 12, 2002&Rad. Arc/BCG&ESI&5.4&0.7&28&1.25 \\ 
&Oct 19, 2001&Tan. Arc&LRIS&3.8&0.7&30&1.0 \\
RXJ 1133&Apr 11-12, 2002&BCG/Rad. Arc&ESI&12.6&0.6-0.7&-24&1.25 \\ 
&Apr 11, 2002&Tan Arc&ESI&3.6&0.7&10&1.25 \\ 
MACS 1206&Apr 13, 2002&BCG&ESI&9.0&0.75&98&1.25 \\ 
&Apr 13, 2002&Tangential arc&ESI&3.6&0.75&273&1.25 \\ 
Abell 1201&Apr 12, 2002&BCG/tan arc&ESI&3.6&0.6&-32&1.25 \\ 
Abell 1201&Apr 12, 2002&BCG/tan arc&ESI&3.6&0.6&-25&1.25 \\ 
Abell 963&Mar 28, 2001&BCG&LRIS&4.8&0.7&-15.5&1.5 \\
\tableline
\end{tabular}
\tablecomments{Summary of the spectroscopic observations.}
\end{center}
\end{table*}

\clearpage

\begin{inlinetable}
\centering
\begin{tabular}{lcc}
\tableline\tableline
Cluster &Spatial Binning (arcsec)&$\sigma$ (\kms)\\ 

\tableline

MACS 1206&-0.30 - 0.30&$257\pm39$\\
&0.30 - 1.37&$245\pm50$\\
&-0.30 - -1.37&$259\pm52$\\
RXJ 1133&0.0 - 0.61 &$333\pm30$\\
&0.61 - 1.22 &$306\pm41$\\
&1.22 - 1.98 &$337\pm67$\\
Abell 1201&-0.35 - 0.35&$231\pm13$\\
PA=-32&-1.06 - -0.35&$257\pm21$\\
&0.35 - 1.06&$232\pm18$\\
&1.06 - 1.76 &$224\pm28$\\
Abell 1201&-0.28 - 0.28 &$238\pm16$\\
PA=-25&-0.99 - -0.28 &$252\pm20$\\
&0.28 - 0.99 &$223\pm15$\\
&0.99 - 1.69 &$207\pm20$\\
Abell 383&-0.49 - 0.07 &$319\pm26$\\
&-1.06 - -0.49 &$228\pm25$\\
&-1.62 - -1.06 &$246\pm32$\\
Abell 963&-0.32 - 0.32 &$299\pm22$\\
&-0.97 - -0.32 &$298\pm29$\\
&-1.61 - -0.97 &$271\pm31$\\
&0.32 - 0.97&$282\pm26$\\
&0.97 - 1.61&$253\pm26$\\

\tableline
\end{tabular}
\end{inlinetable}\\
\noindent{Table~5.--- \small Velocity Dispersion Profiles.  Tabulated
velocity dispersion profiles of the BCGs, not including MS 2137-23, which
was presented in STE02.  All slit widths are $1\farcs25$ except for
Abell 963, which is $1\farcs50$. All spatial values are with respect
to the center of the BCG.}

\clearpage

\begin{figure*}[b]
\begin{center}

\mbox{\rotate
\epsfysize=12cm \epsfbox{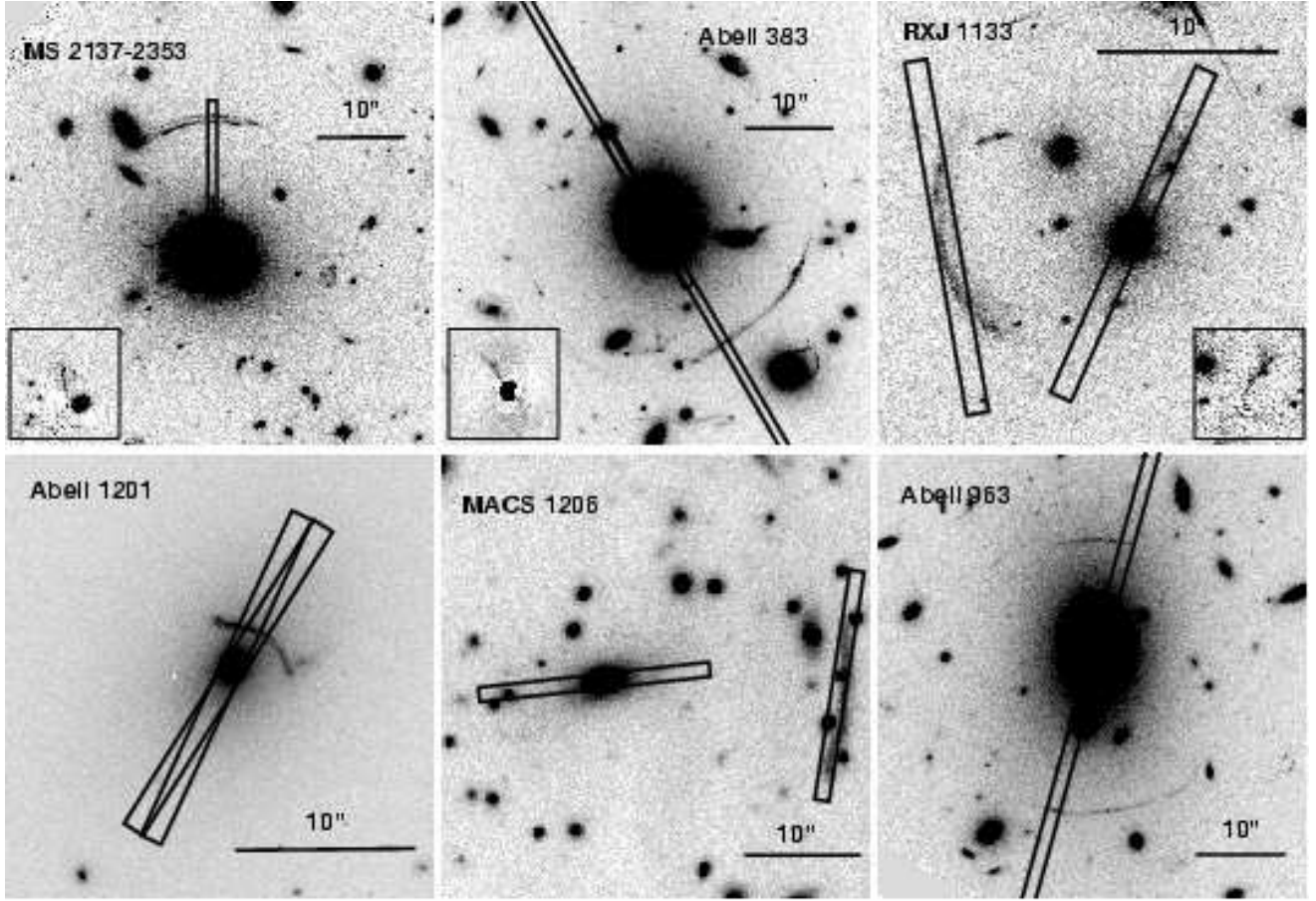}}

\end{center}
\caption{Images of the six clusters in this study.  The top row
features the clusters with both radial and tangential arcs.  The
postage stamp insets show zoomed in BCG subtracted images so that the
radial arcs can be clearly seen.  The bottom row contains those
clusters with tangential arcs only.  The overlaid ``slits'' correspond
to the actual slit positions and sizes that were observed.  See
Table~\ref{tab:specsum} for the spectroscopic observation log.  North
is up and East is to the left in all images. }
\label{fig:image}
\end{figure*}

\clearpage

\begin{figure*}[t]\label{fig:sbplot}
\begin{center}

\mbox{
\mbox{\epsfysize=6cm \epsfbox{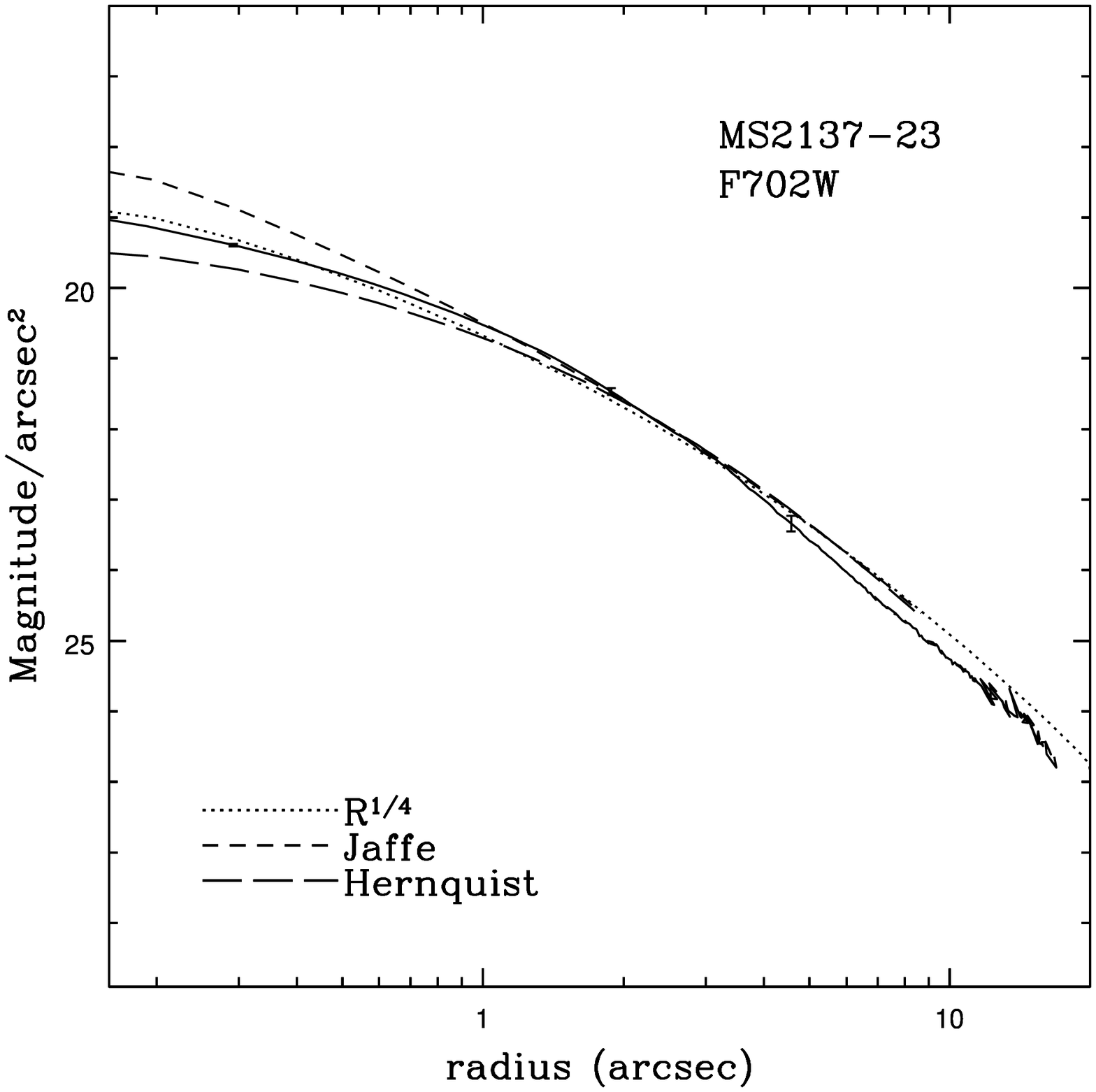}}
\mbox{\epsfysize=6cm \epsfbox{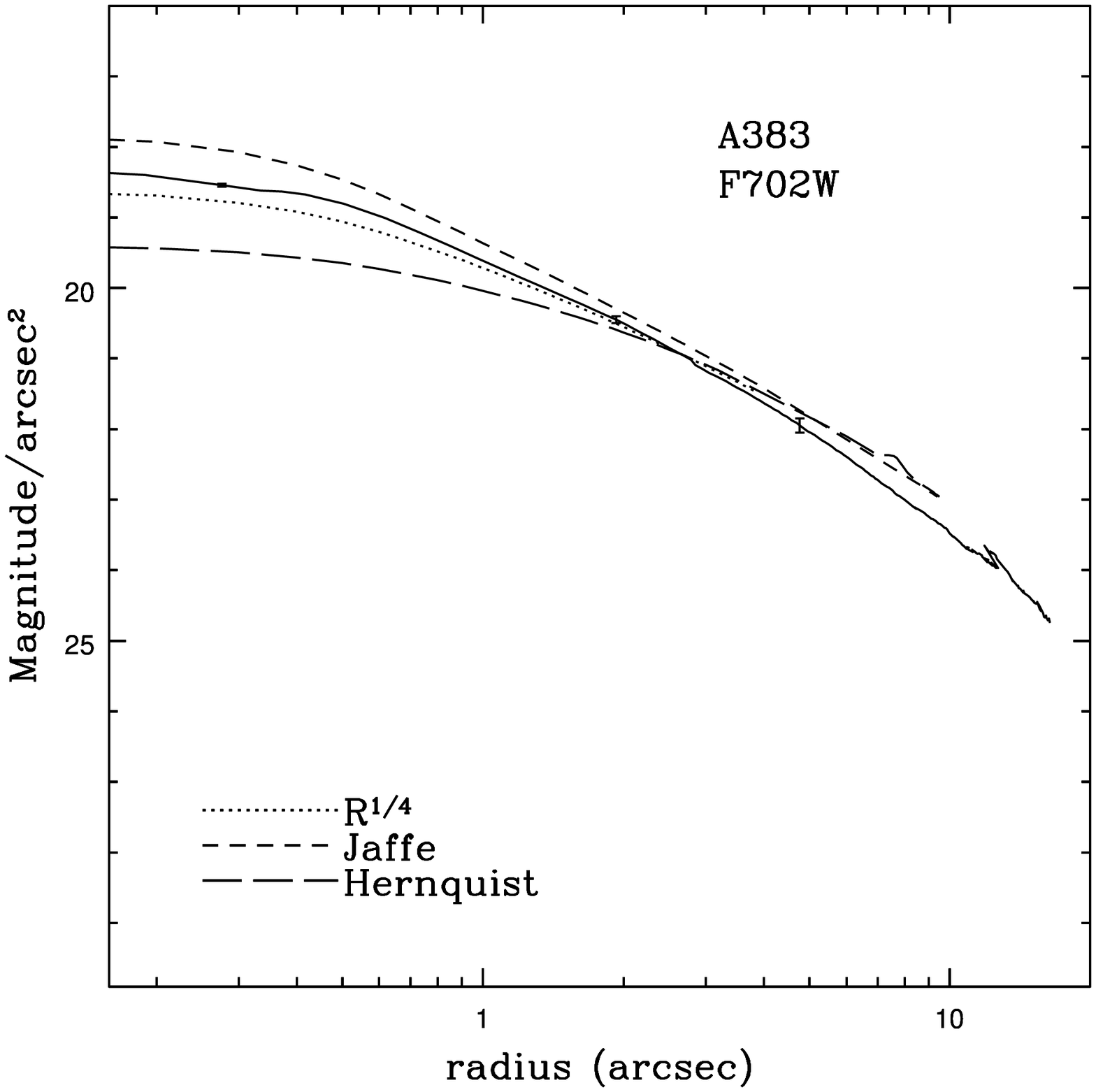}}
\mbox{\epsfysize=6cm \epsfbox{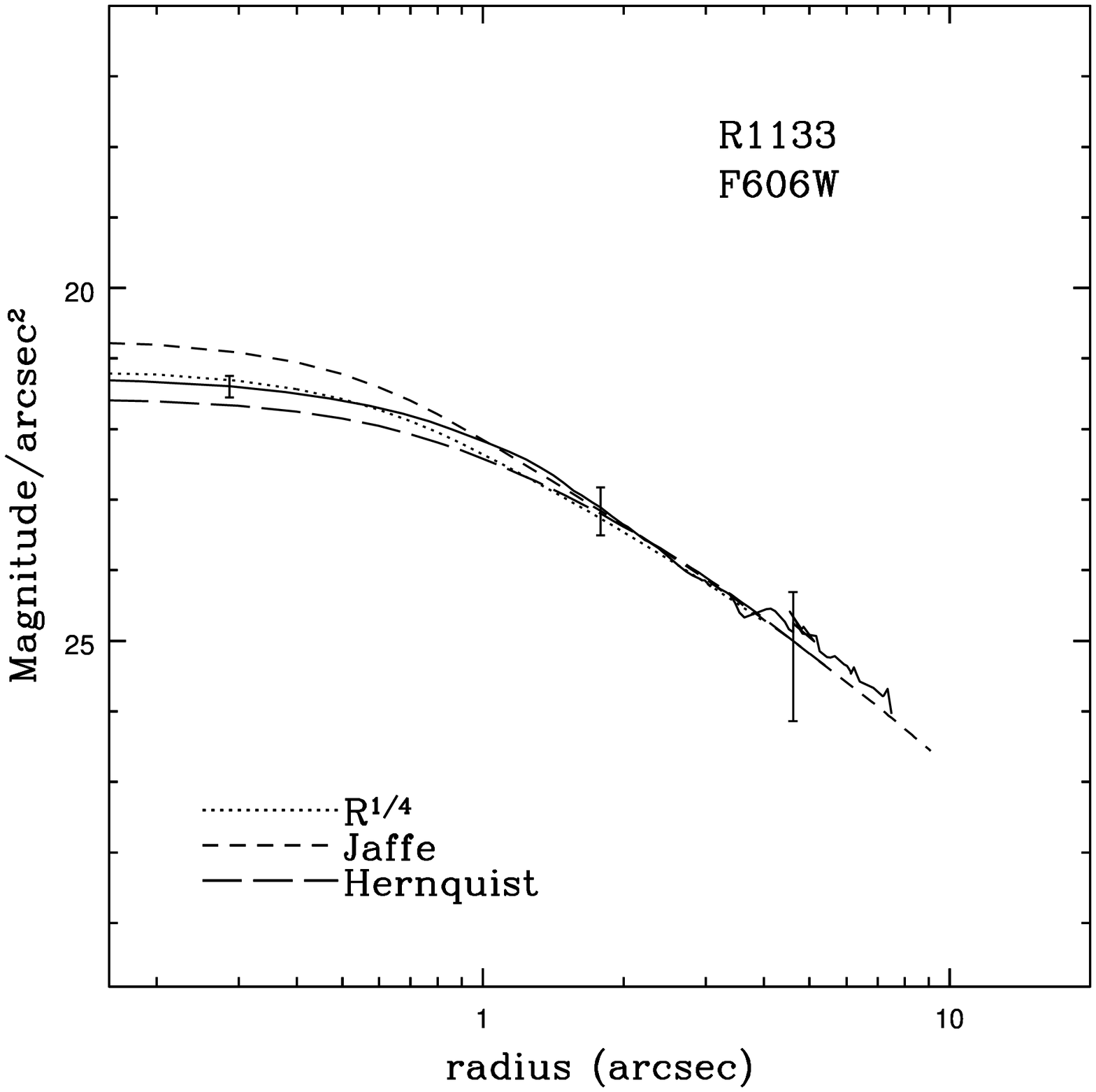}}
}
\mbox{
\mbox{\epsfysize=6cm \epsfbox{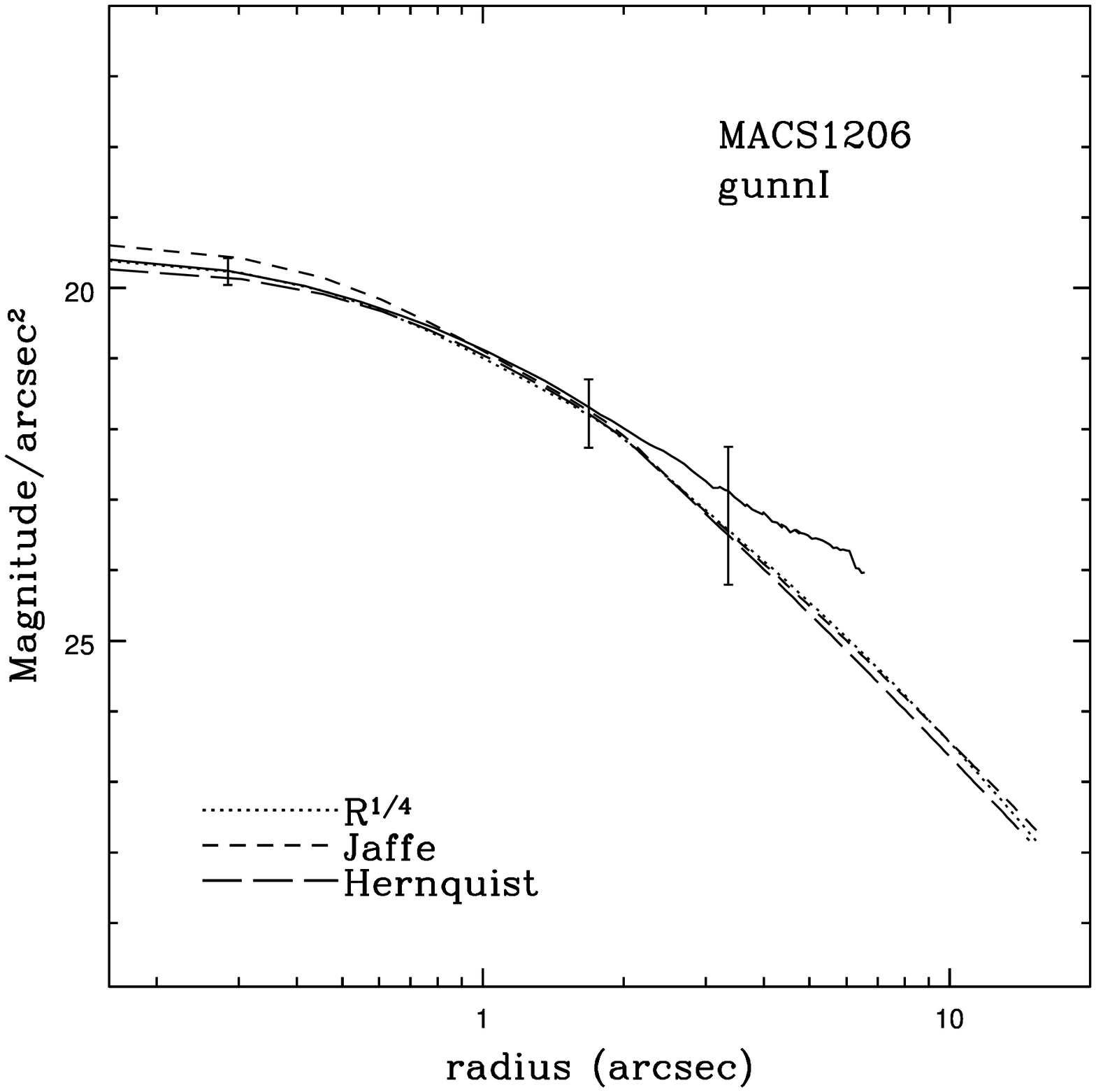}}
\mbox{\epsfysize=6cm \epsfbox{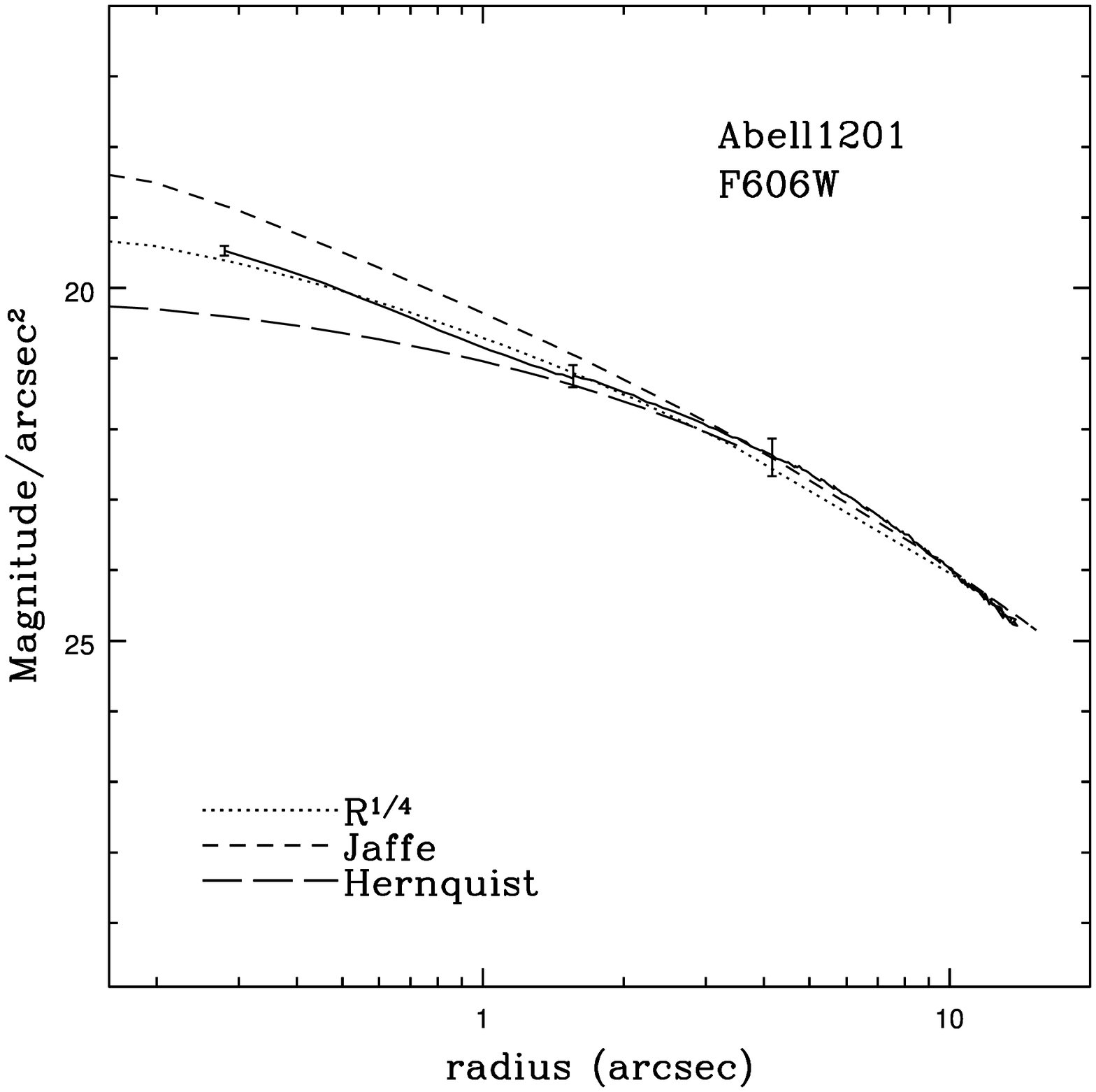}}
\mbox{\epsfysize=6cm \epsfbox{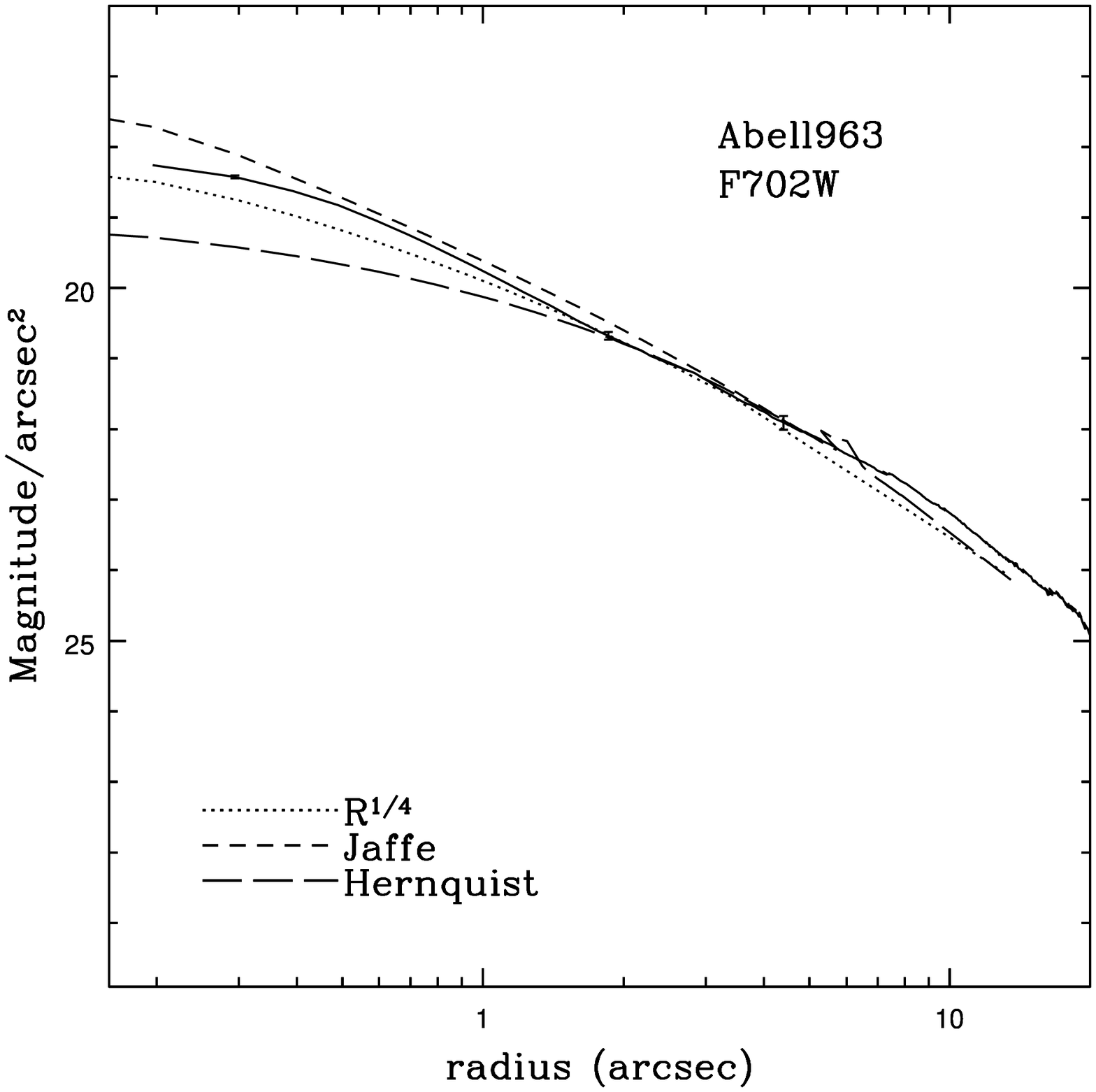}}
}
\caption{Surface brightness profile of the BCGs.  The solid lines are
the measured surface brightness profiles while the other curves are
various parameterizations of the data based on a \dv fit, convolved
with the PSF of the observation.  The uncertainty of the profile is
given at several representative points.  As can be seen, the Jaffe and
Hernquist profile generally bracket the best-fitting \dv at low radii.
See \S 6.3 for a discussion of the effects our chosen luminous mass
component parameterization has on our results.}
\end{center}
\end{figure*}

\clearpage

\begin{figure*}[t] \label{fig:spectra}
\begin{center}

\mbox{
\mbox{\epsfysize=6cm \epsfbox{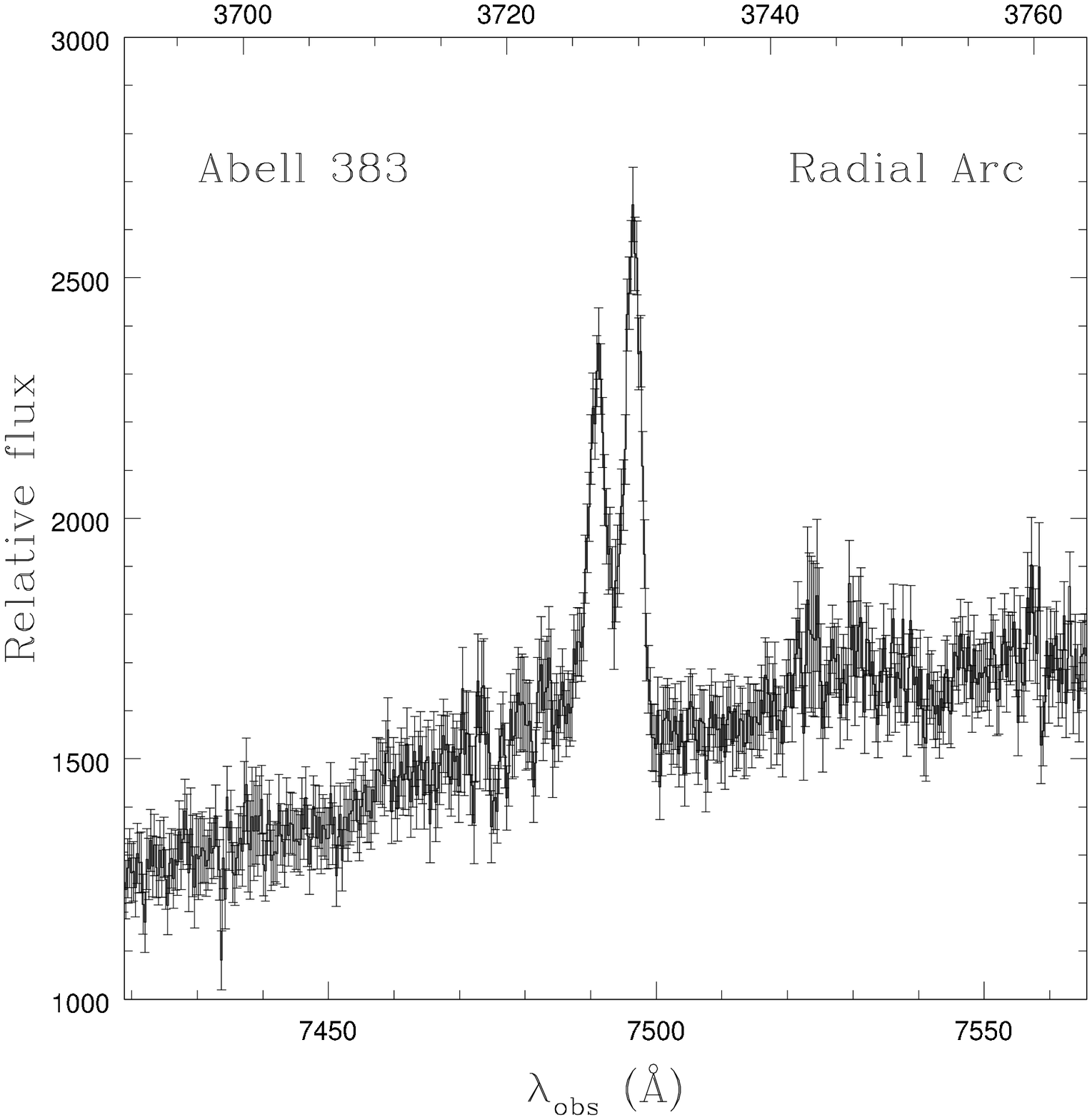}}
\mbox{\epsfysize=6cm \epsfbox{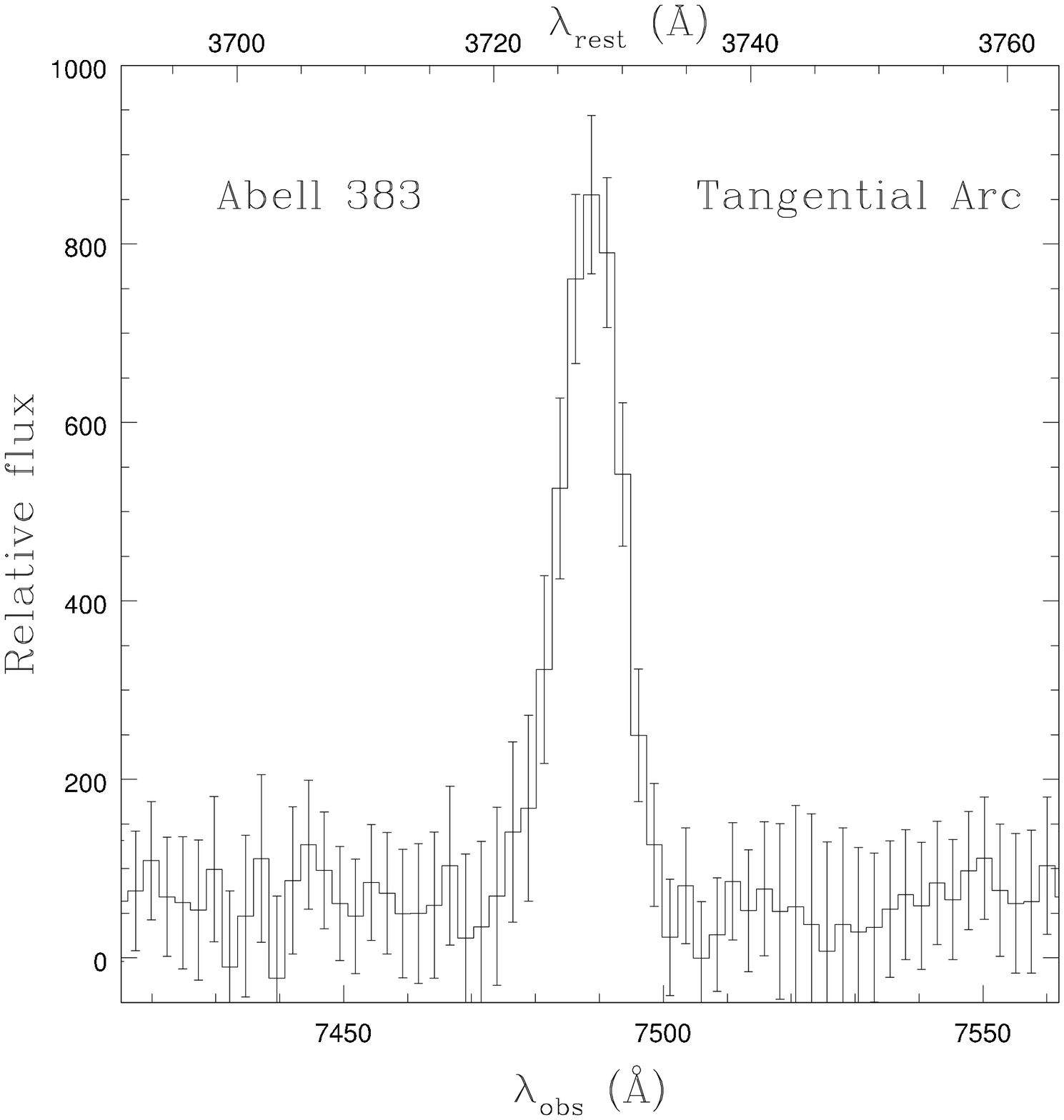}}
\mbox{\epsfysize=6cm \epsfbox{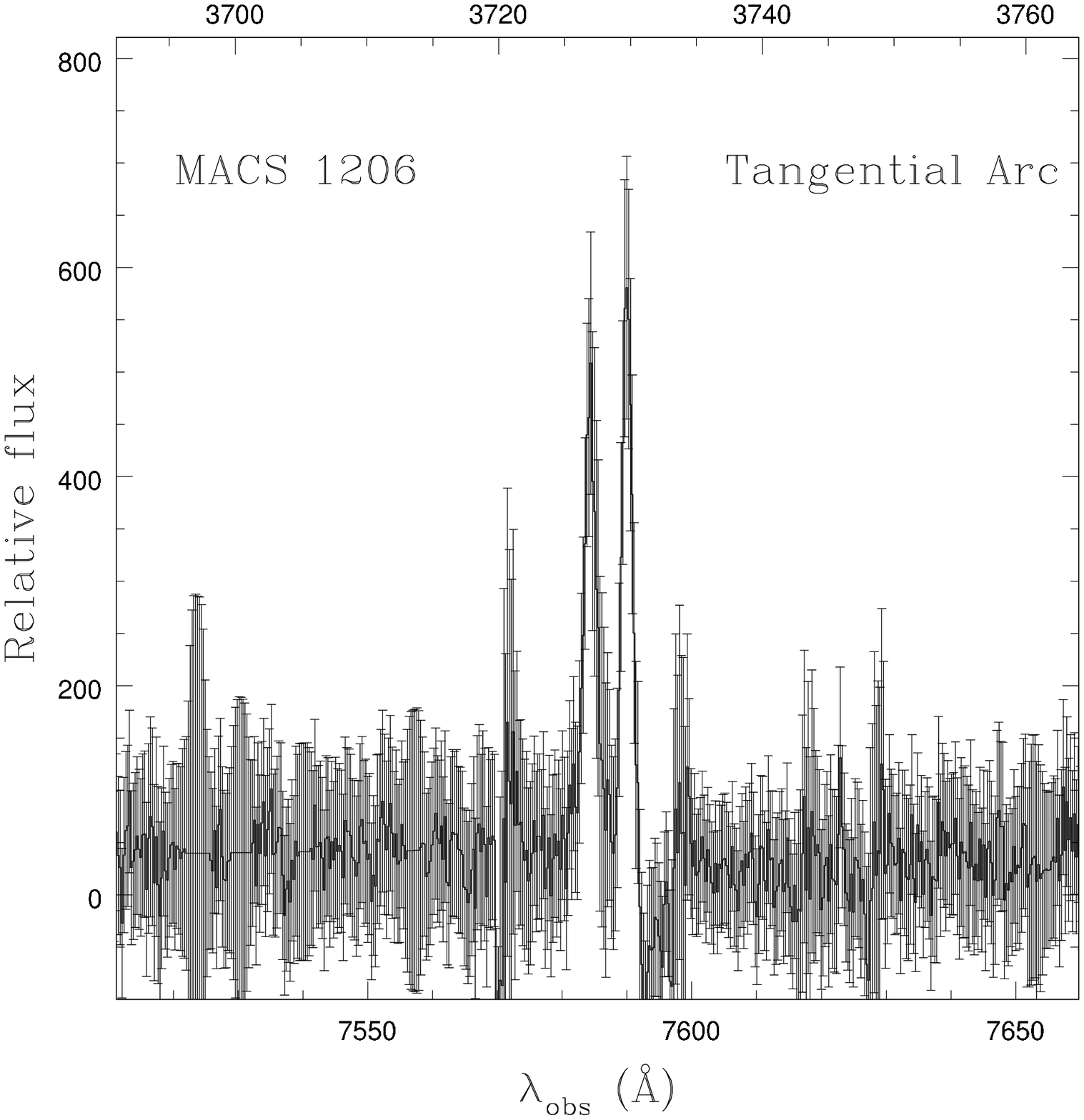}}
}
\mbox{
\mbox{\epsfysize=6cm \epsfbox{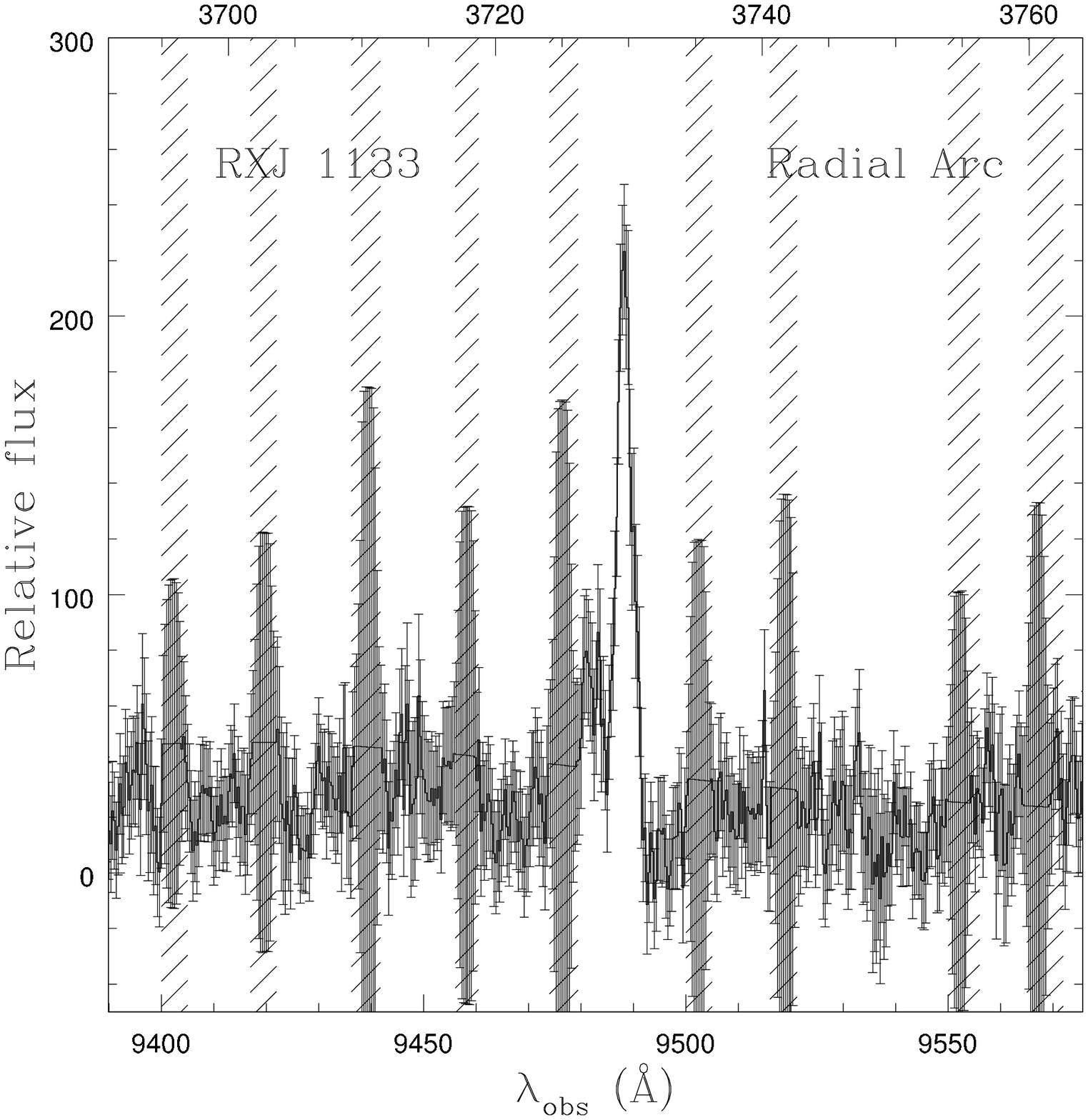}}
\mbox{\epsfysize=6cm \epsfbox{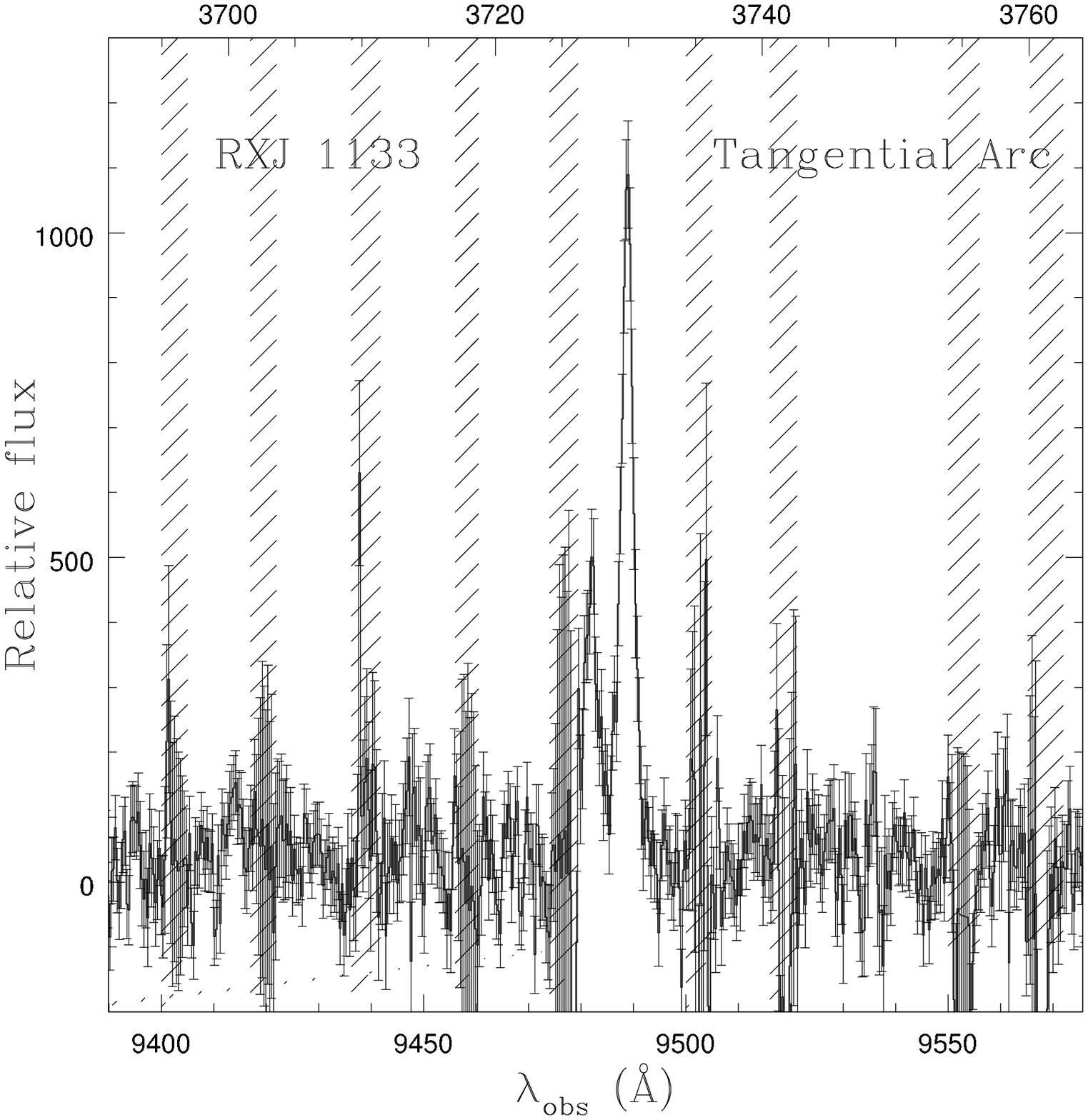}}
\mbox{\epsfysize=6cm \epsfbox{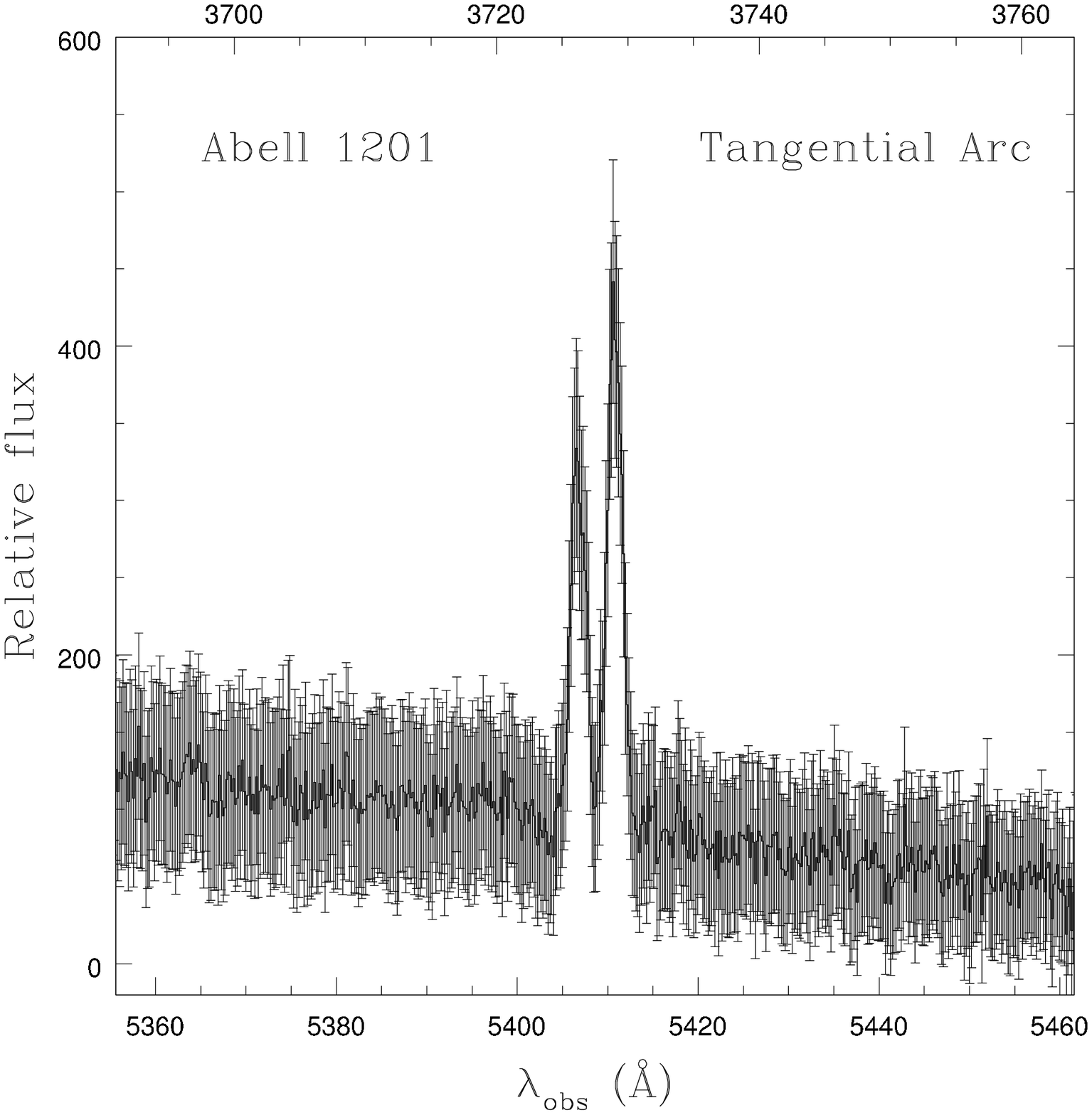}}
}
\caption{New gravitational arc redshift measurements.  All new
redshift measurements of gravitational arcs in this work were
identified by strong [O II] in emission.  Both the radial arc in Abell
383 and the tangential arc in Abell 1201 have strong continuum due to
the nearby presence of the BCG.  See Table~\ref{tab:geo} for a list of
all gravitational arc redshift measurements used in this study. }
\end{center}
\end{figure*}

\clearpage

\begin{figure*}[t]  \label{fig:mlplot}
\begin{center}

\mbox{
\mbox{\epsfysize=5.6cm \epsfbox{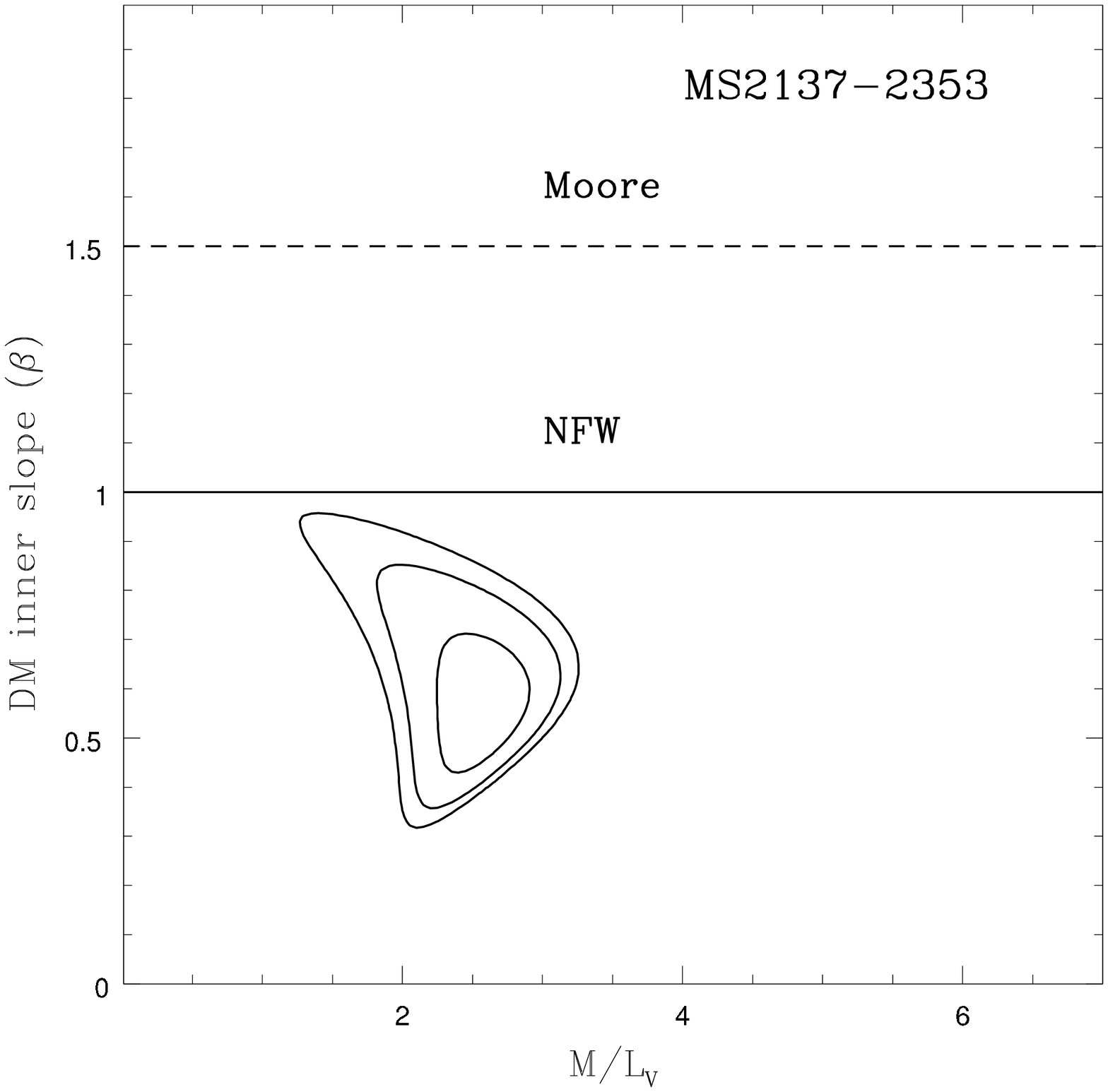}}
\mbox{\epsfysize=5.6cm \epsfbox{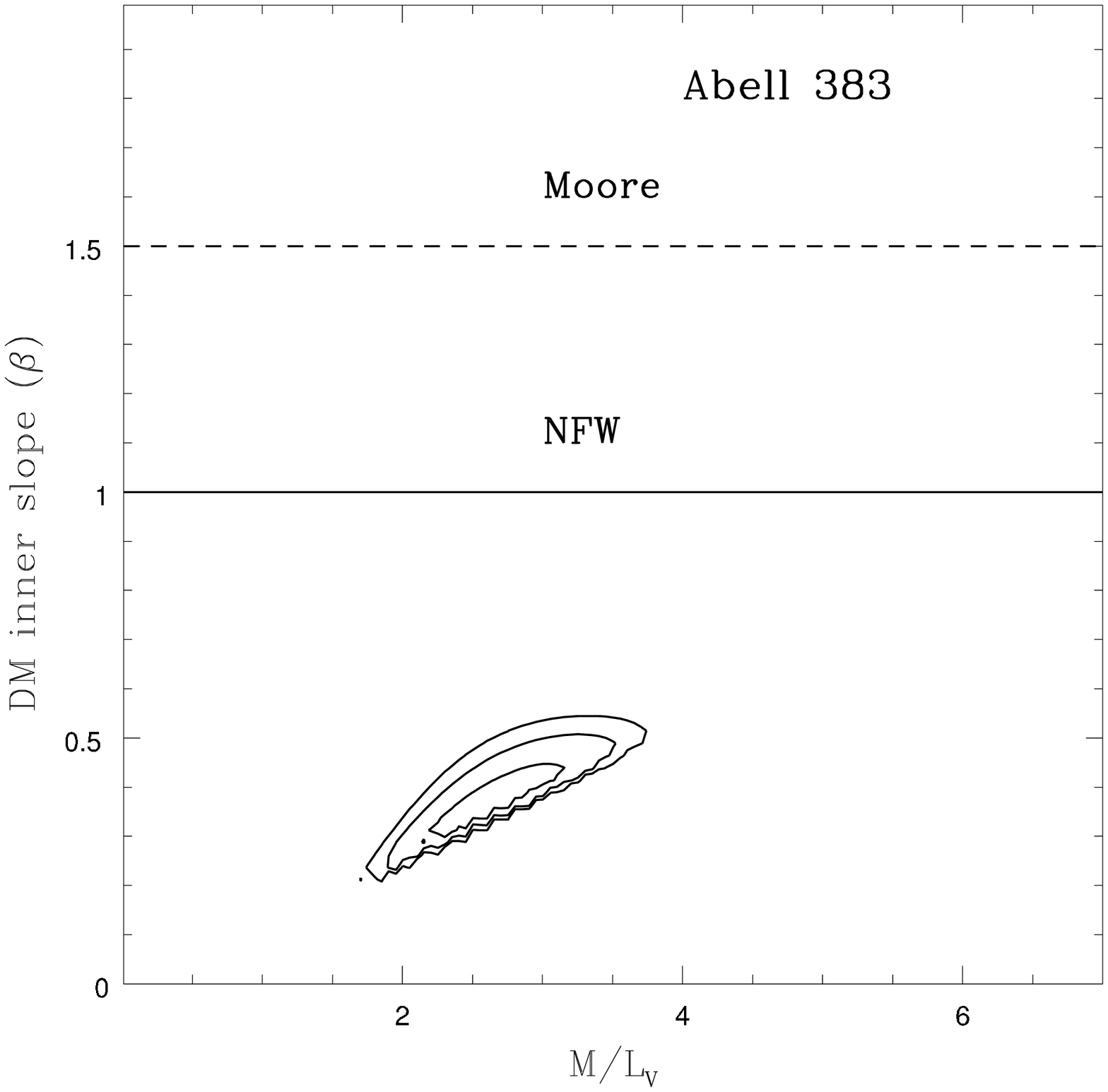}}
\mbox{\epsfysize=5.6cm \epsfbox{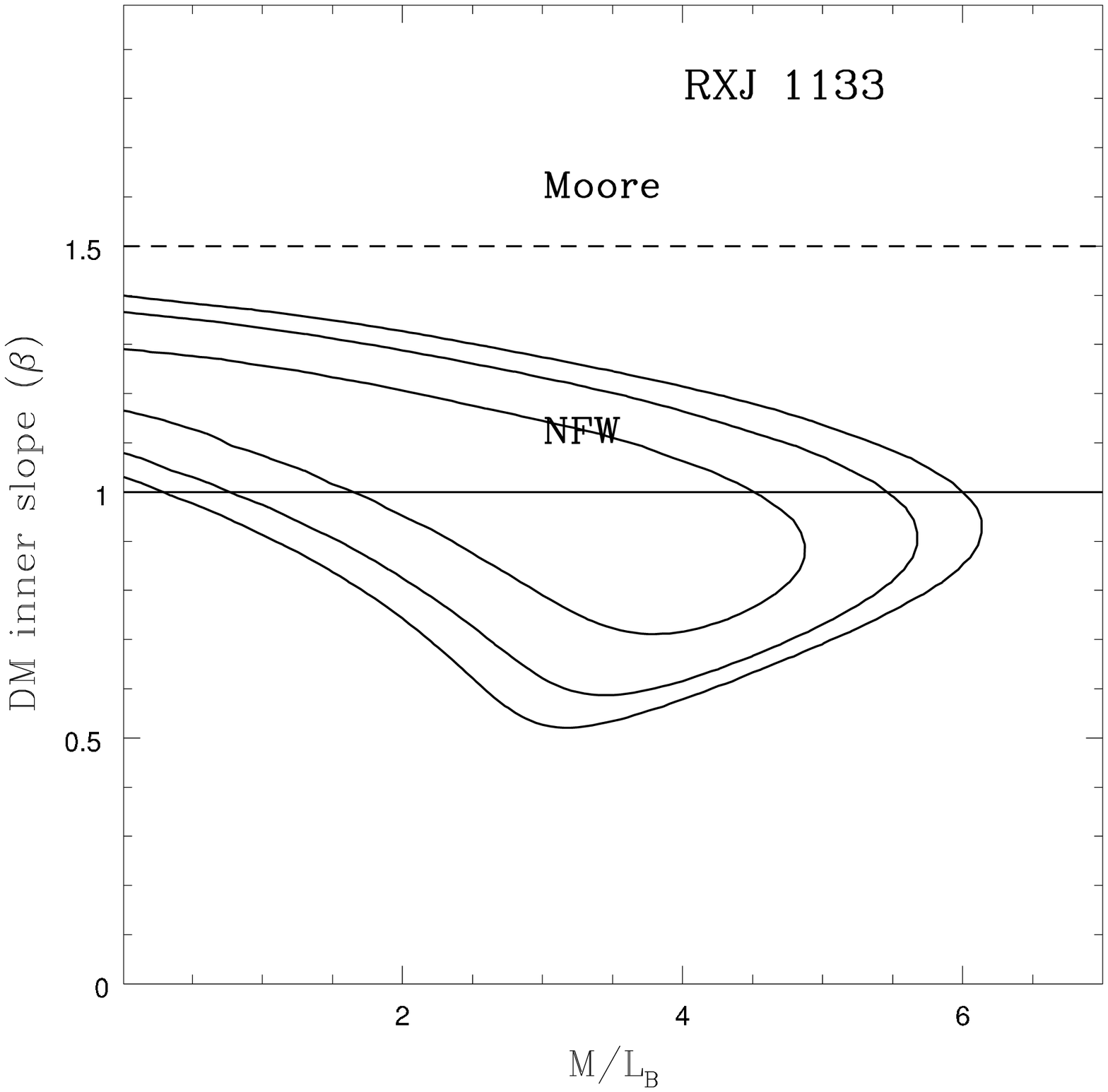}}
}
\mbox{
\mbox{\epsfysize=5.6cm \epsfbox{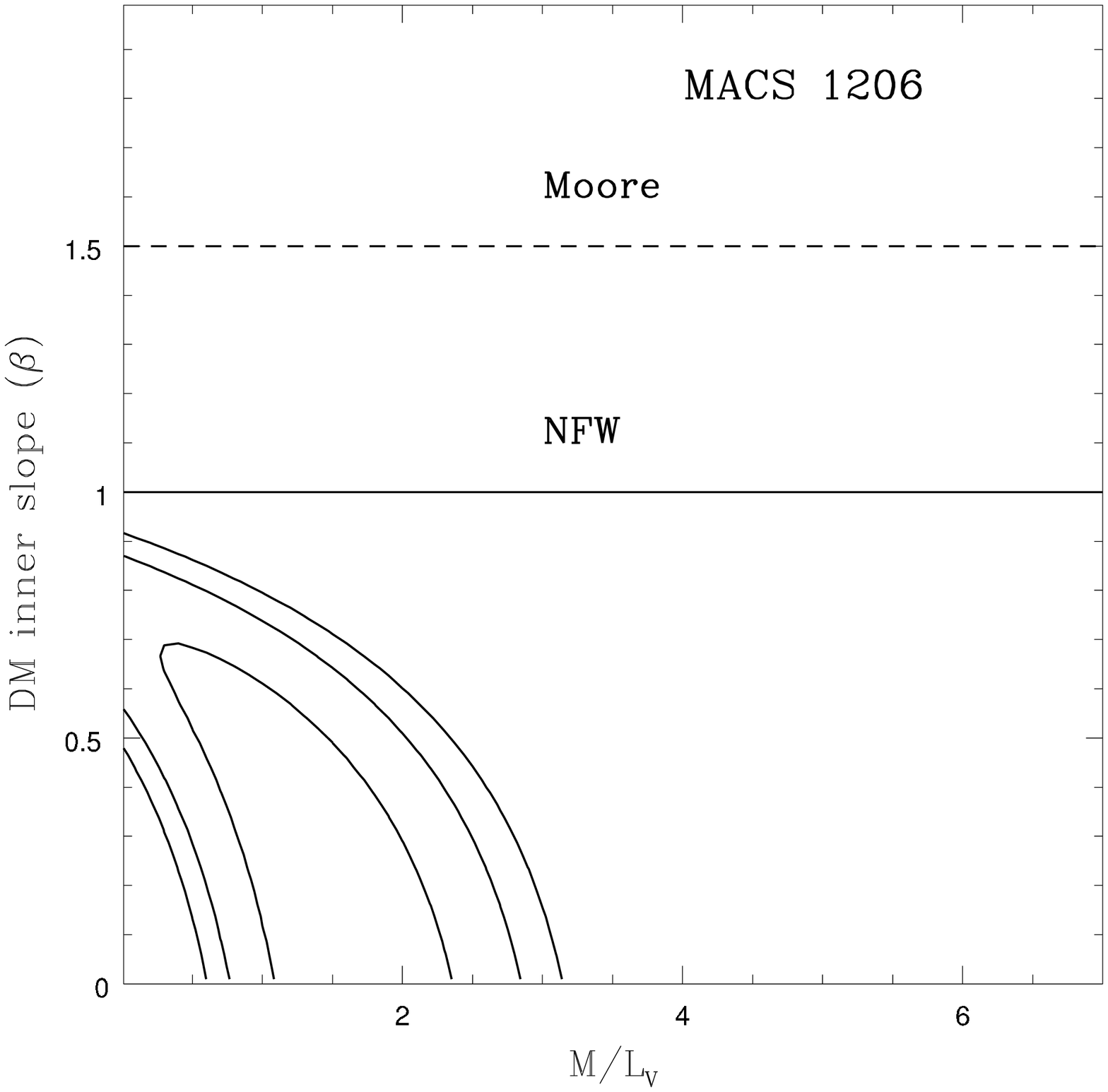}}
\mbox{\epsfysize=5.6cm \epsfbox{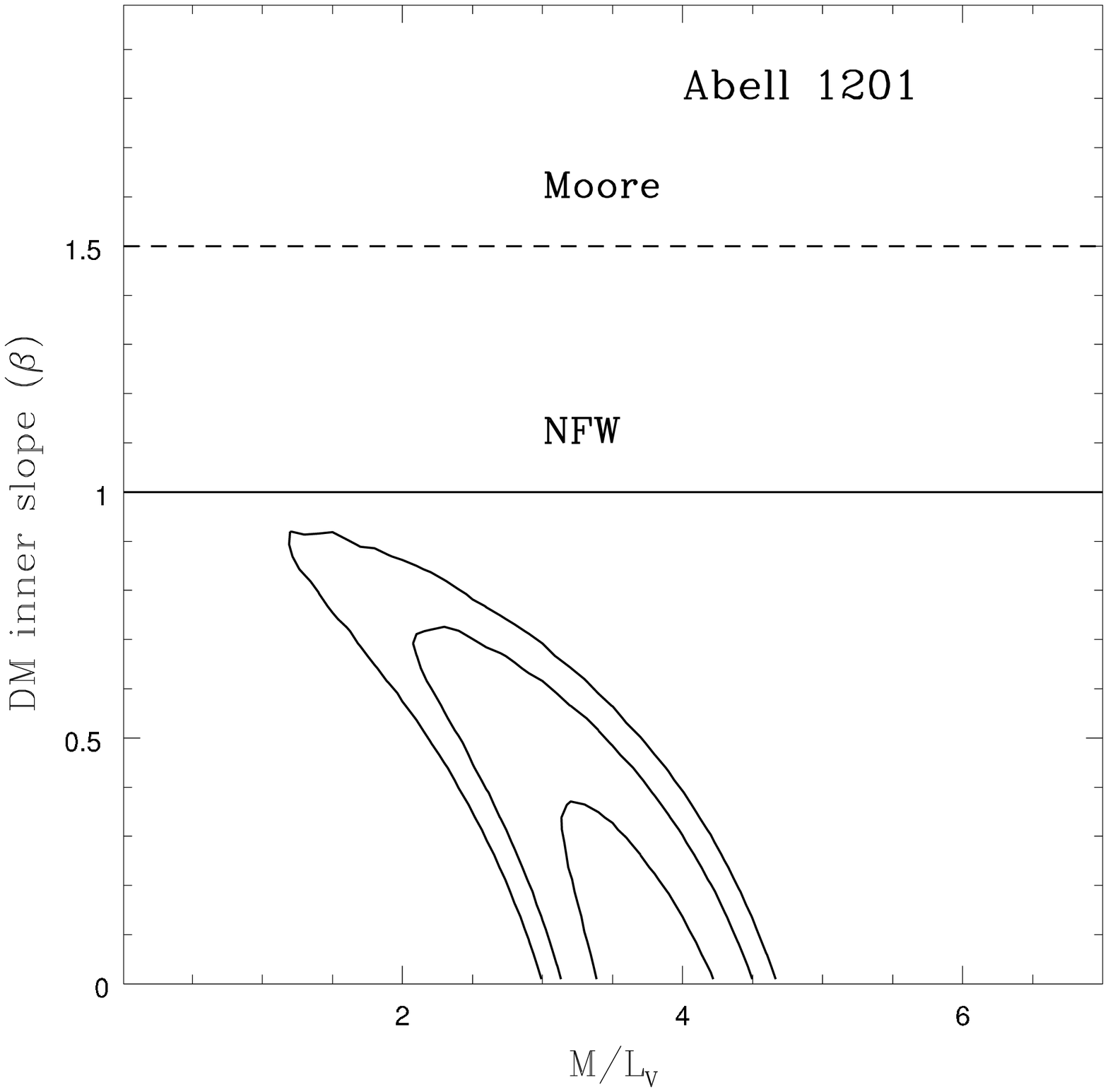}}
\mbox{\epsfysize=5.6cm \epsfbox{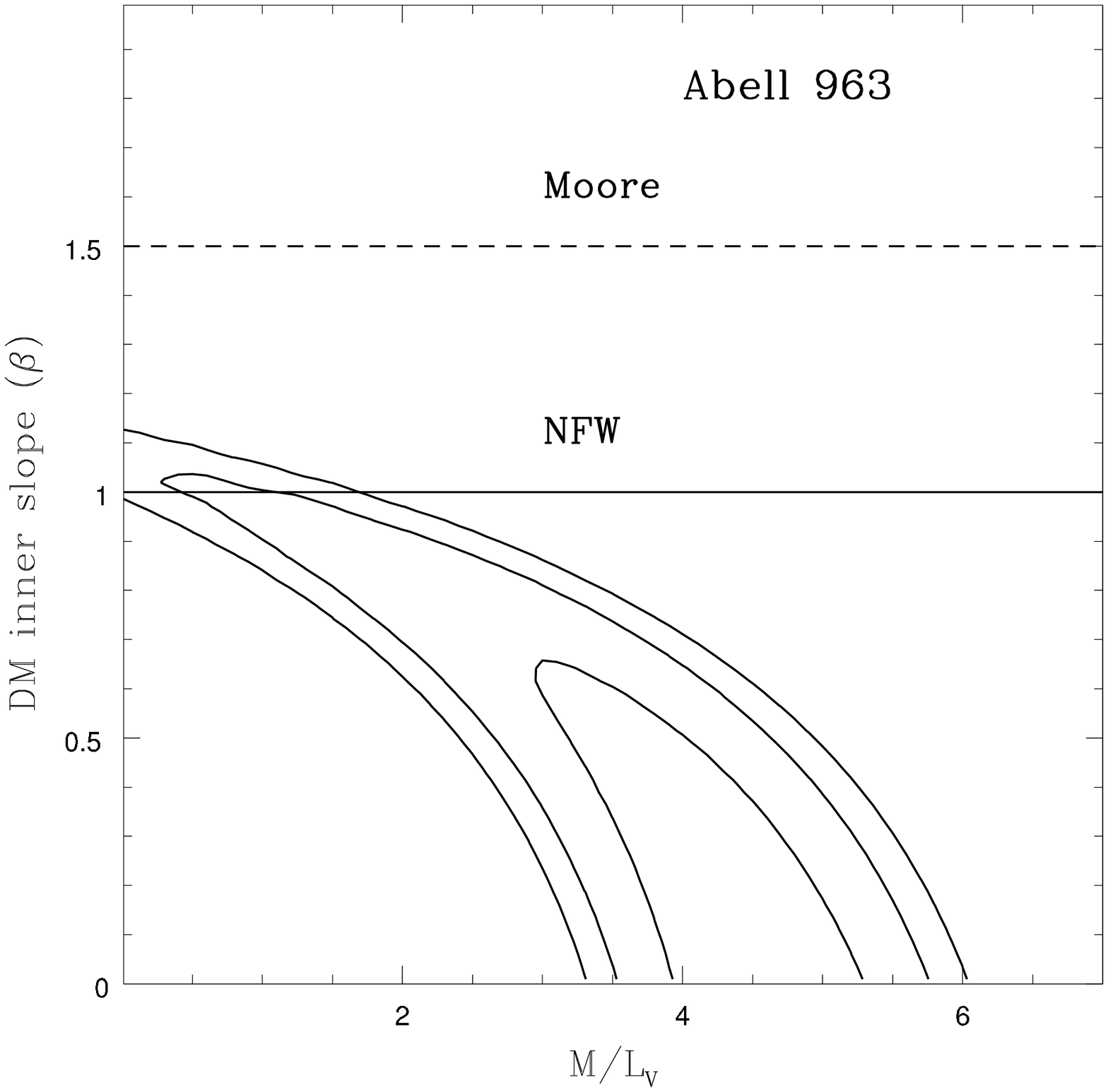}}
}

\caption{Likelihood contours (68\%, 95\% and 99\%) obtained for the
radial arc sample (top row) and the tangential arc sample (bottom row)
with a Jaffe luminous distribution plus a generalized NFW DM
distribution.  These contours were obtained after both the lensing and
dynamical analysis and marginalization with respect to $\delta_{c}$.}
\end{center}
\end{figure*}

\clearpage

\begin{inlinefigure}
\begin{center}
\resizebox{\textwidth}{!}{\includegraphics{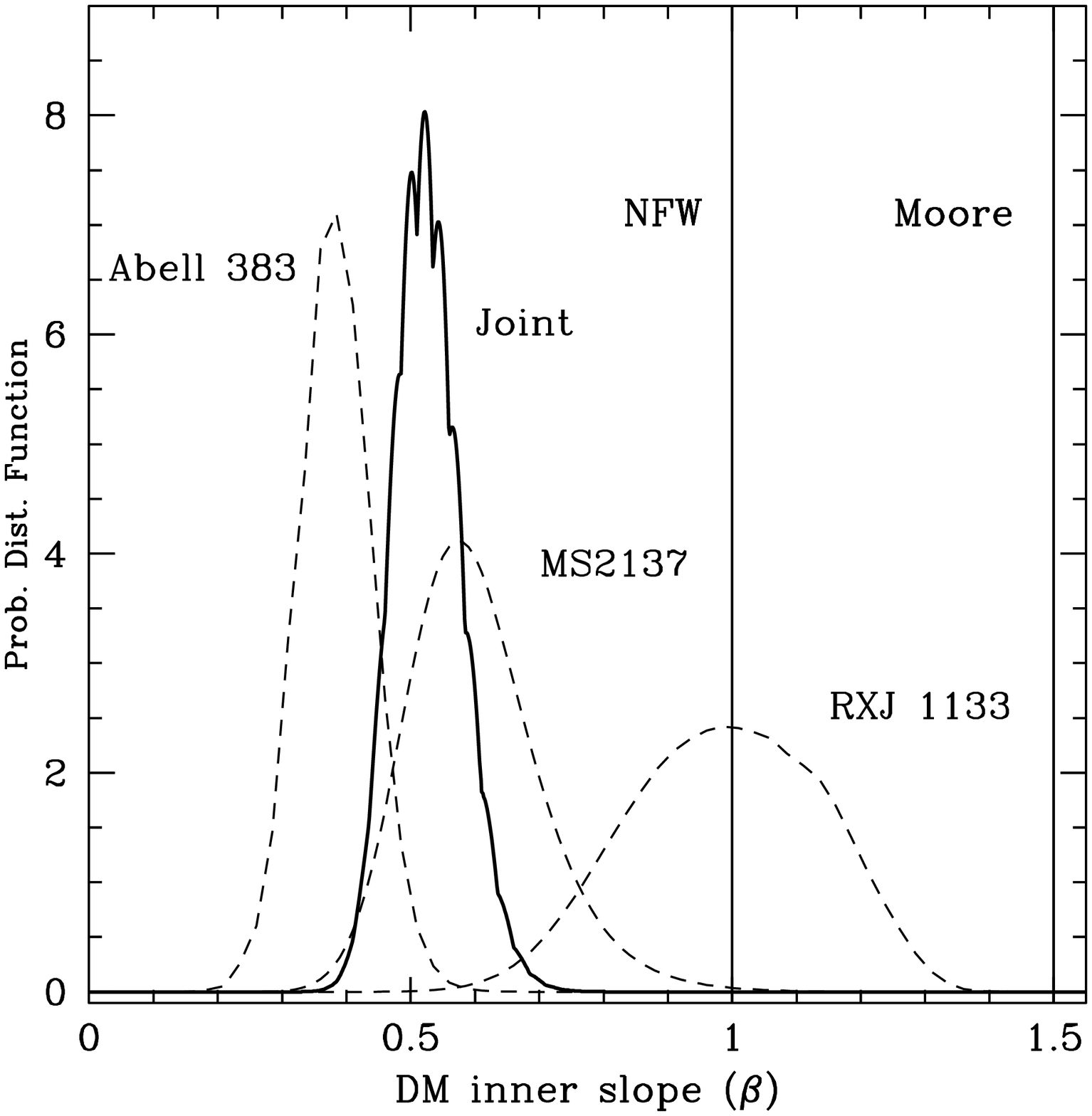}}
\end{center}
\figcaption{Probability distribution function of the DM inner density
slope, $\beta$, for the three radial arc clusters.  Note the wide
scatter in preferred values of $\beta$ from cluster to cluster,
$\Delta\beta\sim$0.3.  The joint distribution was obtained by
multiplying the individual PDFs and normalizing.
\label{fig:radpdf}}
\end{inlinefigure}

\clearpage

\begin{inlinefigure}
\begin{center}
\resizebox{\textwidth}{!}{\includegraphics{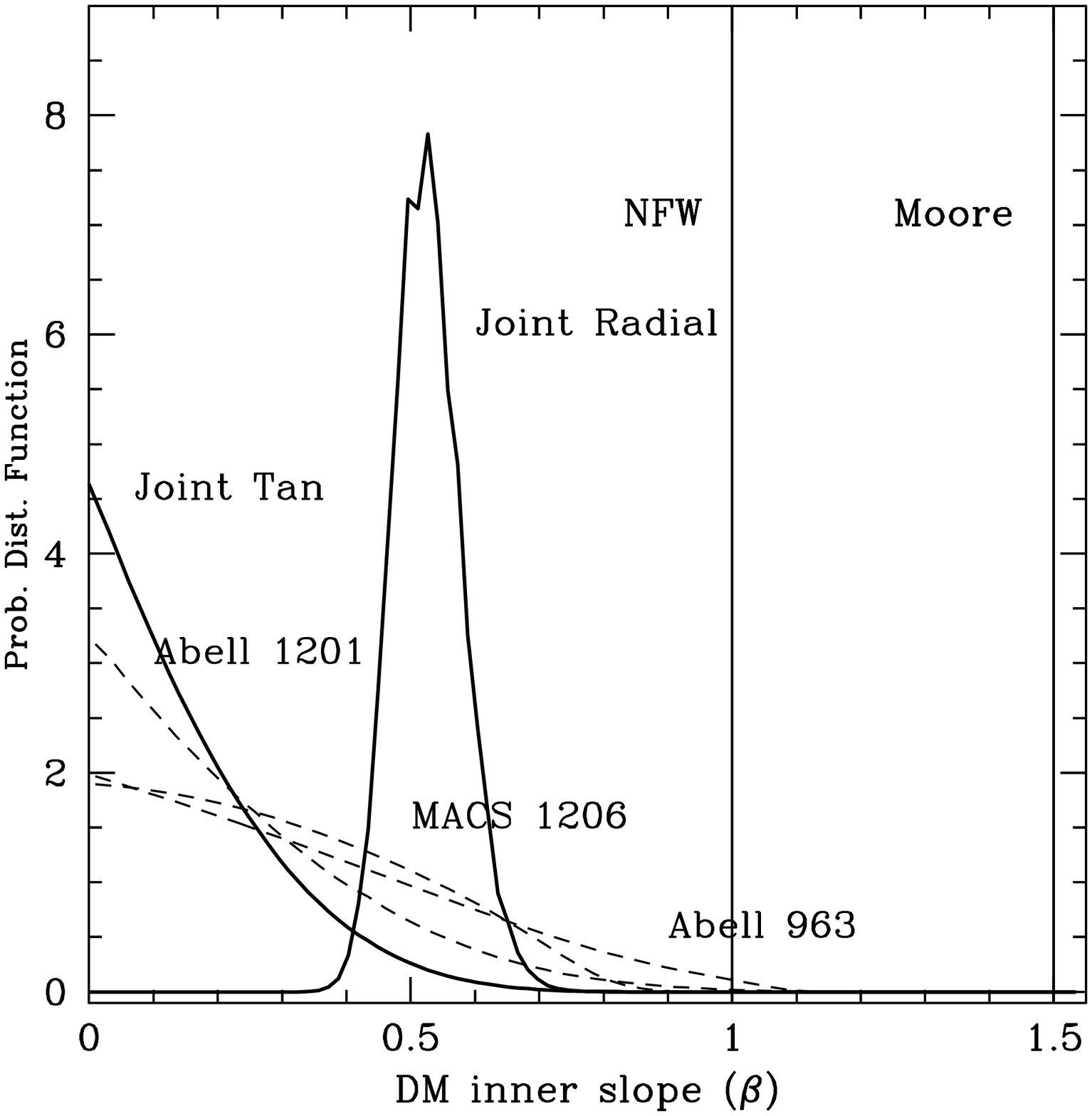}}
\end{center}
\figcaption{Probability distribution function of the DM inner density
slope, $\beta$, for the tangential arc sample.  These effectively
allow us to place an upper limit on $\beta$ for each cluster.  Also
plotted is the joint PDF for the radial arc sample and the tangential
arc sample.  There is no evidence that the radial arc sample is biased
towards lower values of $\beta$.
\label{fig:tanpdf}}
\end{inlinefigure}

\clearpage

\begin{figure*}[t]   \label{fig:vdplot}
\begin{center}

\mbox{
\mbox{\epsfysize=5.8cm \epsfbox{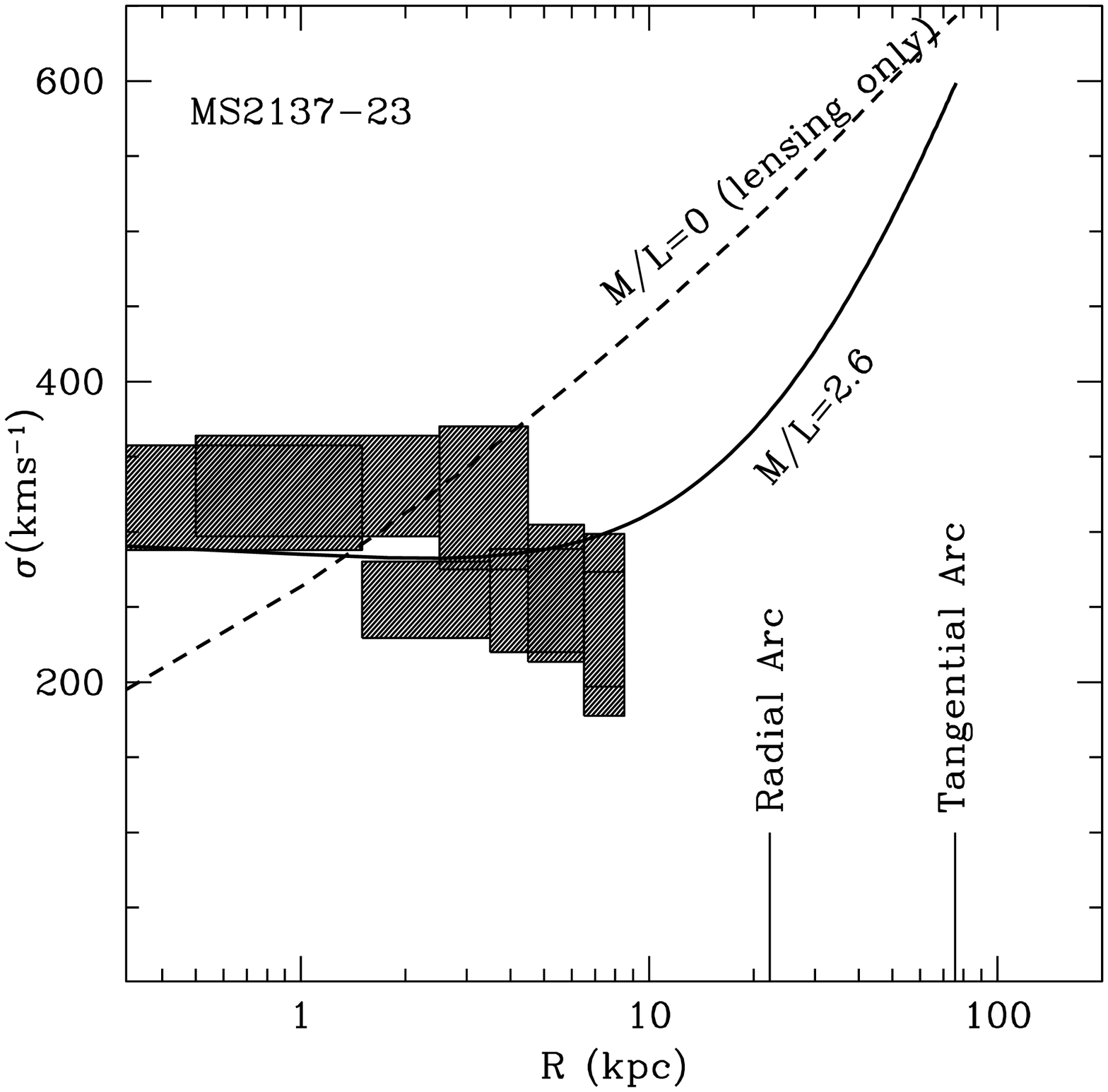}}
\mbox{\epsfysize=5.8cm \epsfbox{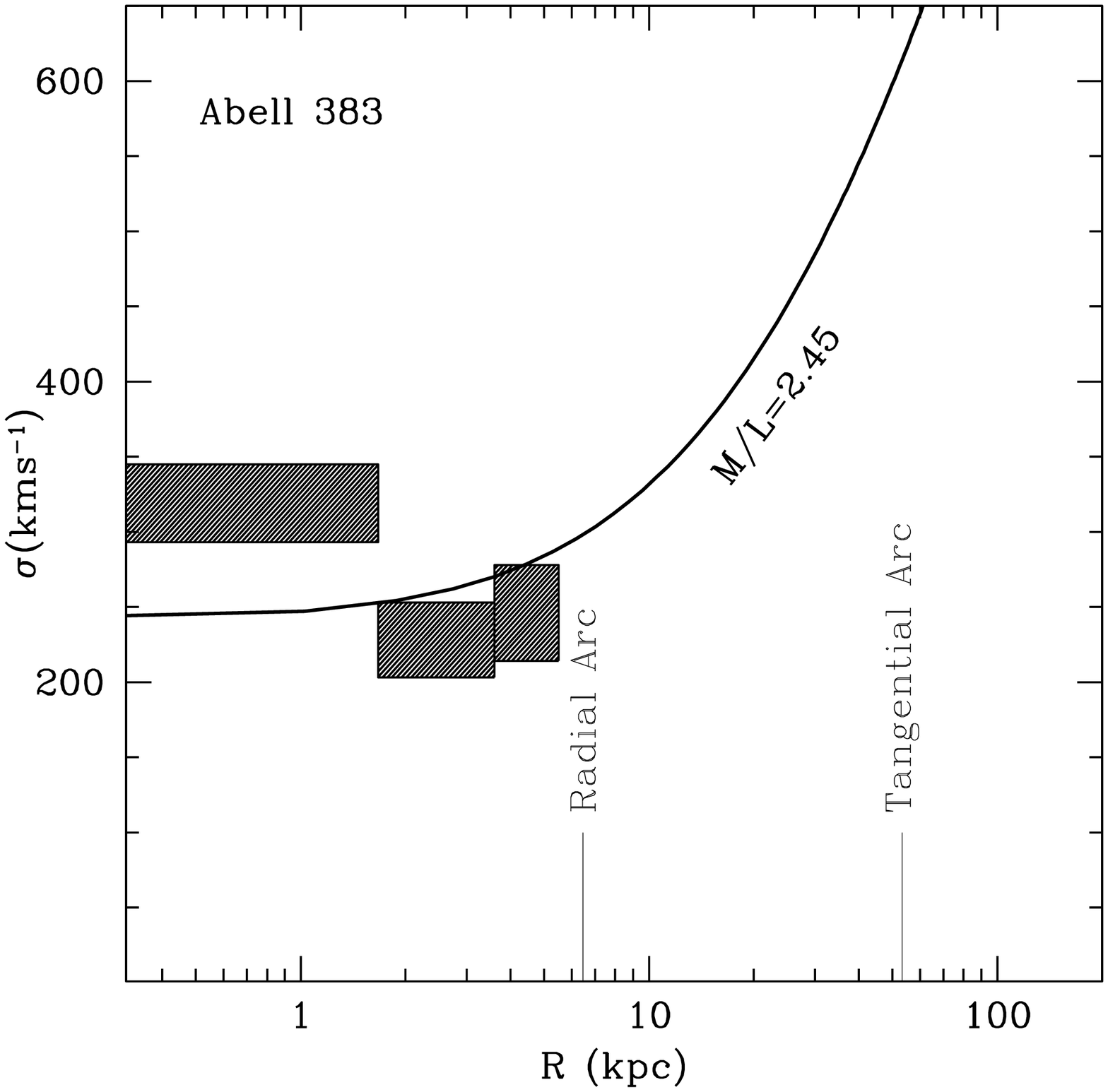}}
\mbox{\epsfysize=5.8cm \epsfbox{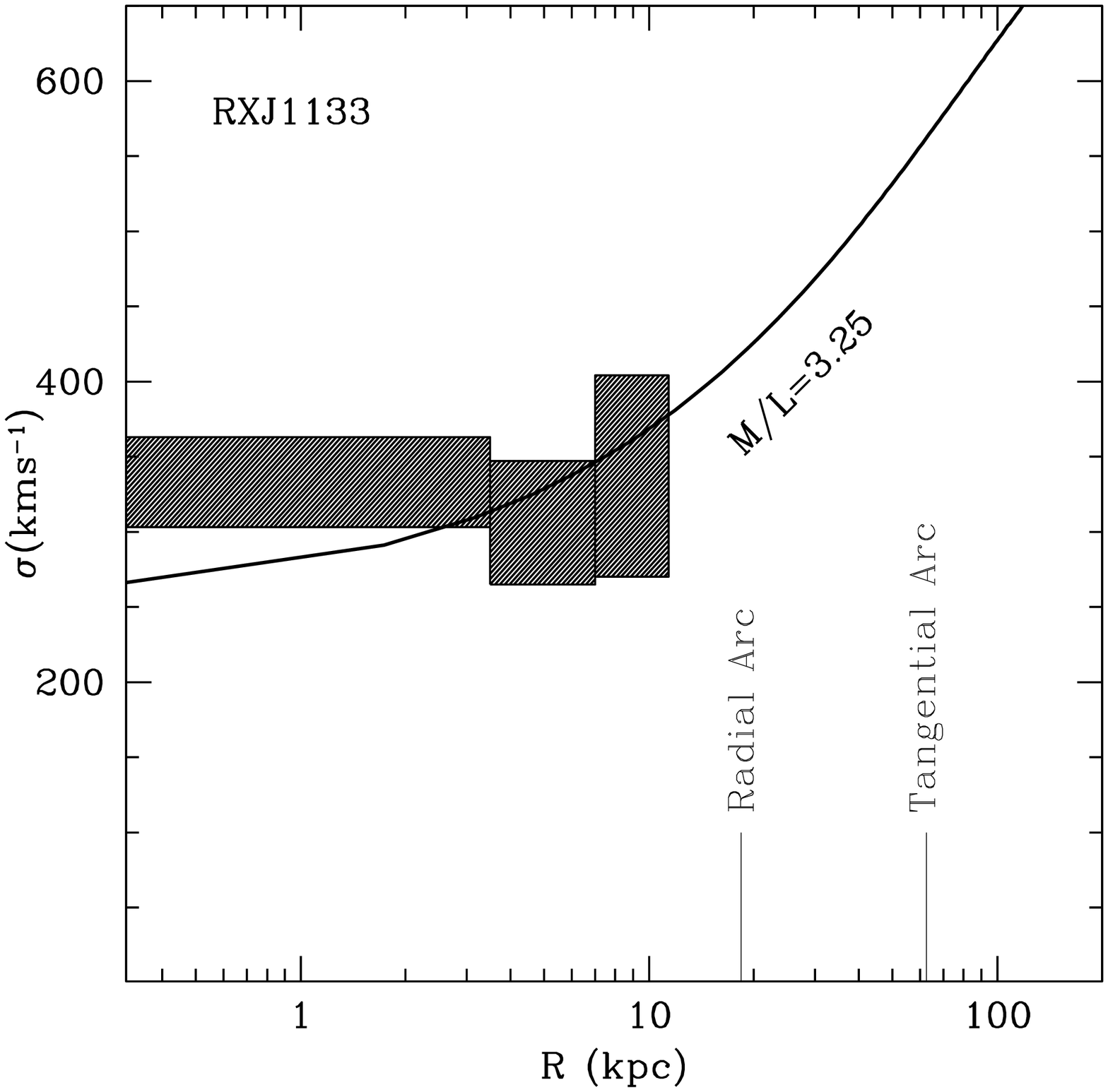}}
}
\mbox{
\mbox{\epsfysize=5.8cm \epsfbox{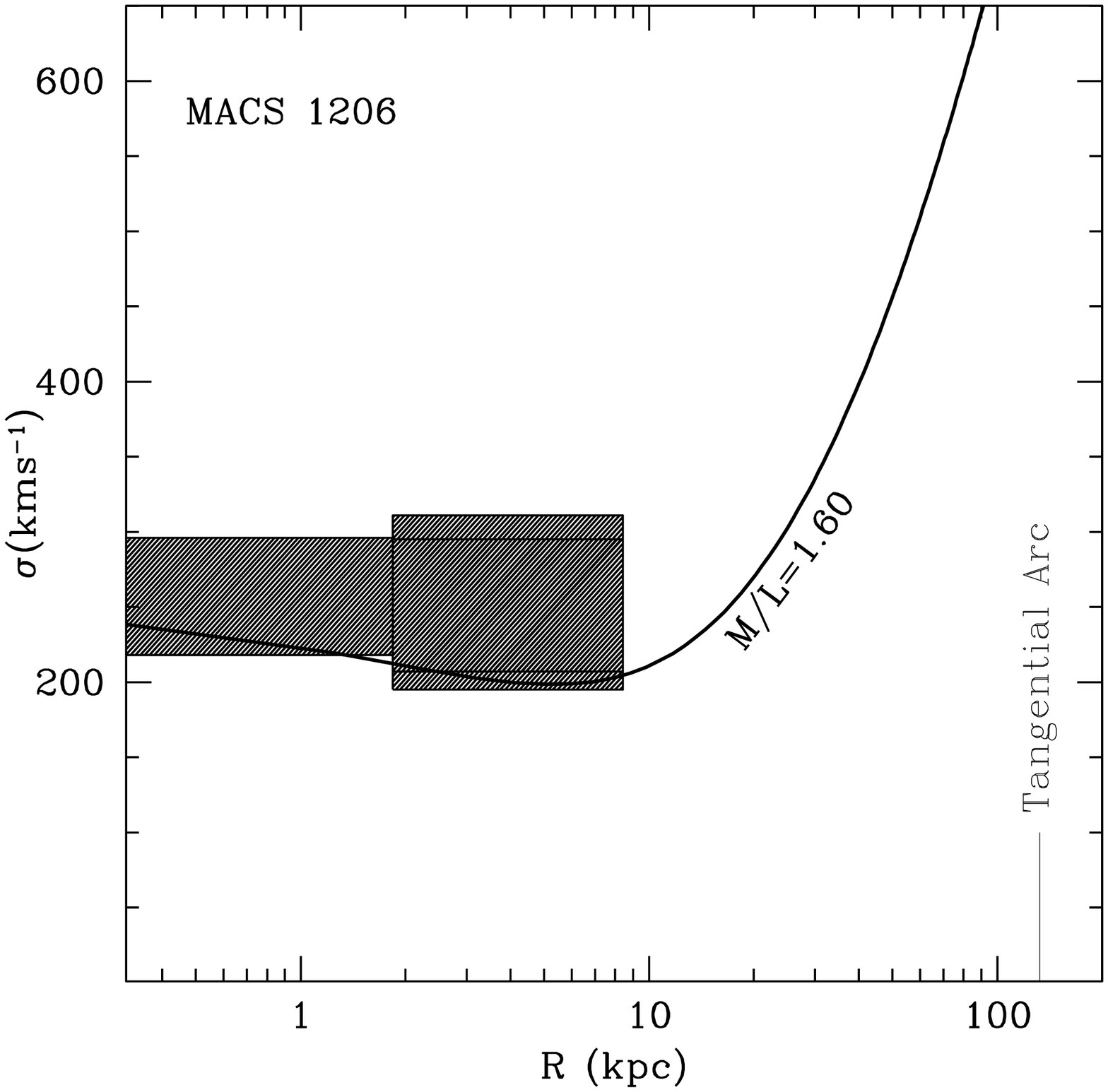}}
\mbox{\epsfysize=5.8cm \epsfbox{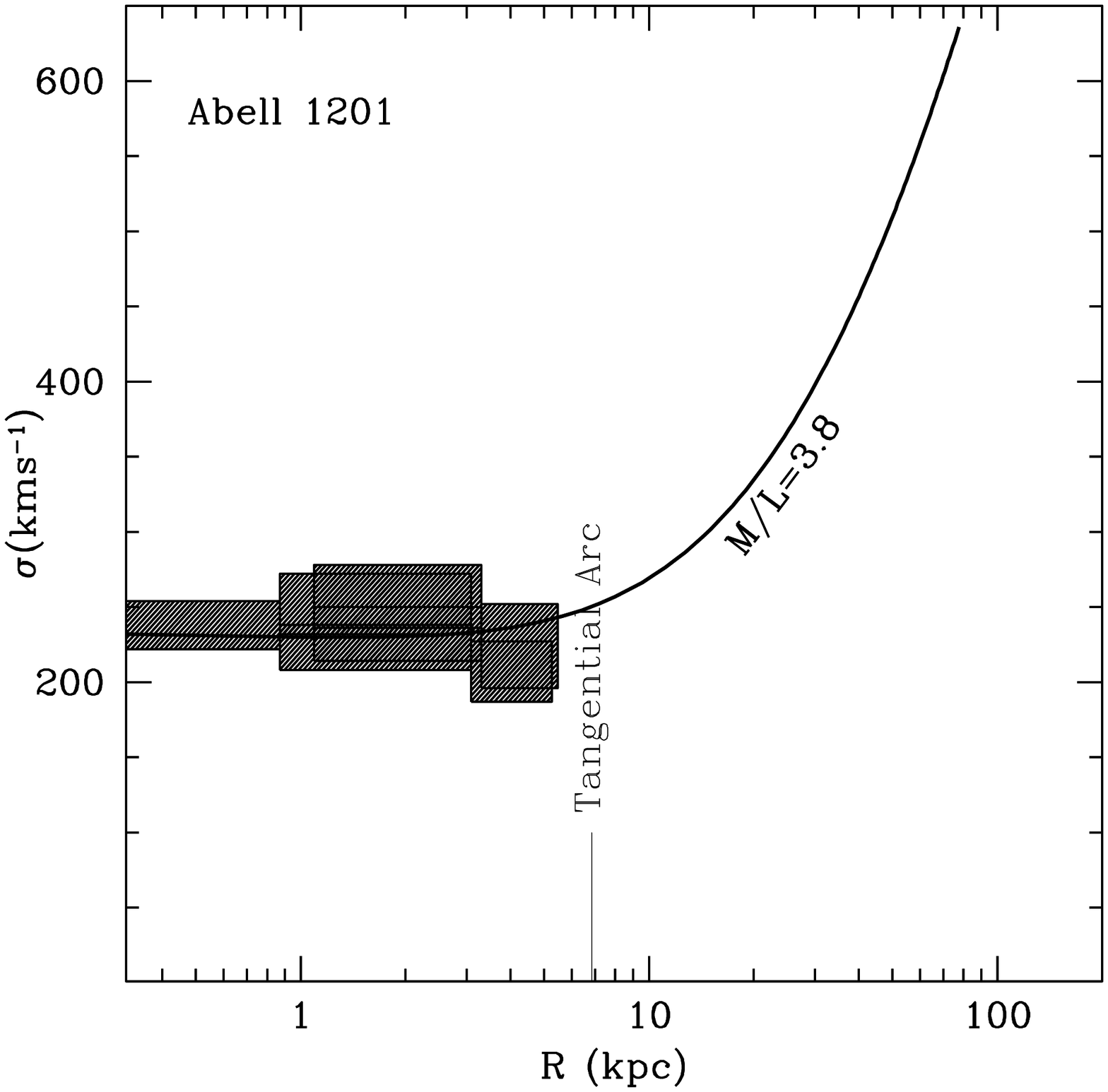}}
\mbox{\epsfysize=5.8cm \epsfbox{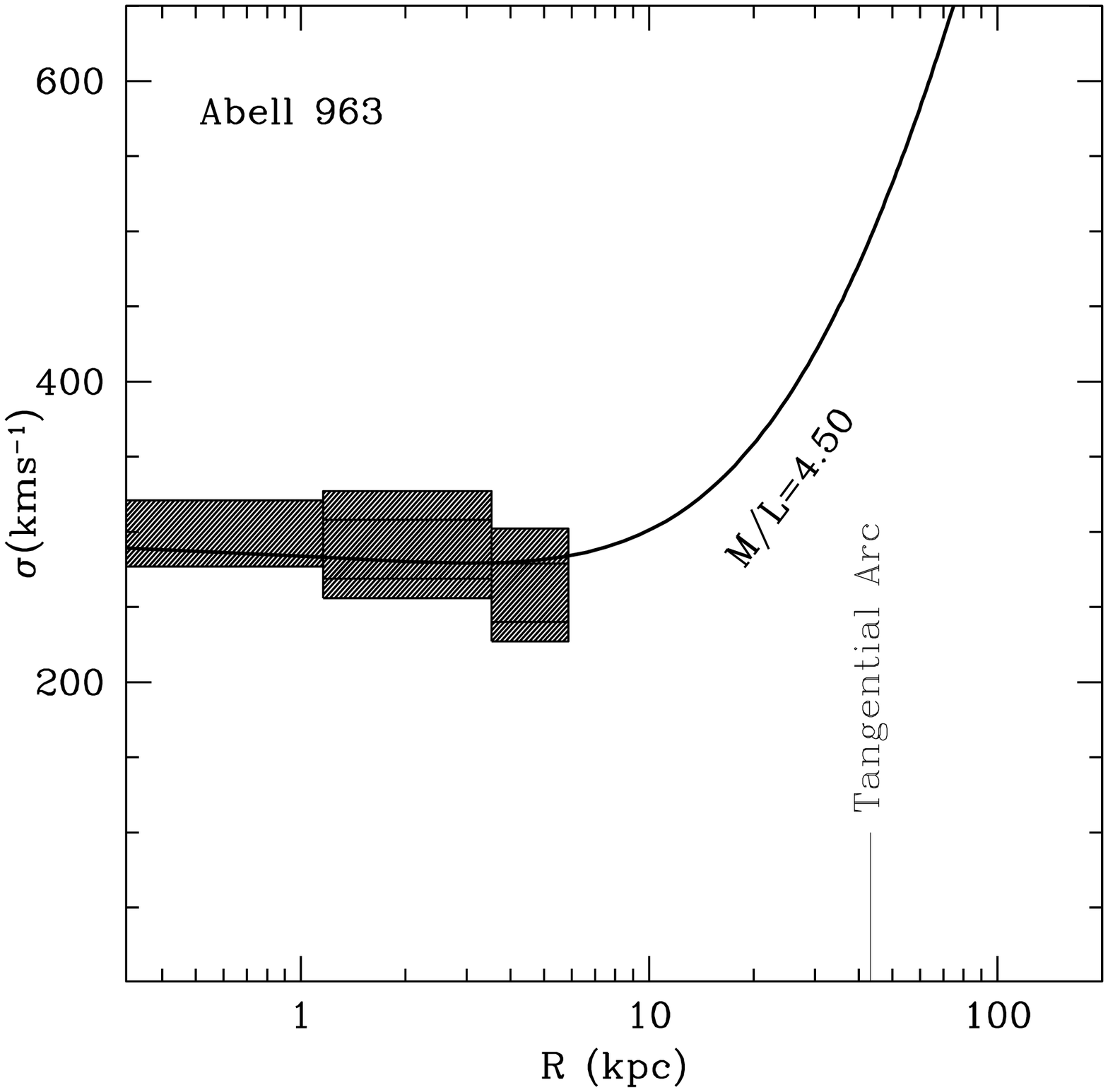}}
}

\caption{ The measured velocity dispersion profile for each BCG
(hatched boxes) along with the best-fitting velocity dispersion
profile calculated from the combined lensing + dynamics analysis
(solid curves).  Note that the solid curves are not exactly equivalent
to those derived from the analysis since they were not binned in
accordance with the slit width, spatial binning of the measurement or
smeared due to the effects of seeing.  The plot of MS2137-23 (top
left) illustrates the power of including the velocity dispersion
profile of the BCG into our analysis.  In this panel we have also
shown a velocity dispersion profile from a mass model that is
compatible with the lensing analysis of that cluster, but does not fit
the velocity dispersion profile ($\beta$=1.30).  }
\end{center}
\end{figure*}

\clearpage
\begin{figure*}[t]   \label{fig:densplot}
\begin{center}

\mbox{
\mbox{\epsfysize=5.8cm \epsfbox{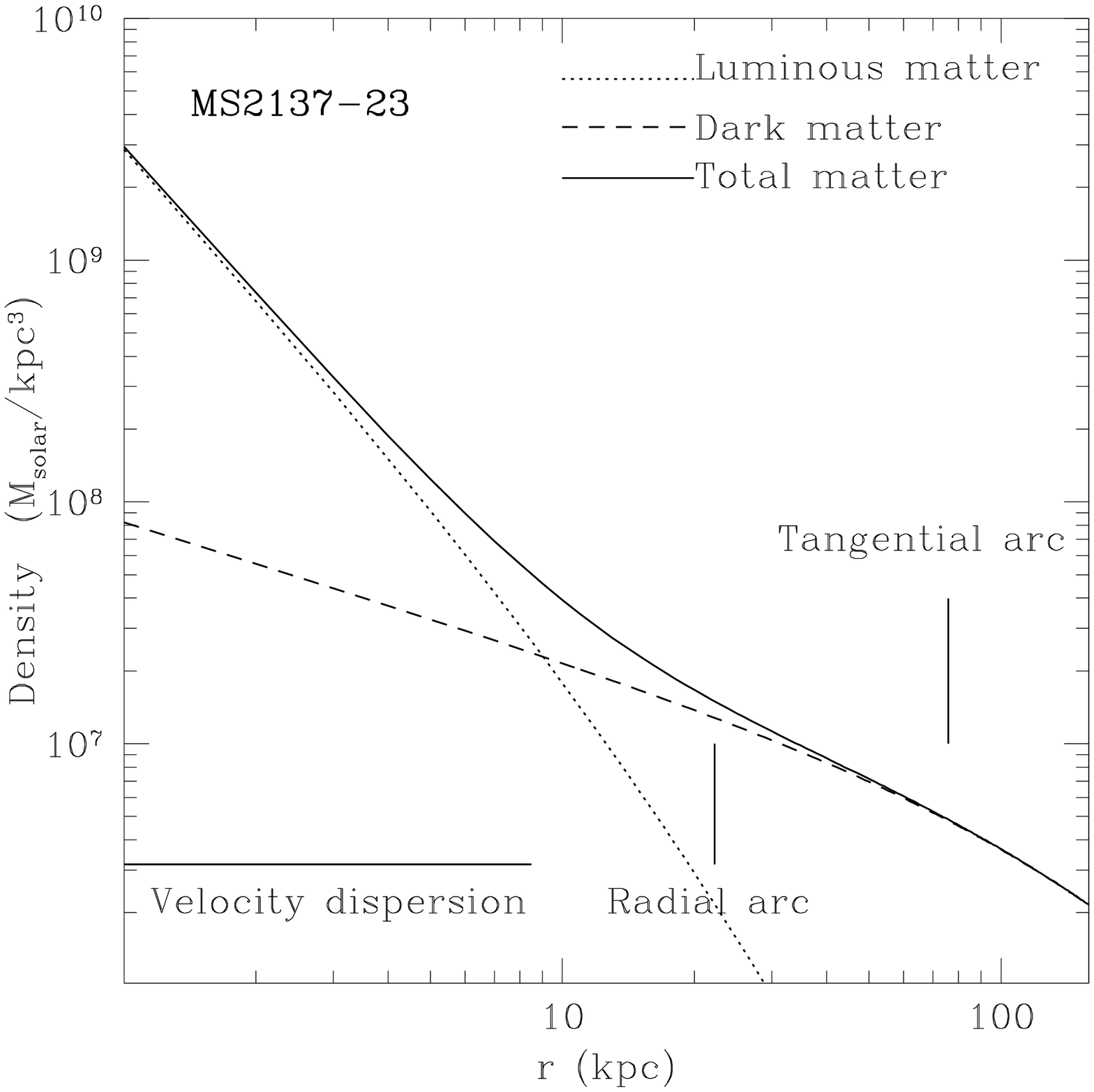}}
\mbox{\epsfysize=5.8cm \epsfbox{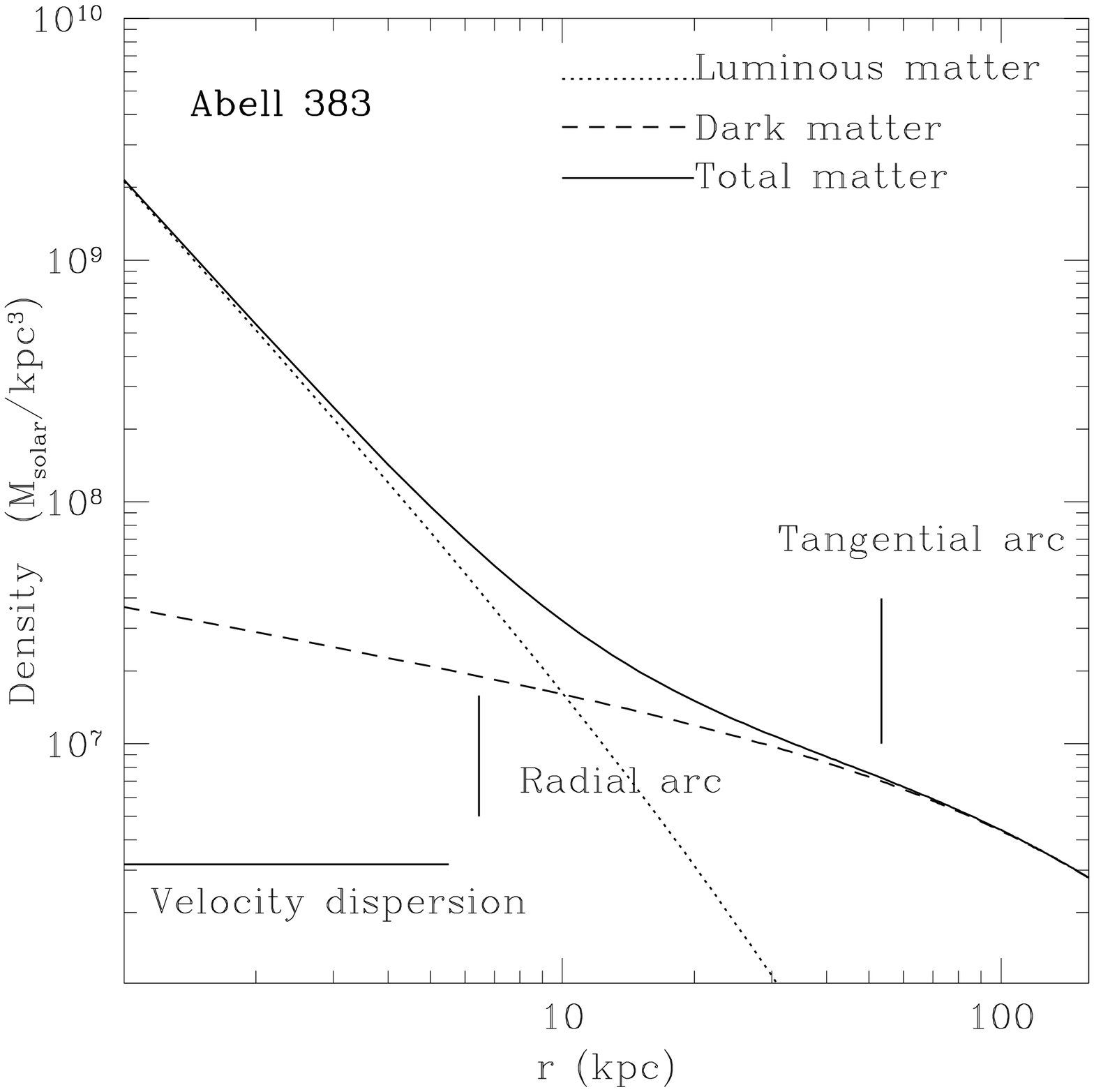}}
\mbox{\epsfysize=5.8cm \epsfbox{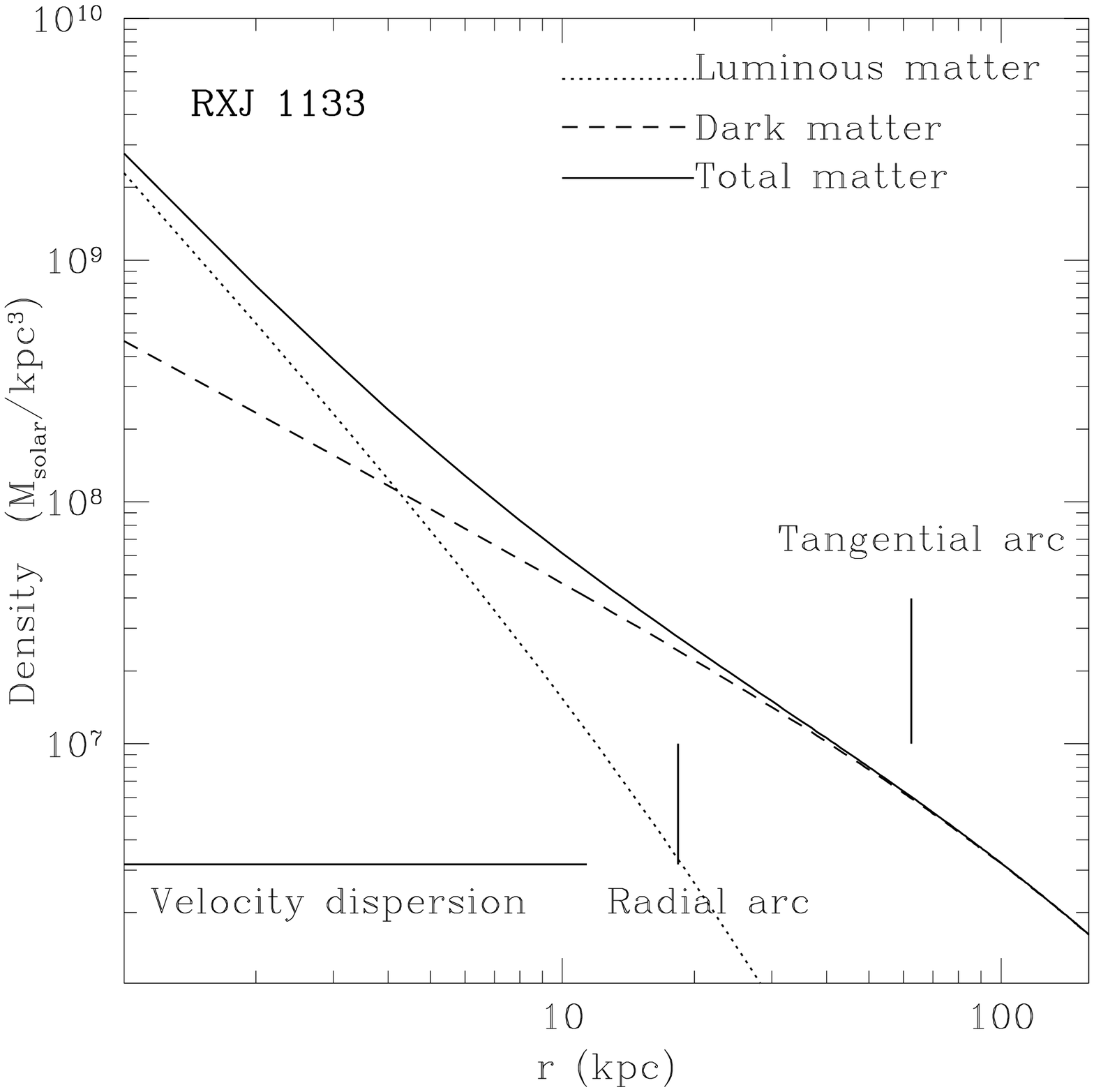}}
}
\mbox{
\mbox{\epsfysize=5.8cm \epsfbox{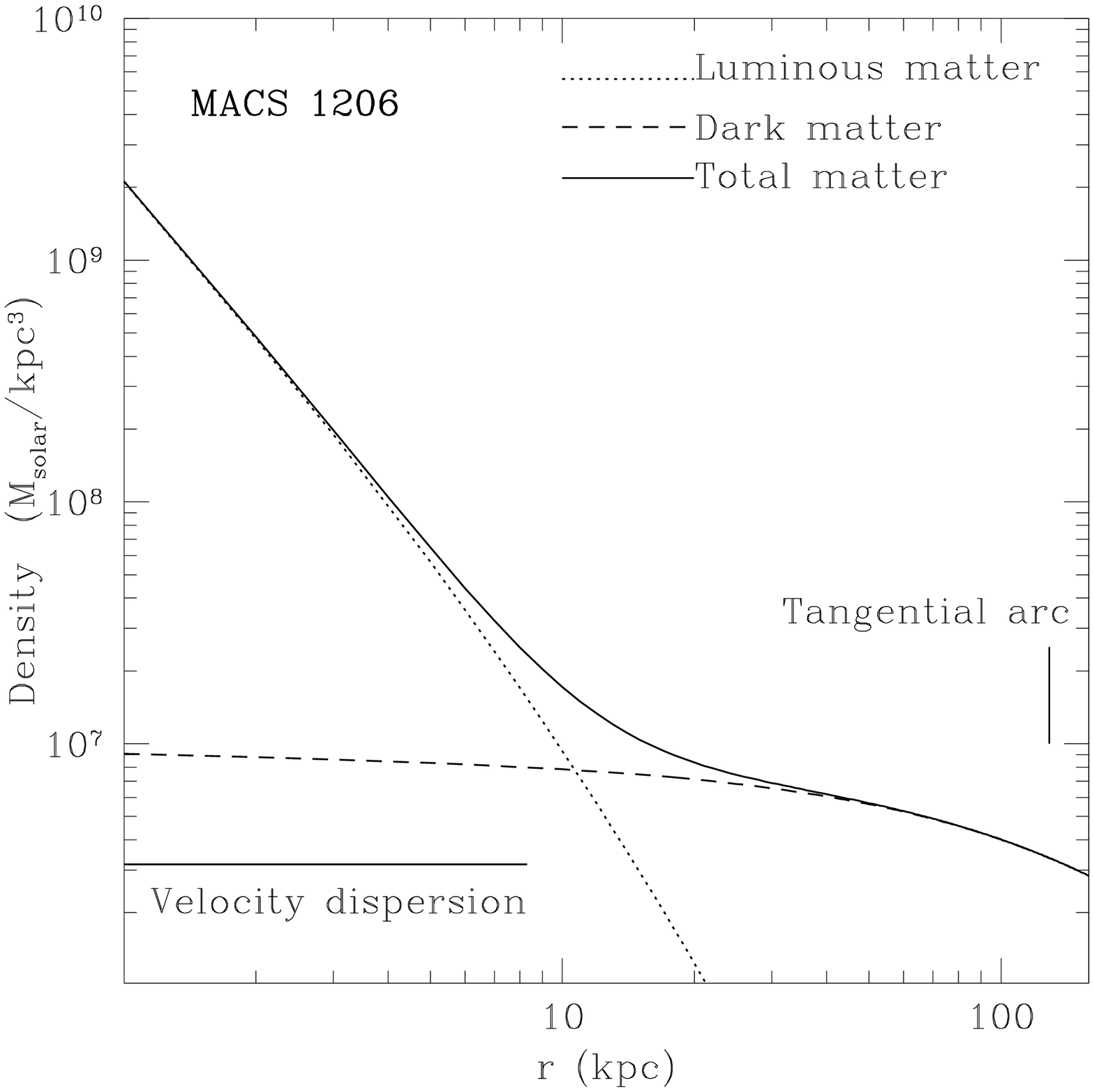}}
\mbox{\epsfysize=5.8cm \epsfbox{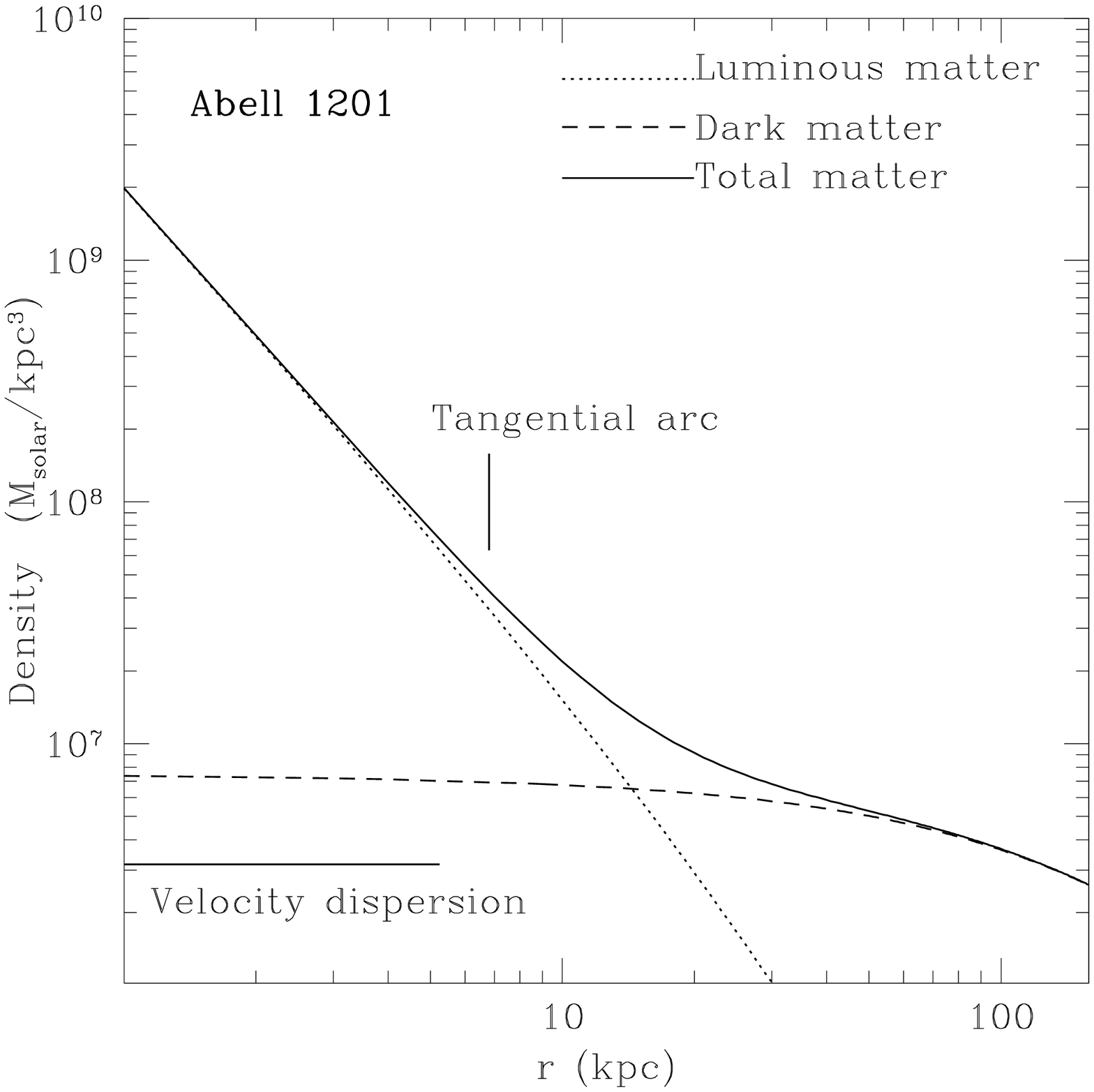}}
\mbox{\epsfysize=5.8cm \epsfbox{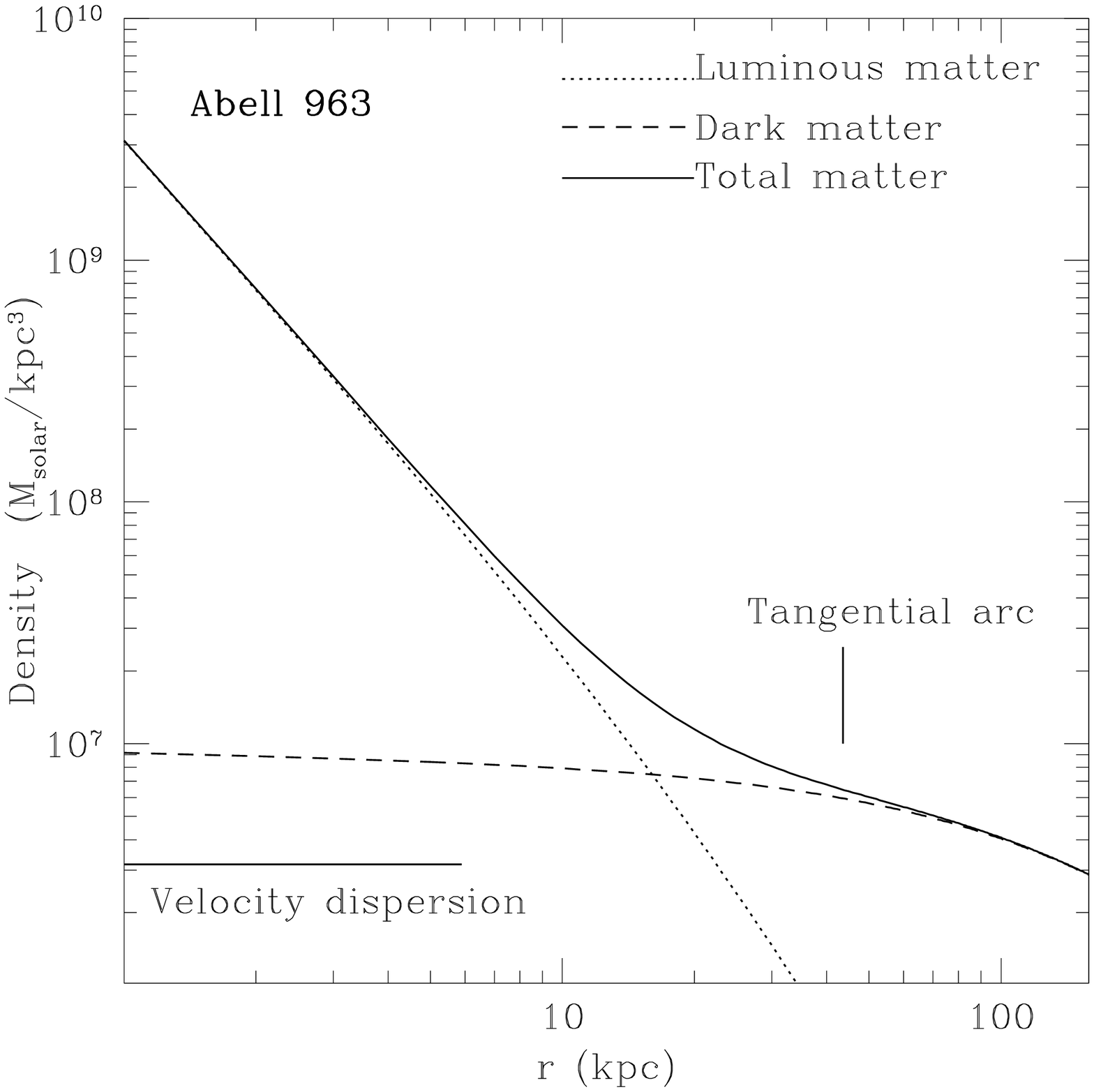}}
}

\caption{Best-fitting total density profile for the entire sample.
The positions of the gravitational arcs and the range over which we
were able to measure the velocity dispersion profile are noted.
Within $\lesssim$ 10 kpc the total density distribution is dominated
by the BCG. }
\end{center}
\end{figure*}

\clearpage

\begin{inlinefigure}
\begin{center}
\resizebox{\textwidth}{!}{\includegraphics{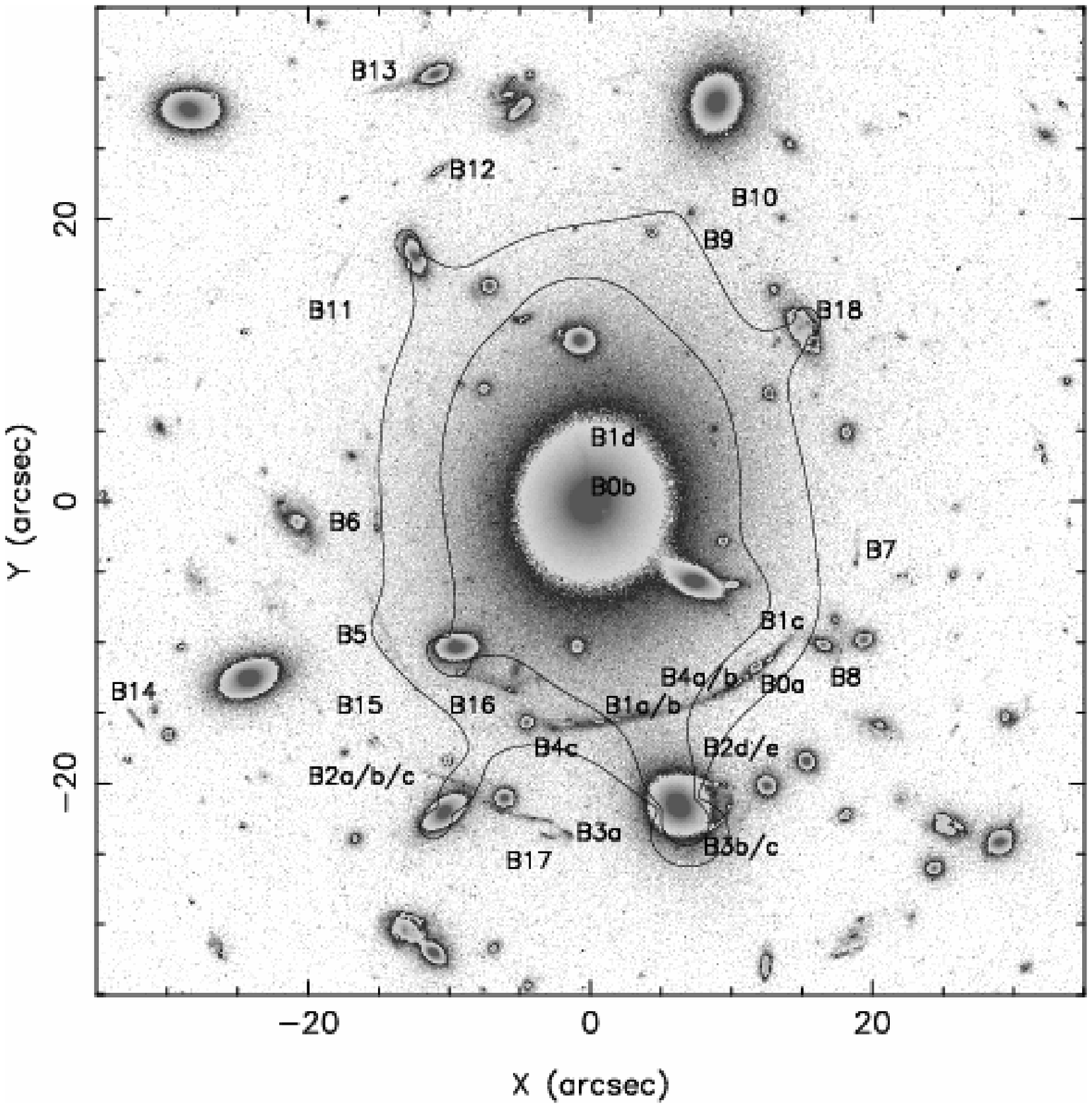}}
\end{center}
\figcaption{The central region of Abell 383 as seen with HST WFPC2.
Overlaid are the $z=1$ and $z=3$ tangential critical lines
calculated from the {\sc lenstool} analysis using the updated S01
model.  The alphanumeric labels are identical to those in Fig. 1 of
S01 and identify several of the lensing and cluster galaxy components
used to construct the model.
\label{fig:lenstool}}
\end{inlinefigure}

\begin{inlinefigure}
\begin{center}
\rotate \resizebox{\textwidth}{!}{\includegraphics{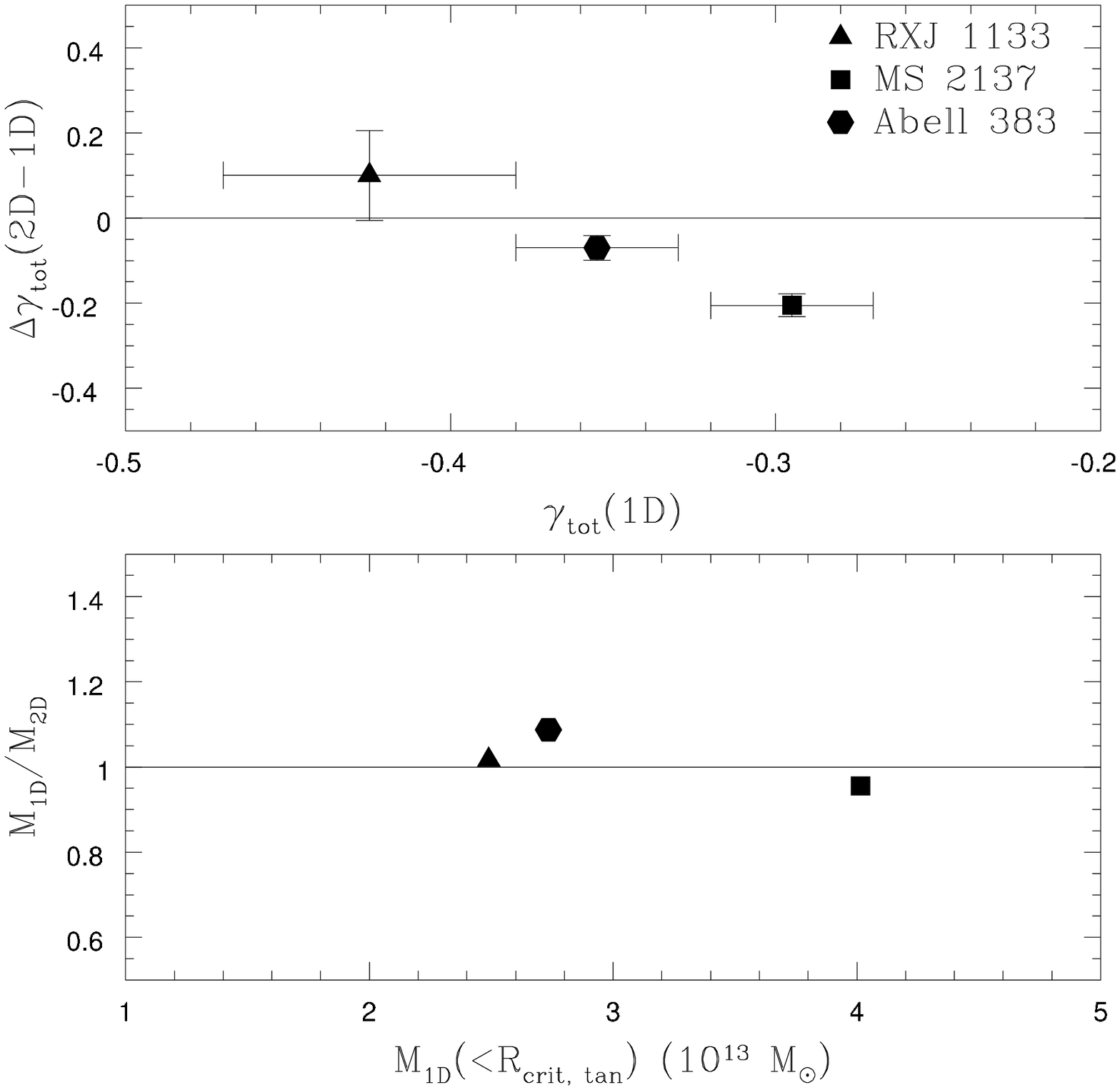}}
\end{center}
\figcaption{Comparison between the 1D models and the 2D check
performed with the {\sc lenstool} software package. We use $\gamma$ to
parameterize the logarithmic slope of the surface density profile.  At
the radial critical line, $\gamma$ should be identical for the two
methods.  (Top panel) The difference in the logarithmic slope between
the two methods versus the 1D logarithmic slope.  The most discrepant
cluster, MS2137-23, would at most effect the DM halo by
$\Delta\beta\sim$0.2 in a direction further away from that predicted
by simulations.  (Bottom panel) Ratio of the mass enclosed at the
tangential critical line, $M(<R=R_{\rm tangential})$, for the two
methods versus the mass enclosed for the 1D method.  There are no
deviations greater than $\sim$8\%.  The uncertainties in a given data
point are approximately the same size as the points themselves.
\label{fig:compare}}
\end{inlinefigure}

\clearpage

\begin{inlinefigure}
\begin{center}
\resizebox{\textwidth}{!}{\includegraphics{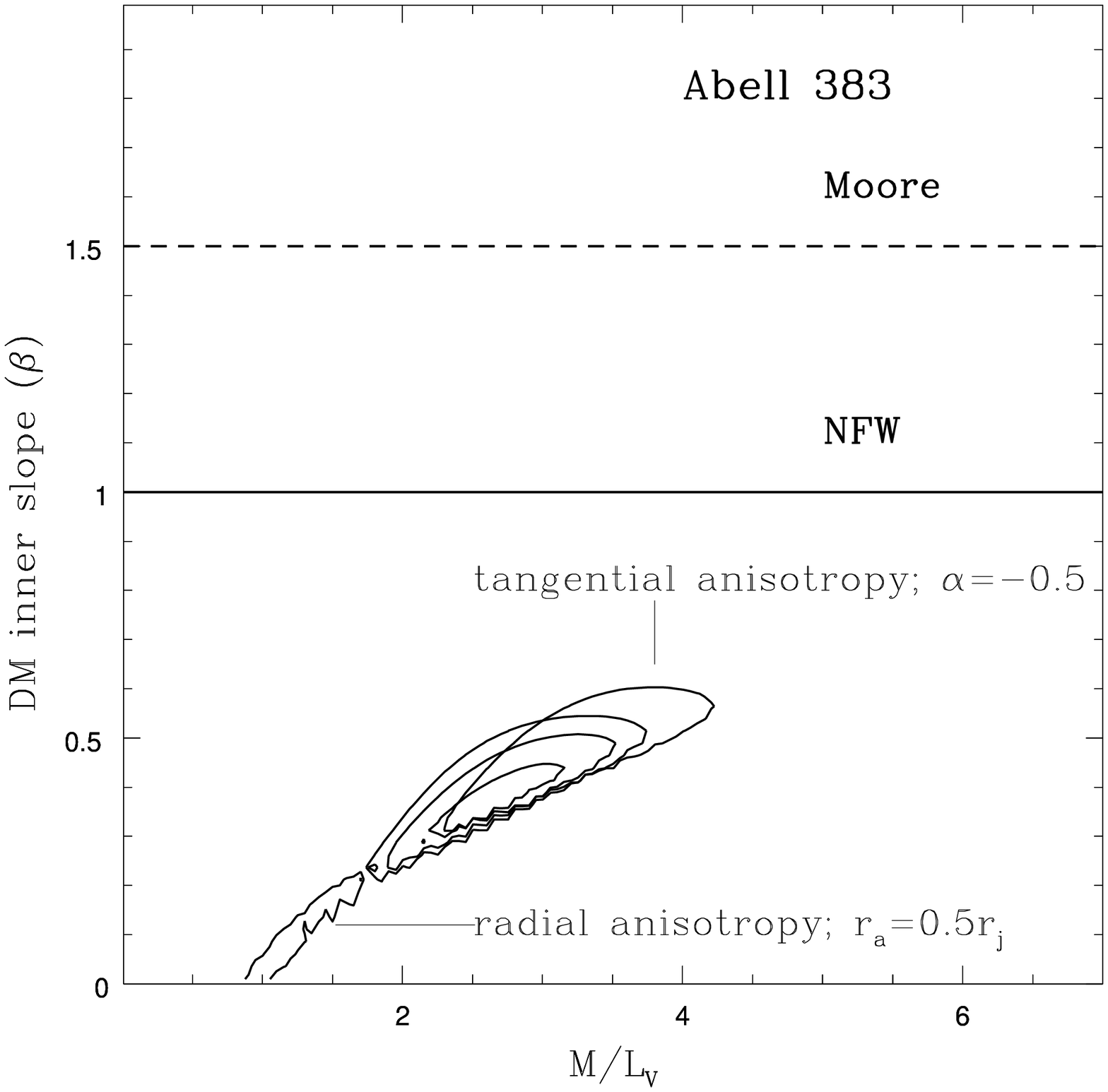}}
\end{center}
\figcaption{The 68\%, 95\% and 99\% confidence contours of Abell 383
along with the 95\% confidence contours for the orbital anisotropy
tests in \S~6.2 that were most discrepant with our original results.
No test causes a shift in the $\beta$ direction greater than
$\Delta\beta\sim$0.2.
\label{fig:systests}}
\end{inlinefigure}

\end{document}